\documentclass[11pt]{article}

\usepackage[preprint]{acl}

\usepackage{times}
\usepackage{latexsym}
\usepackage{amsmath} 
\usepackage{amssymb}
\usepackage{amsthm}
\usepackage{booktabs}
\usepackage{multirow}
\usepackage{makecell}
\usepackage{array}
\usepackage{tabularx}
\usepackage{float}
\usepackage[most]{tcolorbox}
\usepackage{algorithm}
\usepackage{algpseudocode}

\usepackage{xcolor}

\usepackage[T1]{fontenc}

\usepackage[utf8]{inputenc}

\usepackage{microtype}

\usepackage{inconsolata}

\usepackage{graphicx}

\usepackage[table]{xcolor}
\usepackage{arydshln}
\usepackage{pifont}
\usepackage[normalem]{ulem}
\usepackage{enumitem}
\usepackage{listings}

\theoremstyle{definition}
\newtheorem{definition}{Definition}
\theoremstyle{plain}

\definecolor{tabhead}{HTML}{F3E7D3}   
\definecolor{tabstripe}{HTML}{FAF6EF} 
\definecolor{tabours}{HTML}{DFF3E6}   
\definecolor{sigmablue}{HTML}{EAF2FF}
\definecolor{sigmablueframe}{HTML}{4F81BD}
\definecolor{sigmagray}{HTML}{F7F7F7}
\definecolor{sigmagreen}{HTML}{EAF2FF}
\definecolor{skillbg}{HTML}{F7FAFF}
\definecolor{skillframe}{HTML}{6B8FC9}
\definecolor{skilltitle}{HTML}{2E7D32}
\definecolor{skillkey}{HTML}{2B5C9E}
\definecolor{skillval}{HTML}{374151}
\definecolor{skilltag}{HTML}{EEF4FF}
\definecolor{mailbg}{HTML}{F7FAFF}
\definecolor{mailframe}{HTML}{6B8FC9}
\definecolor{mailkey}{HTML}{2B5C9E}
\definecolor{mailval}{HTML}{374151}
\definecolor{mailtag}{HTML}{EEF4FF}
\definecolor{plbg}{HTML}{F7FAFF}
\definecolor{plframe}{HTML}{6B8FC9}
\definecolor{plkey}{HTML}{2B5C9E}
\definecolor{plval}{HTML}{374151}
\definecolor{pltag}{HTML}{EEF4FF}
\definecolor{exbg}{HTML}{F7F7F7}
\definecolor{exframe}{HTML}{EAF2FF}
\definecolor{excodebg}{HTML}{FFFDF6}
\definecolor{promptbg}{HTML}{F8FAFC}
\definecolor{promptframe}{HTML}{6B8FC9}
\definecolor{systembg}{HTML}{EAF2FF}
\definecolor{userbg}{HTML}{F7F7F7}
\definecolor{promptkey}{HTML}{2B5C9E}

\definecolor{caseframe}{HTML}{AEBBD0}

\definecolor{caseblue}{HTML}{EFF6FF}
\definecolor{caseblueframe}{HTML}{A9BEDA}

\definecolor{casegreen}{HTML}{EEF7F1}
\definecolor{casegreenframe}{HTML}{9FC8AE}

\definecolor{casepink}{HTML}{FFF1EC}
\definecolor{casepinkframe}{HTML}{D8A99D}

\definecolor{casepurple}{HTML}{F4F0FA}
\definecolor{casepurpleframe}{HTML}{B8A7D6}

\definecolor{casegray}{HTML}{F7F7F7}
\definecolor{casegrayframe}{HTML}{C7C7C7}

\definecolor{agent0bg}{HTML}{EEF7F1}   
\definecolor{agent0frame}{HTML}{9FC8AE}

\definecolor{agent1bg}{HTML}{EEF5FF}   
\definecolor{agent1frame}{HTML}{9BB6D8}

\definecolor{agent2bg}{HTML}{FFF1EC}   
\definecolor{agent2frame}{HTML}{D8A99D}

\definecolor{agent3bg}{HTML}{F3F0FA}   
\definecolor{agent3frame}{HTML}{B8A7D6}

\definecolor{agent4bg}{HTML}{FFF7E8}   
\definecolor{agent4frame}{HTML}{D8BC7C}

\definecolor{finalbg}{HTML}{F1F0F6}    
\definecolor{finalframe}{HTML}{A9A2BF}

\definecolor{querybg}{HTML}{EAF7F7}      
\definecolor{queryframe}{HTML}{7FB8B8}

\definecolor{tracebg}{HTML}{FFF4E3}      
\definecolor{traceframe}{HTML}{D6A45F}

\definecolor{avatarbg}{HTML}{F7F8FA}
\definecolor{avatarframe}{HTML}{B8C7E6}
\definecolor{avatartext}{HTML}{2F3A4A}

\newcommand{\caseblock}[1]{%
  \vspace{0.45em}
  \noindent\textbf{\large #1}\par
  \vspace{0.25em}
}

\newcommand{\casecardsplit}{%
  \par\vspace{0.35em}\noindent\rule{\linewidth}{0.45pt}\vspace{0.35em}
}

\newcolumntype{L}{>{\raggedright\arraybackslash\bfseries}p{0.27\linewidth}}
\newcolumntype{R}{>{\raggedright\arraybackslash}X}

\newtcolorbox{normalcasebox}[1]{
  enhanced,
  breakable,
  colback=gray!4,
  colframe=gray!55,
  coltitle=black,
  fonttitle=\bfseries,
  title={#1},
  boxrule=0.6pt,
  arc=1.5mm,
  left=1.5mm,
  right=1.5mm,
  top=1.2mm,
  bottom=1.2mm,
  before skip=0.8em,
  after skip=0.8em
}

\newcommand{\caseavatarword}[1]{%
  \tcbox[
    on line,
    enhanced,
    colback=avatarbg,
    colframe=avatarframe,
    coltext=avatartext,
    boxrule=0.5pt,
    arc=2.2mm,
    left=1.6mm,
    right=1.6mm,
    top=0.9mm,
    bottom=0.9mm,
    width=6.8em,
    height=3.0em,
    valign=center,
    halign=center,
    fontupper=\small\bfseries
  ]{#1}%
}

\newcommand{\caseavatar}[1]{%
  \tcbox[
    on line,
    enhanced,
    colback=avatarbg!85!white,
    colframe=avatarframe!30,
    boxrule=0pt,
    borderline={0.35pt}{0pt}{avatarframe!30,dash pattern=on 1.5pt off 1.5pt},
    arc=2.2mm,
    left=1.2mm,
    right=1.2mm,
    top=0.8mm,
    bottom=0.8mm,
    width=6.8em,
    height=3.0em,
    valign=center,
    halign=center
  ]{%
    \includegraphics[
      width=7.2em,
      height=3.2em,
      keepaspectratio
    ]{#1}%
  }%
}

\newtcolorbox{casefigurebox}[1]{%
  enhanced,
  breakable,
  colback=white,
  colframe=caseframe!70,
  boxrule=0.45pt,
  arc=2mm,
  left=4pt,
  right=4pt,
  top=4pt,
  bottom=4pt,
  title={\textbf{#1}},
  fonttitle=\small\bfseries,
  coltitle=black,
  colbacktitle=casegray,
  boxed title style={
    arc=1.5mm,
    boxrule=0.35pt,
    colframe=caseframe!60,
    colback=casegray
  },
  before skip=0.8em,
  after skip=0.6em
}

\newcommand{\casecaption}[2]{%
  \refstepcounter{figure}\label{#2}%
  \noindent{\small\textbf{Figure~\thefigure:} #1}\par\vspace{1.2em}%
}

\newcommand{\caseentryword}[5]{%
  \noindent
  \hspace*{0.13\linewidth}%
  \begin{minipage}[t]{0.845\linewidth}
    \begin{tcolorbox}[
      enhanced,
      colback=#2!92!white,
      colframe=#3!45,
      boxrule=0.3pt,
      arc=2.2mm,
      left=5pt,
      right=5pt,
      top=4.5pt,
      bottom=4.5pt,
      drop shadow={black!6!white},
      borderline west={1.2pt}{0pt}{#3!65},
      overlay={
        \node[
          anchor=north east,
          inner sep=0pt
        ] at ([xshift=-4pt,yshift=0pt]frame.north west)
        {\caseavatarword{#1}};
      }
    ]
    {\small\bfseries #4}\par\vspace{2pt}
    {\footnotesize\setlength{\parskip}{0.25em} #5}
    \end{tcolorbox}
  \end{minipage}
  \par\vspace{1.2mm}%
}

\newcommand{\caseentry}[5]{%
  \noindent
  \hspace*{0.13\linewidth}%
  \begin{minipage}[t]{0.845\linewidth}
    \begin{tcolorbox}[
      enhanced,
      colback=#2!92!white,
      colframe=#3!45,
      boxrule=0.3pt,
      arc=2.2mm,
      left=5pt,
      right=5pt,
      top=4.5pt,
      bottom=4.5pt,
      drop shadow={black!6!white},
      borderline west={1.2pt}{0pt}{#3!65},
      overlay={
        \node[
          anchor=north east,
          inner sep=0pt
        ] at ([xshift=-4pt,yshift=0pt]frame.north west)
        {\caseavatar{#1}};
      }
    ]
    {\small\bfseries #4}\par\vspace{2pt}
    {\footnotesize\setlength{\parskip}{0.25em} #5}
    \end{tcolorbox}
  \end{minipage}
  \par\vspace{1.2mm}%
}

\lstdefinestyle{pseudolabel}{
  basicstyle=\footnotesize\ttfamily,
  backgroundcolor=\color{excodebg},
  breaklines=true,
  columns=fullflexible,
  keepspaces=true,
  showstringspaces=false,
  frame=none
}

\newcolumntype{Y}{>{\raggedright\arraybackslash}X}
\newcommand{\accesssym}[1]{\makebox[0.95em][c]{#1}}
\newcommand{\cmark}{\textcolor{green!60!black}{\scalebox{1.35}{\ding{51}}}}
\newcommand{\xmark}{\textcolor{red!70!black}{\scalebox{1.35}{\ding{55}}}}
\newcommand{\pmark}{\accesssym{\textcolor{orange!80!black}{\ooalign{\hfil\ding{51}\hfil\cr\hfil\raisebox{-0.08ex}{\scalebox{0.72}{\ding{55}}}\hfil\cr}}}}

\newcommand{\skillkey}[1]{\texttt{\textcolor{skillkey}{#1}}}
\newcommand{\skillval}[1]{\texttt{\textcolor{skillval}{#1}}}
\newcommand{\skilltag}[1]{%
  \tcbox[
    on line,
    boxsep=0.5pt,
    left=2pt,
    right=2pt,
    top=1pt,
    bottom=1pt,
    colback=skilltag,
    colframe=blue!18,
    boxrule=0.3pt,
    arc=1mm
  ]{\scriptsize\texttt{#1}}%
}

\newcommand{\mailkey}[1]{\texttt{\textcolor{mailkey}{#1}}}
\newcommand{\mailval}[1]{\texttt{\textcolor{mailval}{#1}}}
\newcommand{\mailtag}[1]{%
  \tcbox[
    on line,
    boxsep=0.5pt,
    left=2pt,
    right=2pt,
    top=1pt,
    bottom=1pt,
    colback=mailtag,
    colframe=blue!18,
    boxrule=0.3pt,
    arc=1mm
  ]{\scriptsize\texttt{#1}}%
}

\newcommand{\plkey}[1]{\texttt{\textcolor{plkey}{#1}}}
\newcommand{\plval}[1]{\texttt{\textcolor{plval}{#1}}}
\newcommand{\pltag}[1]{%
  \tcbox[
    on line,
    boxsep=0.5pt,
    left=2pt,
    right=2pt,
    top=1pt,
    bottom=1pt,
    colback=pltag,
    colframe=plframe!35,
    boxrule=0.3pt,
    arc=1mm
  ]{\scriptsize\texttt{#1}}%
}

\newcommand{\promptkey}[1]{\texttt{\textcolor{promptkey}{#1}}}
\newcommand{\pkey}[1]{\promptkey{#1}}
\newcommand{\promptplaceholder}[1]{\texttt{\{#1\}}}

\newlength{\metriccellw}
\newlength{\scorepartw}
\newlength{\deltapartw}
\newcommand{\gain}[2]{%
  \makebox[\metriccellw][c]{%
    \rule[-0.55ex]{0pt}{2.25ex}%
    \makebox[\scorepartw][r]{#1}%
    \makebox[\deltapartw][l]{%
      \raisebox{-0.45ex}[0pt][0pt]{%
        \tiny\textcolor{red!80!black}{$\uparrow$#2}%
      }%
    }%
  }%
}

\newcommand{\dropgain}[2]{%
  \makebox[\metriccellw][c]{%
    \rule[-0.55ex]{0pt}{2.25ex}%
    \makebox[\scorepartw][r]{#1}%
    \makebox[\deltapartw][l]{%
      \raisebox{-0.45ex}[0pt][0pt]{%
        \tiny\textcolor{cyan!70!black}{$\downarrow$#2}%
      }%
    }%
  }%
}

\newcommand{\best}[1]{%
  \begingroup
  \setlength{\fboxsep}{0.7pt}%
  \colorbox{blue!32}{\textbf{\strut #1}}%
  \endgroup
}

\newcommand{\second}[1]{%
  \begingroup
  \setlength{\fboxsep}{0.7pt}%
  \colorbox{blue!12}{\textcolor{black}{\textbf{\strut #1}}}%
  \endgroup
}
\title{\textsc{Sigma}: Skill-Incidence Graphs for Compositional Multi-Agent Design}

\author{
  \textbf{Kun Zeng\textsuperscript{$\clubsuit$}\thanks{Equal contribution}},
  \textbf{Yu Huo\textsuperscript{$\spadesuit$}\footnotemark[1]},
  \textbf{Siyu Zhang\textsuperscript{$\spadesuit$}},
  \textbf{Yuecheng Zhuo\textsuperscript{$\diamondsuit$}},\\
  \textbf{Yuquan Lu\textsuperscript{$\clubsuit$}},
  \textbf{Haoyue Liu\textsuperscript{$\spadesuit$}},
  \textbf{Siyue Chen\textsuperscript{$\heartsuit$}},
  \textbf{Xiaoying Tang\textsuperscript{$\spadesuit$}\thanks{Corresponding author}}
  \\
  \textsuperscript{$\spadesuit$}School of Science and Engineering, The Chinese University of Hong Kong, Shenzhen\\
  \textsuperscript{$\clubsuit$}Sun Yat-sen University\quad
  \textsuperscript{$\heartsuit$}South China University of Technology\quad
  \textsuperscript{$\diamondsuit$}Taiyuan University of Technology\\
}

\begin{document}
\maketitle 

\begin{abstract}
Existing graph-based multi-agent system (MAS) designers mainly improve collaboration by optimizing communication topologies over predefined agents, roles, or groups. 
However, because each node remains a closed-set entity, these methods struggle to generalize to tasks that require unseen combinations of capabilities. 
We propose \textsc{Sigma}, a skill-incidence graph framework that constructs agents as task-conditioned bundles of reusable skills. 
Given a task and a skill library, \textsc{Sigma} predicts a skill-agent incidence matrix, composes agent node embeddings from selected skills, and decodes a communication topology over the constructed agents. 
During execution, skill-specific mailboxes route messages to the relevant assigned capabilities, making the incidence structure directly operational. 
Across six reasoning and coding benchmarks with three base LLMs, \textsc{Sigma} achieves the best average performance and improves over \textsc{CARD}, the strongest non-compositional topology-based baseline, by $2.06$, $2.36$, and $1.75$ points, respectively.
It also shows stronger robustness to unseen skill libraries, with an average performance drop of only $0.96$ points. 
These results suggest that compositional node construction is a complementary and important axis for multi-agent design beyond communication topology optimization. Code is available at \url{https://anonymous.4open.science/r/SIGMA-2338/}.
\end{abstract}

\begin{figure}[t]
    \centering
    \includegraphics[width=\linewidth]{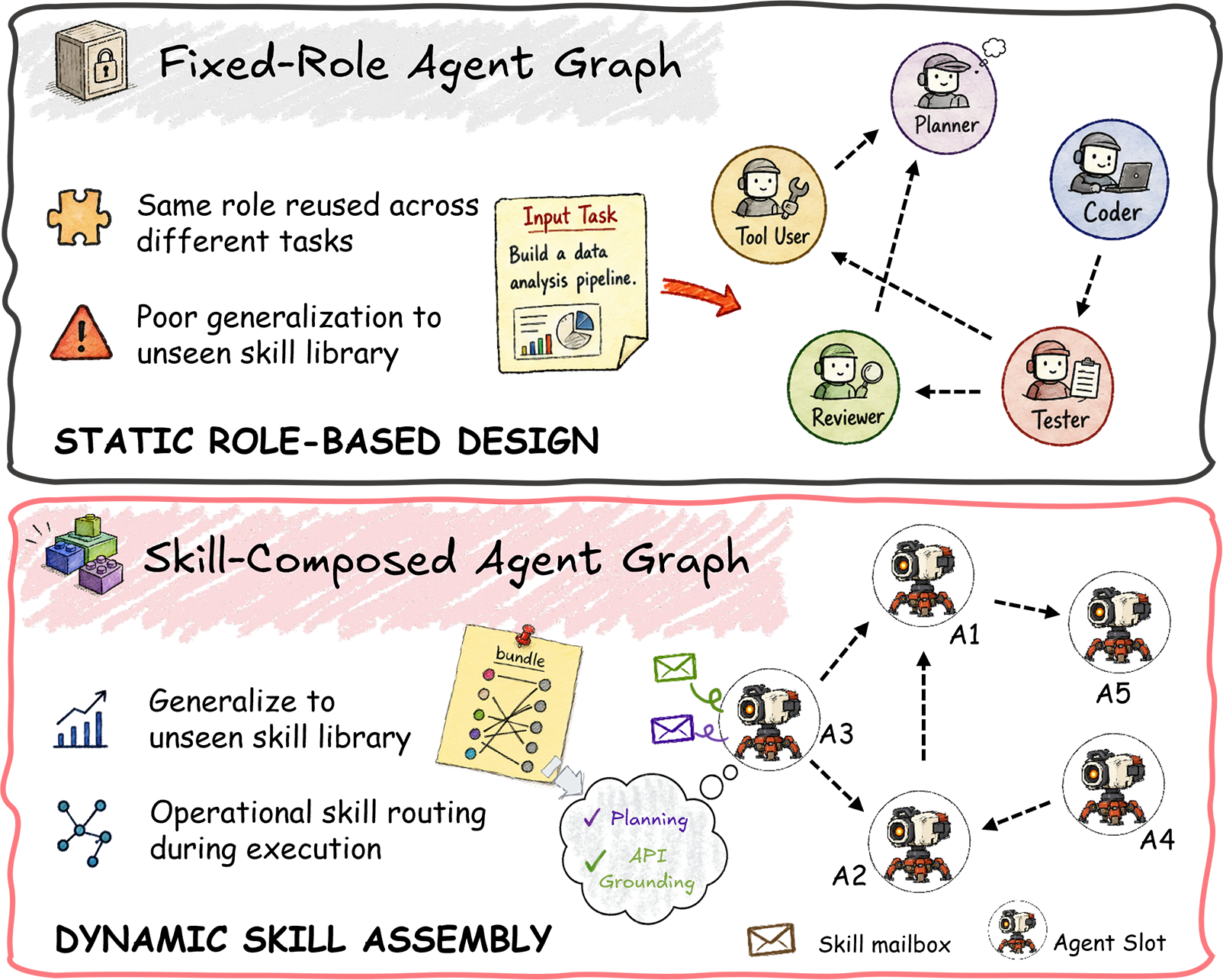}
    \caption{Comparison between fixed-role and skill-composed multi-agent graph design.}
    \label{fig:comparison}
\end{figure}

\section{Introduction}
\label{sec:intro}
Large language model based multi-agent systems (\textbf{MAS}) solve complex tasks by assigning agents specialized roles and communication structures~\citep{li2023camel,wu2024autogen,hong2024metagpt,qian2024chatdev,chen2024agentverse}. 
Recent graph-based designers further automate this process by learning task-adaptive topologies over agent nodes~\citep{liu2024dynamic,zhuge2024gptswarm,zhang2025gdesigner,zhang2025cut}.
Yet most of them still regard each node as a closed-set role, operation, or predefined group: they optimize how existing agents communicate, while leaving what each agent is able to do largely fixed.

This node-level rigidity limits compositional generalization.
Real tasks often require capabilities that cross role boundaries, including tool use and intermediate actions~\citep{karpas2022mrkl,yao2022react}, API grounding and structured tool access~\citep{li2023api,patil2023gorilla,qin2024toolllm}, and reusable procedural skills~\citep{wang2023voyager}.
Such capabilities are not always aligned with a single role. 
Although fixed roles are useful abstractions, they do not explicitly construct new agent identities from reusable capabilities.
Consequently, a topology designer can learn a better graph over predefined agents, but it cannot make the same slot become a retrieval-grounded implementer for one task and a verification-heavy reasoner for another.

In light of this dilemma, we propose the \textbf{LLM-based Skill-Incidence Multi-Agent Protocol (SI-MAP)}, which provides guidance for compositional LLM-MA graph design:

\noindent\fbox{%
    \parbox{0.93\linewidth}{%
       \textbf{Skill-Incidence Multi-Agent Protocol (SI-MAP)}: \textit{Given a task $q$ and skill library $\mathcal{L}$, a compositional LLM-MA designer should satisfy: \textbf{(1) Compositionality}, constructing each agent as a task-conditioned skill bundle; \textbf{(2) Executability}, making assigned skills affect prompts, tools, memory, or routing; \textbf{(3) Topology compatibility}, exposing constructed node representations to a graph decoder; and \textbf{(4) Expandability}, allowing new tool/API skill cards at test time without retraining the topology decoder.}
    }\label{box:simap}
}

\begin{figure}[t]
    \centering
    \includegraphics[width=\linewidth]{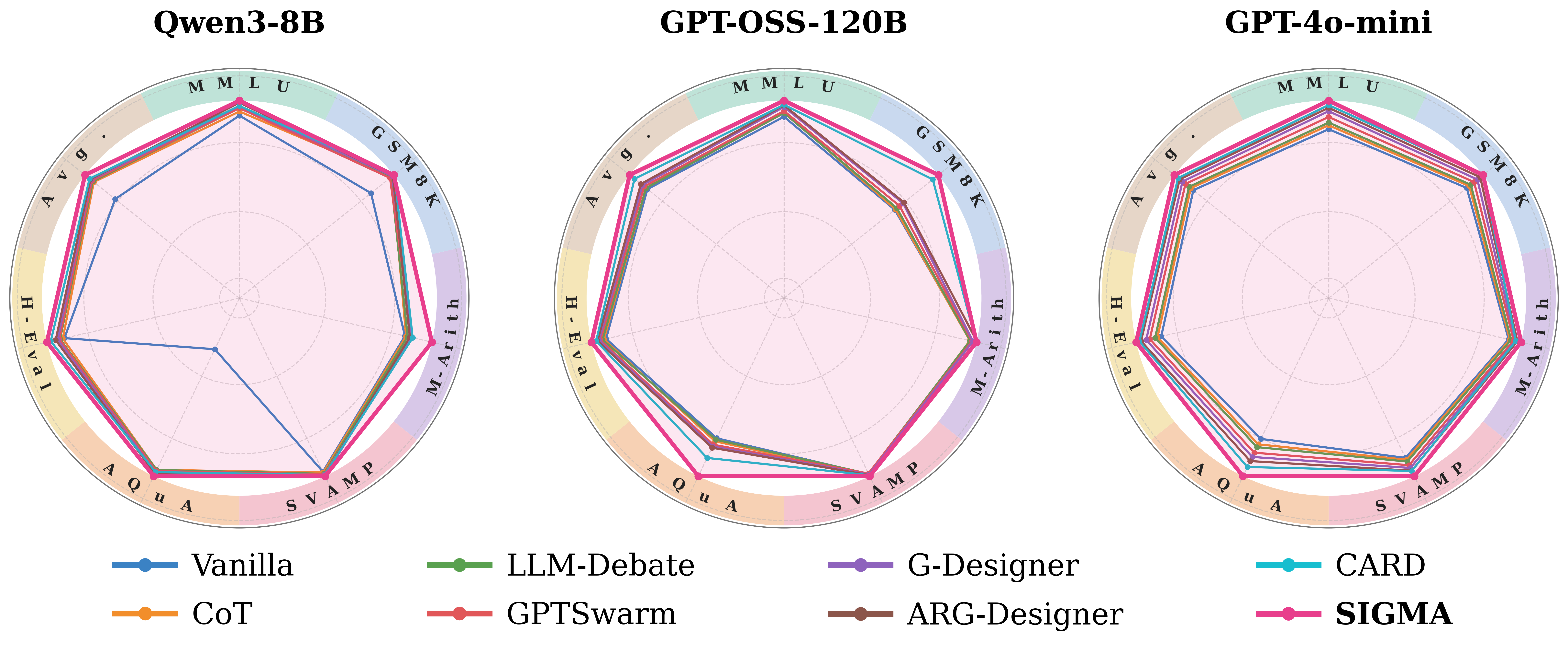}
    \caption{Performance plot across six benchmarks and three base LLMs. }
    \label{fig:radar}
\end{figure}

To instantiate SI-MAP, we introduce \textsc{Sigma}, a compositional and executable LLM-powered multi-agent graph designer. 
\textsc{Sigma} represents reusable capabilities as skill cards, predicts a task-conditioned skill-agent incidence matrix, composes node embeddings from the selected cards, and decodes a communication topology over the constructed agents. 
During execution, the same incidence structure is exposed through skill-specific mailboxes, so skill assignments are operational rather than latent annotations. 
This design deliberately separates two decisions that are coupled in role-based systems: \emph{what} each agent can do is decided by skill incidence, whereas \emph{how} agents exchange information is decided by topology decoding.
The separation lets \textsc{Sigma} keep the matched-crew evaluation protocol of prior graph-based MAS work while testing a different modeling axis, namely whether reusable capabilities can construct more transferable agent nodes.
Figure~\ref{fig:radar} provides a compact preview of our empirical findings.
The formal objects and full algorithms are given in Section~\ref{sec:preliminaries}, Section~\ref{sec:methodology}, and Appendix~\ref{app:method_details}.

Our contribution can be summarized as follows:

\begin{itemize}[leftmargin=*,itemsep=-0.2em,topsep=0.2em]
\item[\ding{182}] \textbf{\textit{Protocol Proposal.}}
We propose SI-MAP, a protocol that shifts graph-based MAS design from connecting fixed agents to constructing executable agent nodes from reusable skills. It formalizes compositionality, executability, topology compatibility, and expandability as requirements for skill-aware multi-agent graph design.

\item[\ding{183}] \textbf{\textit{Practical Solution.}}
We present \textsc{Sigma}, which couples incidence-based node construction, skill-aware topology decoding, and mailbox-based execution under matched multi-agent settings. Unlike methods that only prune or reconnect predefined nodes, \textsc{Sigma} changes the capability composition of each node before topology generation.

\item[\ding{184}] \textbf{\textit{Experimental Validation.}}
We evaluate \textsc{Sigma} against single-agent, debate-based, and topology-based multi-agent baselines under compositional generalization settings.
The evaluation isolates whether skill incidence improves compositional generalization under the same crew size, context budget, and graph-decoding interface. 
As summarized by Figure~\ref{fig:radar}, \textsc{Sigma} achieves consistently strong performance across diverse benchmarks, suggesting that the gains are not tied to a specific backbone or task family.
\end{itemize}

\section{Related Work}

\paragraph{LLM-based multi-agent systems.}
LLM-based multi-agent systems decompose complex tasks into interacting agents with specialized roles, workflows, and communication protocols.
Role-playing frameworks assign agents complementary identities, enabling them to collaborate through instruction following, perspective taking, and iterative communication~\citep{li2023camel,chen2024agentverse}.
Workflow-oriented systems further organize agents into structured pipelines, where different agents are responsible for planning, coding, reviewing, testing, or project management~\citep{wu2024autogen,hong2024metagpt,qian2024chatdev}.
Another work improves reasoning by introducing debate, discussion, or self-reflection among multiple LLM agents, allowing agents to expose errors and refine intermediate answers through interaction~\citep{du2024improving,liang2024encouraging}.
Multi-agent frameworks have also been applied to simulation and decision-making scenarios~\citep{park2023generative}.

\paragraph{Graph-based Multi-agent Design.}
Graph-based MAS methods mainly improve collaboration by adapting the communication topology among predefined agents.
One line of work designs task-adaptive or dynamic connections between agents~\citep{liu2024dynamic,zhuge2024gptswarm}.
Another line treats the graph itself as an optimizable or learnable object, using graph search or graph decoders to produce task-conditioned topologies~\citep{zhang2025gdesigner,wu2026card}.
Recent work further studies conditional topology design, where the communication graph adapts to changing environmental signals such as model capability, tool availability, or knowledge-source variation~\citep{wu2026card}.
Pruning-based methods instead reduce redundant communication by removing unnecessary agents or edges~\citep{zhang2025cut}.
These methods make communication more flexible than hand-written chains or fully connected graphs, but the agent nodes are still largely fixed.

\paragraph{Tool-augmented and skill-based agents.}
Tool-augmented agents extend LLMs with external executable procedures~\citep{karpas2022mrkl,yao2022react,schick2023toolformer,shen2023hugginggpt,li2023api,patil2023gorilla,qin2024toolllm}.
Skill-library agents further show that reusable skills can be stored and adapted across tasks~\citep{wang2023voyager}.

\begin{figure*}[t]
  \includegraphics[width=\linewidth]{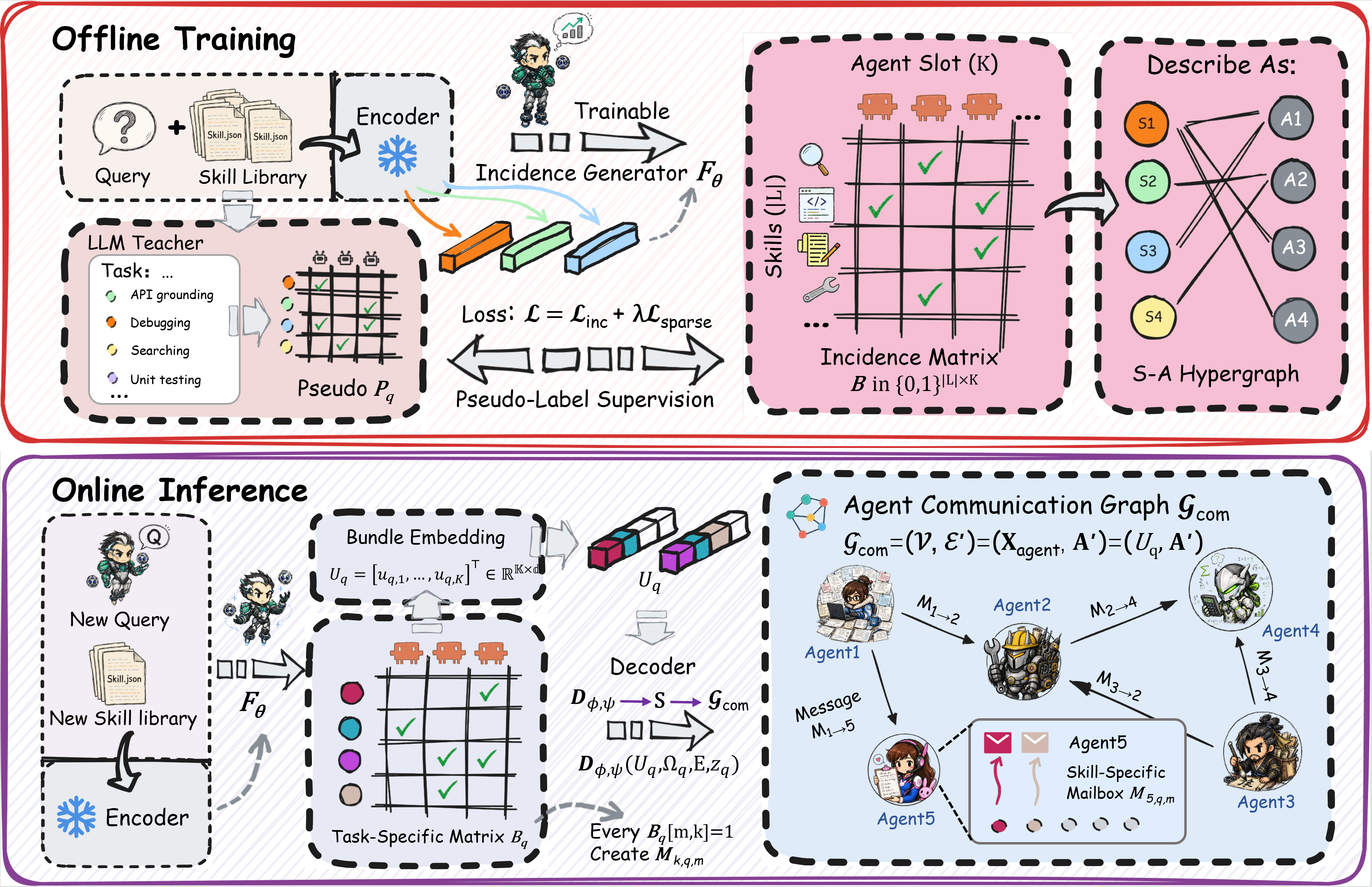}
  \caption{Framework of \textsc{Sigma}. 
Offline, an incidence generator is trained with deterministic pseudo-label supervision to assign reusable skills to agent slots. 
Online, the predicted task-specific incidence matrix is converted into skill-composed agent embeddings, which are used to decode an agent communication graph with skill-specific message routing.}
  \label{fig:framework}
\end{figure*}

\section{Preliminaries}
\label{sec:preliminaries}

\subsection{Graph-Based Multi-Agent Design}
\label{sec:graph_mas}

For a task query $q$, a graph-based MAS with $K$ agent slots is represented as a directed graph
\[
\mathcal{G}_q = (\mathcal{A}, \mathcal{E}_q),
\]
where $\mathcal{A}=\{a_1,\ldots,a_K\}$ denotes agent slots and $\mathcal{E}_q$ denotes directed communication edges.
Its adjacency matrix $\mathbf{A}_q\in\{0,1\}^{K\times K}$ satisfies $\mathbf{A}_q[i,j]=1$ if messages from $a_i$ are passed to $a_j$.
Conventional role-based systems initialize each agent as an atomic entity,
\begin{equation}
a_k = \{\texttt{Base}_k,\texttt{Role}_k,\texttt{State}_k,\texttt{Plugin}_k\},
\label{eq:role_agent}
\end{equation}
where the role/profile and tool configuration determine the node semantics.
This interface supports topology learning, but it leaves the node vocabulary closed-set. 
Given a generated graph $\mathcal{G}_q$, agents communicate for $T$ rounds.
At round $t$, an agent receives the task query and messages produced by its in-neighbors in the previous round:
\begin{equation}
\begin{aligned}
\mathcal{R}_k^{(t)}
&= a_k\left(\mathcal{P}_{k,\mathrm{sys}}^{(t)},
\mathcal{P}_{k,\mathrm{usr}}^{(t)}\right),\\
\mathcal{P}_{k,\mathrm{usr}}^{(t)}
&= \left\{q, \{\mathcal{R}_j^{(t-1)}:
a_j \in \mathcal{N}_{\mathrm{in}}(a_k)\}\right\},
\end{aligned}
\label{eq:agent_response}
\end{equation}
where $\mathcal{N}_{\mathrm{in}}(a_k)$ is the in-neighborhood of $a_k$ and $\mathcal{R}_j^{(0)}$ is initialized as an empty message.
After the final round, an aggregation function returns the system answer:
\begin{equation}
\hat{y}_q \leftarrow
\operatorname{Aggregate}\left(\mathcal{R}_1^{(T)}, \ldots, \mathcal{R}_K^{(T)}\right).
\label{eq:aggregate}
\end{equation}
\textsc{Sigma} keeps this graph-execution interface, summarized in Appendix~\ref{app:graph_execution}, while replacing atomic role nodes with skill-composed nodes.

\subsection{Skill-Agent Incidence}
\label{sec:skill_incidence}

Let $\mathcal{L}=\{s_1,\ldots,s_M\}$ be a library of reusable skills, and let $\mathcal{S}$ denote the corresponding set of skill-card nodes when we form skill-agent bipartite graphs.
\begin{definition}[Skill Card]
\label{def:skill_card}
A skill card is a reusable capability descriptor
\begin{equation}
s_m=(n_m,d_m,\eta_m,\kappa_m,g_m),
\label{eq:skill_card}
\end{equation}
where $n_m$ is the skill name, $d_m$ is its natural-language description, $\eta_m$ is an executable affordance such as a tool, API endpoint, retrieval operation, or reasoning procedure, $\kappa_m$ is an optional input-output contract, and $g_m$ records the grounding source.
Purely cognitive skills are still instantiated as concrete prompt-level procedures or mailbox routing targets, preventing vague role labels from being treated as executable skills.
Each card is embedded by a frozen encoder as $\mathbf{e}_m=\operatorname{Enc}(s_m)\in\mathbb{R}^{d}$.
\end{definition}

For a query $q$, \textsc{Sigma} constructs the $K$ agent slots through a skill-agent incidence matrix
\begin{equation}
\begin{aligned}
\mathbf{B}_q &\in \{0,1\}^{M \times K},\\
\mathbf{B}_q[m,k]&=1
\Longleftrightarrow s_m \mapsto a_k .
\end{aligned}
\label{eq:incidence_matrix}
\end{equation}
The $k$-th column defines the executable skill bundle
\begin{equation}
\mathcal{L}_{q,k}
= \{s_m \in \mathcal{L} \mid \mathbf{B}_q[m,k]=1\}.
\label{eq:skill_bundle}
\end{equation}
Thus, $a_k$ is no longer a single role token; it is a task-conditioned bundle of selected capabilities.

\begin{definition}[Skill-Incidence MAS]
\label{def:skill_incidence_mas}
Given a query $q$, a skill library $\mathcal{L}$, and $K$ agent slots $\mathcal{A}=\{a_1,\ldots,a_K\}$, a skill-incidence multi-agent system is
\begin{equation}
\mathcal{X}_q =
\left(\mathcal{L}, \mathcal{A}, \mathbf{B}_q,
\mathbf{U}_q, \mathcal{G}_q, \mathcal{M}_q\right),
\label{eq:system_object}
\end{equation}
where $\mathbf{B}_q$ is the executable skill-agent incidence matrix, $\mathbf{U}_q$ contains bundle embeddings used by the graph decoder, $\mathcal{G}_q$ is the decoded communication graph, and $\mathcal{M}_q$ is the set of skill-specific mailboxes.
A deployed skill-incidence MAS requires that each nonempty bundle is grounded in skill cards, each node embedding is computed from the selected cards, and every incident pair $(m,k)$ owns an execution-time mailbox.
\end{definition}

This separates capability composition from topology generation: $\mathbf{B}_q$ determines what each agent can do, while $\mathcal{G}_q$ determines how the constructed agents communicate.
The definition describes the discrete system used at inference and execution. During training, \textsc{Sigma} optimizes a continuous relaxation whose hard projection yields $\mathbf{B}_q$; this preserves differentiability without changing the executable object evaluated at test time.

\section{Methodology}
\label{sec:methodology}

\subsection{Overview}
\label{sec:overview}

Figure~\ref{fig:framework} illustrates the overall framework of \textsc{Sigma}.
\textsc{Sigma} treats reusable skills as first-class design objects.
Given a query $q$ and a skill library $\mathcal{L}$, it first predicts a skill-agent incidence structure:
\begin{equation}
\begin{aligned}
(\mathbf{z}_q,\mathbf{E}) &= \operatorname{SkillEncode}(q,\mathcal{L}),\\
(\mathbf{P}_q,\mathbf{B}_q) &= F_\theta(\mathbf{z}_q,\mathbf{E}),\\
\mathcal{H}_q &= (\mathcal{S},\mathcal{A},\mathbf{B}_q).
\end{aligned}
\label{eq:pipeline_incidence}
\end{equation}
It then constructs skill-composed agents, decodes their communication topology, and executes the resulting graph:
\begin{equation}
\begin{aligned}
(\mathbf{U}_q,\boldsymbol{\Omega}_q)
&= \operatorname{Construct}(\mathcal{H}_q,\mathbf{z}_q,\mathbf{E}),\\
\mathcal{G}_q
&= D_{\phi,\psi}(\mathbf{U}_q,\boldsymbol{\Omega}_q,\mathbf{E},\mathbf{z}_q),\\
\hat{y}_q
&= \operatorname{Execute}(\mathcal{G}_q,\mathbf{B}_q,q).
\end{aligned}
\label{eq:pipeline_decode_execute}
\end{equation}
Here $\mathbf{P}_q$ is the soft incidence matrix used for learning, $\mathbf{B}_q$ is the hard executable incidence matrix, $\mathcal{H}_q$ is the induced skill-agent bipartite graph, and $\boldsymbol{\Omega}_q$ stores skill-attention weights.
The learnable components are the incidence generator $F_\theta$ and the skill-aware topology decoder $D_{\phi,\psi}$.
The full decoding algorithm and implementation details are provided in Appendix~\ref{app:method_details}.

\subsection{Incidence Generation and Agent Construction}
\label{sec:incidence_generation}

Each skill card is serialized using the fields in Definition~\ref{def:skill_card} and encoded together with the query by a frozen text encoder:
\begin{equation}
\begin{aligned}
\mathbf{z}_q&=\operatorname{Enc}(q),\\
\mathbf{e}_m&=\operatorname{Enc}(s_m),\\
\mathbf{E}&=[\mathbf{e}_1,\ldots,\mathbf{e}_M]^\top.
\end{aligned}
\label{eq:skill_encoding}
\end{equation}
The frozen encoder lets new tool/API cards enter at test time by extending $\mathcal{L}$, rather than changing the model architecture.

For each skill $s_m$ and slot $a_k$, the incidence generator computes
\begin{equation}
\alpha_{m,k}
= F_\theta([\mathbf{z}_q,\mathbf{e}_m,\mathbf{p}_k,
\mathbf{z}_q\odot\mathbf{e}_m]),
\label{eq:skill_logits}
\end{equation}
where $\mathbf{p}_k$ is a learned embedding for the $k$-th canonical agent slot and $\odot$ denotes element-wise product.
The soft incidence probability is
\begin{equation}
\mathbf{P}_q[m,k]=\sigma(\alpha_{m,k}).
\label{eq:soft_incidence}
\end{equation}
For notational brevity, we write $p_{m,k}=\mathbf{P}_q[m,k]$ below.
Soft incidence provides supervision and a differentiable training path, while the hard matrix $\mathbf{B}_q$ determines executable bundles.
We obtain $\mathbf{B}_q$ through a deterministic sparse projection.
For each slot $k$, \textsc{Sigma} first forms the thresholded candidate set
\begin{equation}
\mathcal{C}_{q,k}
= \{m:\sigma(\alpha_{m,k})>\tau_{\mathrm{skill}}\}.
\label{eq:candidate_set}
\end{equation}
It then selects at most $k_s$ skills, with an argmax fallback when no skill exceeds the threshold:
\begin{equation}
\mathcal{I}_{q,k}
=
\begin{cases}
\operatorname{Top}_{k_s}(\mathcal{C}_{q,k};\alpha_{\cdot,k}),
& \mathcal{C}_{q,k}\neq \emptyset,\\
\{\arg\max_m \alpha_{m,k}\},
& \text{otherwise}.
\end{cases}
\label{eq:incidence_decode}
\end{equation}
We set $\mathbf{B}_q[m,k]=1$ iff $m\in\mathcal{I}_{q,k}$.
If two slots produce the same bundle, the later slot is repaired by replacing the lowest-margin duplicated skill with the highest-scoring unused skill.
This avoids repeated agents under a fixed crew size; Appendix~\ref{app:incidence_decoding} gives the complete algorithm.

The hard incidence matrix defines a skill-agent bipartite graph $\mathcal{H}_q=(\mathcal{S},\mathcal{A},\mathbf{B}_q)$.
For each slot, \textsc{Sigma} computes query-conditioned attention over selected skills:
\begin{equation}
\begin{aligned}
\beta_{m,k}
&=
\frac{\mathbf{B}_q[m,k]\exp(\alpha_{m,k})}
{\sum_{\ell=1}^{M}\mathbf{B}_q[\ell,k]\exp(\alpha_{\ell,k})}.
\end{aligned}
\label{eq:skill_attention}
\end{equation}
During training, an analogous soft support is used for differentiability.
Let $\mathcal{J}_{q,k}=\operatorname{Top}_{r_c}(\{1,\ldots,M\};\alpha_{\cdot,k})$ be the top-ranked candidate set with $r_c\geq k_s$.
The soft attention is
\begin{equation}
\tilde{\beta}_{m,k}
=
\frac{\mathbf{1}[m\in\mathcal{J}_{q,k}]
p_{m,k}\exp(\alpha_{m,k})}
{\sum_{\ell\in\mathcal{J}_{q,k}}
p_{\ell,k}\exp(\alpha_{\ell,k})}.
\label{eq:soft_skill_attention}
\end{equation}
Below $\omega_{m,k}$ denotes $\tilde{\beta}_{m,k}$ during training and $\beta_{m,k}$ during inference, and $\boldsymbol{\Omega}_q=[\omega_{m,k}]$.
The node embedding combines slot identity, task context, and selected skill cards:
\begin{equation}
\mathbf{u}_{q,k}
= \operatorname{Norm}
\left(
\mathbf{p}_k+\mathbf{z}_q+
\sum_{m=1}^{M}\omega_{m,k}\mathbf{e}_m
\right).
\label{eq:agent_construction}
\end{equation}
The resulting node matrix is
\begin{equation}
\mathbf{U}_q=
[\mathbf{u}_{q,1},\ldots,\mathbf{u}_{q,K}]^\top
\in\mathbb{R}^{K\times d}.
\label{eq:node_matrix}
\end{equation}
It is consumed by the topology decoder under the same crew size $K$ as role-based graph baselines.
Thus, a node is not a pooled role embedding; its semantics are determined by the task and by executable skills assigned to the slot.

\subsection{Skill-Aware Topology Decoding}
\label{sec:topology_decoder}

\textsc{Sigma} decodes communication edges from both node-level affinity and skill-bundle compatibility.
For each ordered pair $(i,j)$, the agent-level affinity is
\begin{equation}
r_{ij}
= \operatorname{MLP}_{\phi}
([\mathbf{u}_{q,i},\mathbf{u}_{q,j},\mathbf{z}_q,
\mathbf{u}_{q,i}\odot\mathbf{u}_{q,j}]).
\label{eq:agent_affinity}
\end{equation}
The skill-bundle compatibility aggregates pairwise complementarity between selected skills:
\begin{equation}
\begin{aligned}
g_{mnq}
&= \operatorname{MLP}_{\psi}([\mathbf{e}_m,\mathbf{e}_n,\mathbf{z}_q,
\mathbf{e}_m\odot\mathbf{e}_n]),\\
c_{ij}
&= \sum_{m,n}
\omega_{m,i}\omega_{n,j}g_{mnq}.
\end{aligned}
\label{eq:skill_compatibility}
\end{equation}
The compatibility term captures task-specific complementarity such as retrieval skills feeding API grounding, planning skills feeding code writing, or testing skills feeding debugging.
The sum is evaluated only over sparse active supports: $\mathcal{J}_{q,i}\times\mathcal{J}_{q,j}$ during training and the Cartesian product of hard-selected skills during inference.
Thus, the active compatibility cost depends on bundle size rather than the full library.
The final edge logit combines both terms with a weak workflow prior:
\begin{equation}
\ell_{ij}
= r_{ij}+\lambda_c c_{ij}+\lambda_0\mathbf{A}_{0}[i,j],
\quad i\neq j,
\label{eq:edge_logit}
\end{equation}
where $\mathbf{A}_{0}$ is a simple anchor topology such as a chain, star, or dataset-provided workflow prior.
Setting $\lambda_c=0$ recovers a matched decoder that ignores explicit skill-pair compatibility.
At inference time, edge probabilities are thresholded as
\begin{equation}
\begin{aligned}
\mathcal{G}_q&=(\mathcal{A},\mathcal{E}_q),\\
\mathcal{E}_q
&=\{(a_i,a_j):\sigma(\ell_{ij})>\gamma_e,\ i\neq j\}.
\end{aligned}
\label{eq:decoded_graph}
\end{equation}
If a benchmark requires a directed acyclic schedule, a canonical slot-order mask is applied before thresholding.
This decoder remains topology-compatible with prior graph-based MAS methods because it consumes the constructed node matrix $\mathbf{U}_q$ and outputs the same type of directed communication graph.

\subsection{Execution and Learning}
\label{sec:mailbox}

\textsc{Sigma} uses $\mathbf{B}_q$ during execution as well as representation learning.
For each incident pair $(m,k)$, the executor maintains a skill-specific mailbox; incoming messages are routed to the most relevant assigned skills using frozen embedding similarity, and only the corresponding mailbox summaries enter the agent prompt.
Thus, skill incidence affects prompts, tools, memory organization, and message routing instead of remaining a hidden variable.
Formally, for each incident pair $(m,k)$ with $\mathbf{B}_q[m,k]=1$, \textsc{Sigma} creates
\begin{equation}
\mathcal{M}_{q,k,m}^{(t)}
= \{r: r \rightarrow (a_k,s_m),\; \mathrm{time}(r)<t\},
\label{eq:mailbox}
\end{equation}
where $r \rightarrow (a_k,s_m)$ indicates that message $r$ is routed to skill $s_m$ inside slot $a_k$.
A frozen router scores incoming message $r$ against each incident skill by
\begin{equation}
\gamma_m(r)=\operatorname{sim}(\operatorname{Enc}(r),\mathbf{e}_m),
\label{eq:routing}
\end{equation}
and routes it to the top-$r_s$ assigned skills.
At round $t$, the system prompt contains the selected skill cards and state, while the user prompt contains the task and mailbox summaries:
\begin{equation}
\begin{aligned}
\mathbf{s}_{q,k}^{(t)}
&= \operatorname{Summarize}(\mathcal{M}_{q,k}^{(t)}),\\
\mathcal{P}_{k,\mathrm{sys}}^{(t)}
&= \{\mathcal{L}_{q,k}, \texttt{State}_k^{(t)}\},\\
\mathcal{P}_{k,\mathrm{usr}}^{(t)}
&= \left\{q,\mathbf{s}_{q,k}^{(t)}\right\}.
\end{aligned}
\label{eq:sigma_prompt}
\end{equation}
Each mailbox summary is capped under the same total context budget used by the baselines.
Appendix~\ref{app:mailbox_execution} gives the full prompt format and fallback routing.

Training uses pseudo-labels $\mathbf{B}_q^\star$ built only from training-split role descriptions, tool/API documents, and execution traces.
For each training slot, a deterministic labeler retrieves candidate skills and selects at most $k_s$ cards, yielding canonical incidence columns without using held-out queries, answers, traces, or inserted tool/API cards.
The incidence generator is trained with binary cross-entropy:
\begin{equation}
\mathcal{L}_{\mathrm{inc}}
=
\sum_q \sum_{m=1}^{M}\sum_{k=1}^{K}
\operatorname{BCE}\left(\mathbf{B}_q^\star[m,k], p_{m,k}\right).
\label{eq:incidence_loss}
\end{equation}
When annotated communication graphs are available, the topology decoder is trained with an edge reconstruction loss:
\begin{equation}
\mathcal{L}_{\mathrm{edge}}
=
\sum_q\sum_{i\neq j}
\operatorname{BCE}(\mathbf{A}_q^\star[i,j],\sigma(\ell_{ij})).
\label{eq:edge_loss}
\end{equation}
In our benchmark setting, datasets do not provide ground-truth communication edges; therefore $\mathbf{A}_q^\star$ is instantiated from the same weak anchor topology $\mathbf{A}_0$ used across controlled graph baselines, rather than from held-out answers or evaluation traces.
We further use a soft bundle-size cap and graph sparsity penalty,
\begin{equation}
\begin{aligned}
\mathcal{L}_{\mathrm{sparse}}
=& \sum_q \sum_{k=1}^{K}
\max\left(0,\sum_{m=1}^{M} p_{m,k} - k_s\right)\\
&+\lambda_g\sum_q\sum_{i\neq j}\sigma(\ell_{ij}).
\end{aligned}
\label{eq:sparse_loss}
\end{equation}
The final objective is
\begin{equation}
\mathcal{L}
= \mathcal{L}_{\mathrm{inc}}
+ \lambda_e\mathcal{L}_{\mathrm{edge}}
+ \lambda_s \mathcal{L}_{\mathrm{sparse}}.
\label{eq:total_loss}
\end{equation}
If annotated target graphs are unavailable, $\mathcal{L}_{\mathrm{edge}}$ is computed against this shared structural prior or omitted in the matched-decoder control.
Pseudo-label construction and complexity analysis are detailed in Appendix~\ref{app:pseudo_labels} and Appendix~\ref{app:implementation_details}.
At test time, \textsc{Sigma} can use the frozen training library or append new cards derived from held-out tool/API documents, because new skills enter through the same text-card encoder.

\begin{table*}[t]
\centering
\scriptsize
\setlength{\tabcolsep}{3.2pt}
\renewcommand{\arraystretch}{0.92}
\caption{
Performance comparison with three different base LLMs.
\textbf{Mul.}, \textbf{Topo.}, and \textbf{Comp.} indicate whether the method
supports multi-agent execution, communication-topology design, and skill-based composition, respectively.
Dark blue cells denote the \best{best} and light blue cells \second{second-best} results.
Cross, mixed check--cross, and check marks indicate no, partial, and full support.
}
\label{tab:main_results}
\resizebox{\textwidth}{!}{
\begin{tabular}{l@{\hspace{0.8em}}ccc@{\hspace{1.0em}}ccccccc}
\toprule
\rowcolor{tabhead}
\textbf{Method}
& \textbf{Mul.}
& \textbf{Topo.}
& \textbf{Comp.}
& \textbf{MMLU}
& \textbf{GSM8K}
& \textbf{MultiArith}
& \textbf{SVAMP}
& \textbf{AQuA}
& \textbf{HumanEval}
& \textbf{Avg.} \\
\midrule

\multicolumn{11}{c}{
    \raisebox{-0.15em}{\includegraphics[height=1.1em]{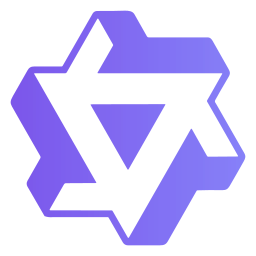}}
    \hspace{0.35em}
    \textit{Qwen3-8B}
} \\
\midrule
Vanilla
& \xmark & \xmark & \xmark
& 57.71 & 80.76 & 87.37 & 96.68 & 35.83 & 82.23 & 73.43 \\
\hdashline

CoT
& \xmark & \xmark & \xmark
& \gain{58.74}{1.03}
& \gain{90.75}{9.99}
& \gain{87.83}{0.46}
& \gain{97.01}{0.33}
& \gain{81.10}{45.27}
& \gain{82.85}{0.62}
& \gain{83.05}{9.62} \\

LLM-Debate
& \cmark & \xmark & \xmark
& \gain{59.61}{1.90}
& \gain{90.12}{9.36}
& \gain{88.46}{1.09}
& \gain{97.62}{0.94}
& \gain{81.14}{45.31}
& \gain{83.85}{1.62}
& \gain{83.47}{10.04} \\

GPTSwarm
& \cmark & \pmark & \xmark
& \gain{59.58}{1.87}
& \gain{89.78}{9.02}
& \gain{90.27}{2.90}
& \gain{97.85}{1.17}
& \gain{81.52}{45.69}
& \gain{84.14}{1.91}
& \gain{83.86}{10.43} \\

$\mathtt{G}$-$\mathtt{Designer}$
& \cmark & \cmark & \xmark
& \gain{60.07}{2.36}
& \gain{90.85}{10.09}
& \gain{90.14}{2.77}
& \gain{\second{98.45}}{1.77}
& \gain{82.08}{46.25}
& \gain{84.73}{2.50}
& \gain{84.39}{10.96} \\

ARG-Designer
& \cmark & \cmark & \xmark
& \gain{\second{60.92}}{3.21}
& \gain{91.14}{10.38}
& \gain{89.38}{2.01}
& \gain{98.21}{1.53}
& \gain{\second{82.47}}{46.64}
& \gain{85.47}{3.24}
& \gain{84.60}{11.17} \\

\textsc{CARD}
& \cmark & \cmark & \xmark
& \gain{60.13}{2.42}
& \gain{\best{91.87}}{11.11}
& \gain{\second{90.54}}{3.17}
& \gain{98.13}{1.45}
& \gain{81.92}{46.09}
& \gain{\second{87.28}}{5.05}
& \gain{\second{84.98}}{11.55} \\

\rowcolor{tabours}
\textbf{\textsc{Sigma}}
& \cmark & \cmark & \cmark
& \gain{\best{61.44}}{3.73}
& \gain{\second{91.56}}{10.80}
& \gain{\best{98.39}}{11.02}
& \gain{\best{98.68}}{2.00}
& \gain{\best{83.46}}{47.63}
& \gain{\best{88.71}}{6.48}
& \gain{\best{87.04}}{13.61} \\

\midrule
\multicolumn{11}{c}{
    \raisebox{-0.15em}{\includegraphics[height=1.1em]{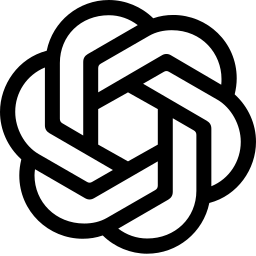}}
    \hspace{0.35em}
    \textit{GPT-OSS-120B}
} \\
\midrule

Vanilla
& \xmark & \xmark & \xmark
& 81.58 & 84.49 & 95.90 & 95.07 & 74.77 & 90.04 & 86.98 \\
\hdashline

CoT
& \xmark & \xmark & \xmark
& \gain{83.39}{1.81}
& \gain{84.81}{0.32}
& \dropgain{95.63}{0.27}
& \gain{95.19}{0.12}
& \gain{76.02}{1.25}
& \gain{90.87}{0.83}
& \gain{87.65}{0.68} \\

LLM-Debate
& \cmark & \xmark & \xmark
& \gain{82.92}{1.34}
& \gain{85.53}{1.04}
& \dropgain{95.72}{0.18}
& \dropgain{94.89}{0.18}
& \gain{75.38}{0.61}
& \gain{91.43}{1.39}
& \gain{87.65}{0.67} \\

GPTSwarm
& \cmark & \pmark & \xmark
& \gain{83.59}{2.01}
& \gain{86.88}{2.39}
& \gain{96.50}{0.60}
& \dropgain{94.77}{0.30}
& \gain{77.41}{2.64}
& \gain{92.30}{2.26}
& \gain{88.58}{1.60} \\

$\mathtt{G}$-$\mathtt{Designer}$
& \cmark & \cmark & \xmark
& \gain{85.06}{3.48}
& \gain{88.62}{4.13}
& \gain{96.86}{0.96}
& \gain{95.21}{0.14}
& \gain{78.13}{3.36}
& \gain{92.88}{2.84}
& \gain{89.46}{2.49} \\

ARG-Designer
& \cmark & \cmark & \xmark
& \gain{85.63}{4.05}
& \gain{91.08}{6.59}
& \gain{\best{98.67}}{2.77}
& \dropgain{94.81}{0.26}
& \gain{78.62}{3.85}
& \gain{93.47}{3.43}
& \gain{90.38}{3.41} \\

\textsc{CARD}
& \cmark & \cmark & \xmark
& \gain{\second{85.78}}{4.20}
& \gain{\second{93.44}}{8.95}
& \gain{98.21}{2.31}
& \gain{\second{95.48}}{0.41}
& \gain{\second{82.76}}{7.99}
& \gain{\second{93.98}}{3.94}
& \gain{\second{91.61}}{4.63} \\

\rowcolor{tabours}
\textbf{\textsc{Sigma}}
& \cmark & \cmark & \cmark
& \gain{\best{87.28}}{5.70}
& \gain{\best{96.25}}{11.76}
& \gain{\second{98.34}}{2.44}
& \gain{\best{95.94}}{0.87}
& \gain{\best{90.16}}{15.39}
& \gain{\best{95.83}}{5.79}
& \gain{\best{93.97}}{6.99} \\

\midrule
\multicolumn{11}{c}{
    \raisebox{-0.15em}{\includegraphics[height=1.1em]{picture/logos/openai.png}}
    \hspace{0.35em}
    \textit{GPT-4o-mini}
} \\
\midrule

Vanilla
& \xmark & \xmark & \xmark
& 67.61 & 85.33 & 93.94 & 86.12 & 69.80 & 77.29 & 80.02 \\
\hdashline

CoT
& \xmark & \xmark & \xmark
& \gain{68.84}{1.23}
& \gain{86.51}{1.18}
& \gain{94.46}{0.52}
& \gain{86.89}{0.77}
& \gain{71.85}{2.05}
& \gain{78.64}{1.35}
& \gain{81.20}{1.18} \\

LLM-Debate
& \cmark & \xmark & \xmark
& \gain{69.66}{2.05}
& \gain{87.29}{1.96}
& \gain{94.97}{1.03}
& \gain{87.67}{1.55}
& \gain{72.93}{3.13}
& \gain{79.48}{2.19}
& \gain{82.00}{1.99} \\

GPTSwarm
& \cmark & \pmark & \xmark
& \gain{71.40}{3.79}
& \gain{88.88}{3.55}
& \gain{96.01}{2.07}
& \gain{89.21}{3.09}
& \gain{75.06}{5.26}
& \gain{81.39}{4.10}
& \gain{83.66}{3.64} \\

$\mathtt{G}$-$\mathtt{Designer}$
& \cmark & \cmark & \xmark
& \gain{73.19}{5.58}
& \gain{90.46}{5.13}
& \gain{96.78}{2.84}
& \gain{90.37}{4.25}
& \gain{76.58}{6.78}
& \gain{82.67}{5.38}
& \gain{85.01}{4.99} \\

ARG-Designer
& \cmark & \cmark & \xmark
& \gain{74.23}{6.62}
& \gain{91.84}{6.51}
& \gain{\second{97.56}}{3.62}
& \gain{91.53}{5.41}
& \gain{78.16}{8.36}
& \gain{84.54}{7.25}
& \gain{86.31}{6.30} \\

\textsc{CARD}
& \cmark & \cmark & \xmark
& \gain{\second{75.07}}{7.46}
& \gain{\second{93.13}}{7.80}
& \gain{96.94}{3.00}
& \gain{\second{91.62}}{5.50}
& \gain{\second{80.40}}{10.60}
& \gain{\second{85.11}}{7.82}
& \gain{\second{87.05}}{7.03} \\

\rowcolor{tabours}
\textbf{\textsc{Sigma}}
& \cmark & \cmark & \cmark
& \gain{\best{76.47}}{8.86}
& \gain{\best{93.20}}{7.87}
& \gain{\best{99.11}}{5.17}
& \gain{\best{93.85}}{7.73}
& \gain{\best{83.86}}{14.06}
& \gain{\best{86.29}}{9.00}
& \gain{\best{88.80}}{8.78} \\

\bottomrule
\end{tabular}
}
\end{table*}

\section{Experiments}
\subsection{Setup}
\paragraph{Benchmarks.}
We evaluate \textsc{Sigma} on six representative benchmarks covering code generation,
general knowledge, mathematical reasoning, and symbolic problem solving:
\ding{182} \textbf{Code Generation}: HumanEval~\citep{chen2021evaluating};
\ding{183} \textbf{General Reasoning}: MMLU~\citep{hendrycks2020measuring};
\ding{184} \textbf{Mathematical Reasoning}: GSM8K~\citep{cobbe2021training}, SVAMP~\citep{patel-etal-2021-nlp}, MultiArith~\citep{roy2015solving}, and AQuA~\citep{ling2017program}.
For all benchmarks, we report task accuracy as the primary metric.
To examine whether the proposed method is robust across model scales and families,
we conduct experiments with three base LLMs: \textit{Qwen3-8B}~\citep{yang2025qwen3}, \textit{GPT-OSS-120B}~\citep{agarwal2025gpt},
and \textit{GPT-4o-mini}~\citep{hurst2024gpt}.
The average score across all benchmarks is also reported as an overall measure of
general task-solving ability.
All methods use the same benchmark splits, deterministic decoding, and dataset-specific answer extractor; Appendix~\ref{app:baseline_prompts} reports the concrete Vanilla and CoT prompt templates used in the evaluation.
For MMLU, we follow the cost-controlled evaluation protocol used by $\mathtt{G}$-$\mathtt{Designer}$ and ARG-Designer: all methods are evaluated on the same fixed 153-question shuffled subset, while training/optimization uses separate examples and never uses held-out answers.

\paragraph{Baselines.}
We compare \textsc{Sigma} with both single-agent and multi-agent baselines. 
For single-agent settings, we use \textbf{Vanilla} prompting and \textbf{CoT}~\citep{wei2022cot}. 
For multi-agent settings, we include \textbf{LLM-Debate}~\citep{du2024improving}, \textbf{GPTSwarm}~\citep{zhuge2024gptswarm}, \textbf{ARG-Designer}~\citep{li2026assemble}, \textsc{CARD}~\citep{wu2026card}, and $\mathtt{G}$-$\mathtt{Designer}$~\citep{zhang2025gdesigner}. 
These baselines evaluate whether \textsc{Sigma}'s gains come from skill-composed node construction rather than stronger prompting, additional agents, or topology optimization alone.
Table~\ref{tab:app_training_config} also displays more training and experiment details of \textsc{Sigma}.

\begin{figure}[t]
    \centering
    \includegraphics[width=\linewidth]{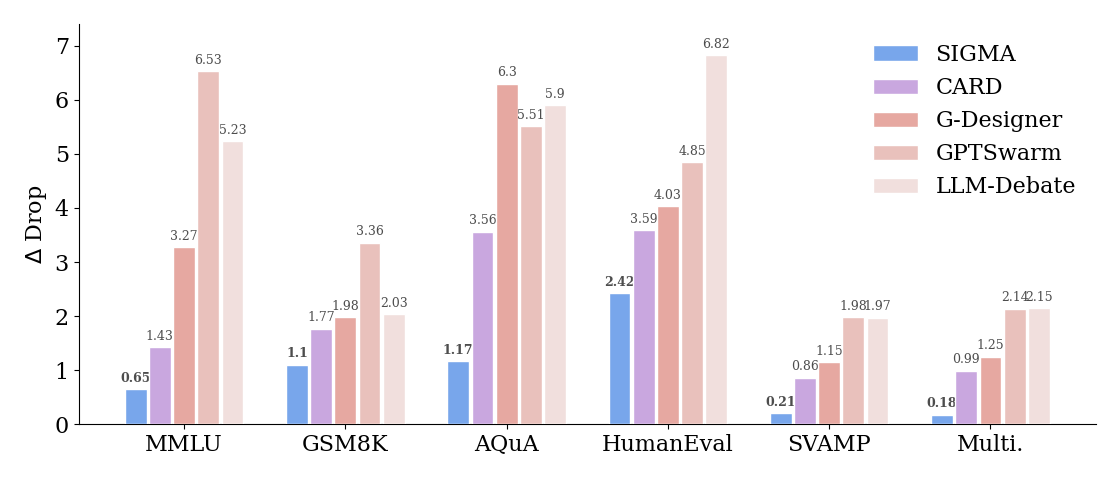}
    \caption{Performance drop from source skill libraries to unseen skill libraries. Lower values indicate stronger robustness to library changes.}
    \label{fig:unseen_delta}
\end{figure}

\subsection{Main Results}

\paragraph{Superior Overall Performance.}
The results in Table~\ref{tab:main_results} show that \textsc{Sigma} consistently improves multi-agent collaboration across different model scales and task domains.
Across \textit{Qwen3-8B}, \textit{GPT-OSS-120B}, and \textit{GPT-4o-mini}, \textsc{Sigma} achieves the highest average scores of $87.04$, $93.97$, and $88.80$, respectively.
Compared with \textsc{CARD}, the strongest non-compositional topology-based baseline in terms of average performance, \textsc{Sigma} brings consistent average improvements of $2.06{\uparrow}$, $2.36{\uparrow}$, and $1.75{\uparrow}$ points.
It ranks first on $16$ of the $18$ benchmark--model combinations, with especially clear gains on AQuA and HumanEval, where structured reasoning and executable construction matter.
These results indicate that the gains of \textsc{Sigma} are robust across baselines and benchmarks.
To further isolate whether the improvements come merely from exposing task-relevant skill cards to the model, we also introduce \textsc{Single-Agent+Skills}, a controlled baseline that receives the same union of selected skill cards as \textsc{Sigma} but removes multi-agent execution, topology decoding, and mailbox routing.
As a controlled baseline, \textsc{Single-Agent+Skills} improves over CoT but remains below \textsc{Sigma} on average and on most datasets in Appendix Table~\ref{tab:single_agent_skills_control}, suggesting that the gains come not only from exposing skill cards, but also from topology-aware multi-agent execution and mailbox routing.

\paragraph{Generalization to unfamiliar skill libraries.}
When the source skill library is replaced with unseen skill cards at test time, \textsc{Sigma} suffers the smallest average degradation.
As shown in Table~\ref{tab:unseen_skill_library} and Figure~\ref{fig:unseen_delta}, its performance drops by only $0.96$ points, lower than \textsc{CARD} ($2.03$), $\mathtt{G}$-$\mathtt{Designer}$ ($3.00$), GPTSwarm ($4.06$), and LLM-Debate ($4.02$).
This robustness is especially evident on MMLU, AQuA, and HumanEval, where topology-centric baselines exhibit much larger performance drops under library shifts.
The result indicates that skill incidence can compose unseen cards into agent slots instead of relying on a fixed capability inventory.

\begin{figure}[t]
  \includegraphics[width=\linewidth]{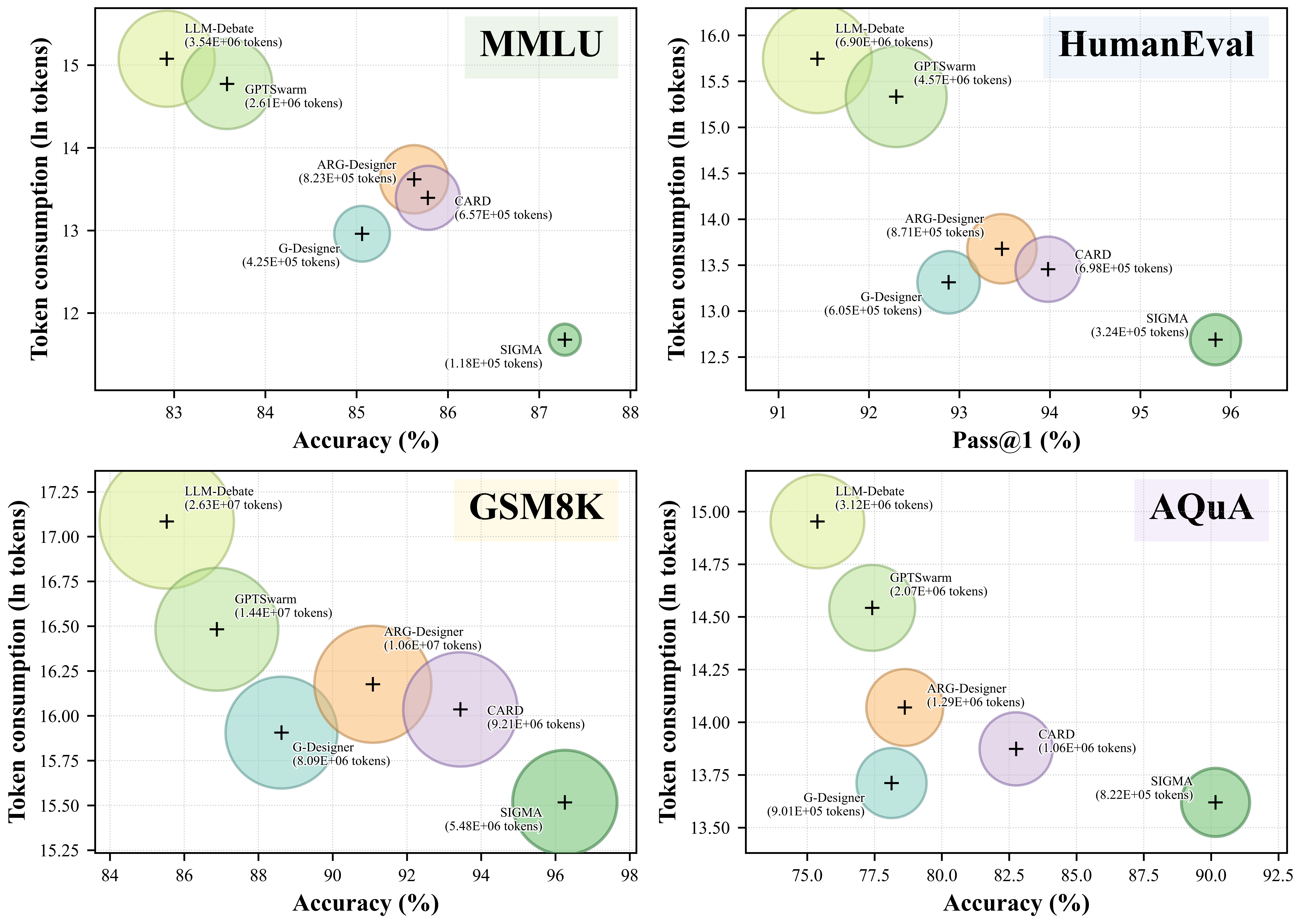}
  \caption{Visualization of the accuracy and token trade-off across benchmarks.}
  \label{fig:token}
\end{figure}

\paragraph{Efficiency.}
Figure~\ref{fig:token} shows that \textsc{Sigma} reaches the best accuracy with the fewest tokens among compared multi-agent methods on MMLU, HumanEval, GSM8K, and AQuA.
Compared with GPTSwarm, it reduces token consumption by $95.5\%$, $61.9\%$, $92.9\%$, and $60.3\%$ on these benchmarks, respectively; compared with LLM-Debate, the maximum reduction reaches $96.7\%$.
Thus, \textsc{Sigma} improves collaboration efficiency by composing task-specific agents rather than increasing communication volume; detailed statistics are in Appendix~\ref{app:detailed_token_runtime}.

\subsection{Ablation Study}

Table~\ref{tab:ablation_unseen_library} ablates the main components of \textsc{Sigma}.
$\blacksquare$\textbf{\emph{w/o Inc.}} replaces the learned skill-agent incidence predictor with flat skill assignment.
$\blacksquare$\textbf{\emph{w/o Skill Dec.}}, $\blacksquare$\textbf{\emph{w/o Mailbox}}, and $\blacksquare$\textbf{\emph{w/o Anchor Prior}} remove skill-aware decoding, mailbox routing, and the weak workflow prior, respectively.
The consistent drops show that \textsc{Sigma}'s generalization comes from the joint effect of compositional node construction, skill-aware topology decoding, and executable skill routing.

\begin{table}[H]
\centering
\small
\setlength{\tabcolsep}{5.0pt}
\renewcommand{\arraystretch}{1.10}
\caption{Ablation study under source and unseen skill libraries, using \textit{GPT-OSS-120B}}
\label{tab:ablation_unseen_library}
\arrayrulecolor{black}
\begin{tabular}{@{}l|cc|cc@{}}
\toprule

\textbf{Variant}
& \multicolumn{2}{c|}{\textbf{MMLU}}
& \multicolumn{2}{c}{\textbf{HumanEval}} \\
\cline{2-5}

& \textbf{Source} & \textbf{Unseen}
& \textbf{Source} & \textbf{Unseen} \\
\midrule
\rowcolor{tabours}
\textsc{Sigma}
& \textbf{85.62} & \textbf{84.31}
& \textbf{95.97} & \textbf{95.16} \\
\midrule
\rowcolor{tabstripe}
w/o Inc.
& 80.39 & 80.39
& 91.23 & 90.77 \\

w/o Skill Dec.
& 83.01 & 82.35
& 92.33 & 92.41 \\

\rowcolor{tabstripe}
w/o Mailbox
& 81.04 & 80.39
& 93.07 & 92.18 \\

w/o Anchor Prior
& 83.01 & 83.66
& 93.74 & 92.93 \\
\bottomrule
\end{tabular}
\arrayrulecolor{black}
\end{table}

\section{Conclusion}
We presented \textsc{Sigma}, a skill-incidence framework for compositional multi-agent design.
Rather than treating agents as fixed role nodes, \textsc{Sigma} constructs each agent as a task-conditioned bundle of reusable skills and decodes communication over these skill-composed nodes.
Skill-specific mailboxes further make the incidence structure executable during multi-agent interaction.
By separating capability composition from topology generation, \textsc{Sigma} enables fixed agent slots to express diverse task-specific identities and generalize to unseen skill combinations.
Our results highlight that effective multi-agent design should consider not only how agents communicate, but also how their internal capabilities are composed.

\clearpage

\section*{Limitations}
We note several limitations.
\ding{182} \textsc{Sigma} depends on the quality and coverage of the skill library, making noisy or incomplete skill cards a potential source of error.
\ding{183} Its deterministic pseudo-labels and canonical slot ordering may inherit biases from labeling heuristics.
\ding{184} The current framework assumes a fixed number of agent slots, leaving dynamic crew-size prediction as future work.
\ding{185} Our evaluation focuses on benchmark-style reasoning and coding tasks, while real-world tool-use environments may involve noisier APIs, longer horizons, and skills with side effects.
\ding{186} Finally, although skill-specific mailboxes make skill incidence executable, adaptive routing and conflict resolution remain open directions.

\textsc{Sigma} does not attempt to solve the globally optimal skill-agent assignment problem by exhaustive search. 
Instead, it learns a task-conditioned incidence generator and applies a sparse deterministic projection, trading global optimality guarantees for scalability under a fixed bundle-size budget.

\bibliography{custom}

\clearpage

\appendix

\section{Appendix Roadmap}
\label{app:roadmap}

This appendix provides additional details for \textsc{Sigma}, including method design,
skill-card serialization, incidence decoding, mailbox execution, pseudo-label construction,
evaluation protocols, implementation settings, and additional robustness results.

\begin{table}[H]
\centering
\small
\begin{tcolorbox}[
  enhanced,
  width=\linewidth,
  colback=sigmagray,
  colframe=black!15,
  boxrule=0.5pt,
  arc=2mm,
  left=3mm,
  right=3mm,
  top=2mm,
  bottom=2mm,
  title=\textbf{Appendix Roadmap},
  coltitle=black,
  colbacktitle=white,
  fonttitle=\bfseries,
  attach boxed title to top left={xshift=2mm,yshift=-1.5mm},
  boxed title style={
    colback=white,
    colframe=black!12,
    boxrule=0.4pt,
    arc=1mm
  }
]

\begin{tcbraster}[
  raster columns=1,
  raster equal height=rows,
  raster column skip=2mm,
  raster row skip=2mm
]

\begin{tcolorbox}[
  colback=white,
  colframe=sigmablueframe!45,
  boxrule=0.5pt,
  arc=1.5mm,
  left=1.5mm,
  right=1.5mm,
  top=1mm,
  bottom=1mm
]
\textbf{Appendix~\ref{app:method_details}: Method Details}\\
Additional method details, including graph execution, skill-card serialization, hard incidence decoding, continuous relaxation, and end-to-end inference.
\end{tcolorbox}

\begin{tcolorbox}[
  colback=white,
  colframe=sigmablueframe!45,
  boxrule=0.5pt,
  arc=1.5mm,
  left=1.5mm,
  right=1.5mm,
  top=1mm,
  bottom=1mm
]
\textbf{Appendix~\ref{app:mailbox_execution}: Mailbox Execution}\\
Skill-mailbox execution, mailbox summary format, and fallback routing.
\end{tcolorbox}

\begin{tcolorbox}[
  colback=white,
  colframe=sigmablueframe!45,
  boxrule=0.5pt,
  arc=1.5mm,
  left=1.5mm,
  right=1.5mm,
  top=1mm,
  bottom=1mm
]
\textbf{Appendix~\ref{app:pseudo_labels}: Pseudo Labels}\\
Deterministic pseudo-label construction and canonical slot ordering.
\end{tcolorbox}

\begin{tcolorbox}[
  colback=white,
  colframe=sigmablueframe!45,
  boxrule=0.5pt,
  arc=1.5mm,
  left=1.5mm,
  right=1.5mm,
  top=1mm,
  bottom=1mm
]
\textbf{Appendix~\ref{app:experimental_protocol}: Experimental Protocol}\\
Matched-crew evaluation, held-out skill composition, test-time skill insertion, baselines, and metrics.
\end{tcolorbox}

\begin{tcolorbox}[
  colback=white,
  colframe=sigmablueframe!45,
  boxrule=0.5pt,
  arc=1.5mm,
  left=1.5mm,
  right=1.5mm,
  top=1mm,
  bottom=1mm
]
\textbf{Appendix~\ref{app:implementation_details}: Implementation Details}\\
Hyperparameters, training configuration, training objective, and complexity analysis.
\end{tcolorbox}

\begin{tcolorbox}[
  colback=white,
  colframe=sigmablueframe!45,
  boxrule=0.5pt,
  arc=1.5mm,
  left=1.5mm,
  right=1.5mm,
  top=1mm,
  bottom=1mm
]
\textbf{Appendix~\ref{app:additional_results}: Additional Results}\\
Additional results, including robustness to unfamiliar skill libraries.
\end{tcolorbox}

\begin{tcolorbox}[
  colback=white,
  colframe=sigmablueframe!45,
  boxrule=0.5pt,
  arc=1.5mm,
  left=1.5mm,
  right=1.5mm,
  top=1mm,
  bottom=1mm
]
\textbf{Appendix~\ref{app:prompts}: Prompt Templates}\\
Prompt templates for SIGMA agent execution, Vanilla/CoT baselines, and dataset-specific answer extraction.
\end{tcolorbox}

\begin{tcolorbox}[
  colback=white,
  colframe=sigmablueframe!45,
  boxrule=0.5pt,
  arc=1.5mm,
  left=1.5mm,
  right=1.5mm,
  top=1mm,
  bottom=1mm
]
\textbf{Appendix~\ref{app:additional_detail}: Additional Details}\\
More details about datasets, skill libraries, and models used in experiments.
\end{tcolorbox}

\begin{tcolorbox}[
  colback=white,
  colframe=sigmablueframe!45,
  boxrule=0.5pt,
  arc=1.5mm,
  left=1.5mm,
  right=1.5mm,
  top=1mm,
  bottom=1mm
]
\textbf{Appendix~\ref{app:failure_analysis}: Failure Cases}\\
Analysis of some failure cases during experiments.
\end{tcolorbox}

\begin{tcolorbox}[
  colback=white,
  colframe=sigmablueframe!45,
  boxrule=0.5pt,
  arc=1.5mm,
  left=1.5mm,
  right=1.5mm,
  top=1mm,
  bottom=1mm
]
\textbf{Appendix~\ref{app:impact}: Impact Statement}\\
Impact of \textsc{Sigma}.
\end{tcolorbox}

\begin{tcolorbox}[
  colback=white,
  colframe=sigmablueframe!45,
  boxrule=0.5pt,
  arc=1.5mm,
  left=1.5mm,
  right=1.5mm,
  top=1mm,
  bottom=1mm
]
\textbf{Appendix~\ref{app:case_study}: Case Study}\\
Detailed, card-style replay of representative HumanEval cases from the raw execution logs.
\end{tcolorbox}

\end{tcbraster}
\end{tcolorbox}

\label{tab:appendix_roadmap}
\end{table}

\section{Additional Method Details}
\label{app:method_details}

\subsection{SI-MAP Design Checklist}
\label{app:simap_checklist}

The Skill-Incidence Multi-Agent Protocol (\textsc{SI-MAP}) requires that a graph-based multi-agent
designer satisfies four operational properties. Table~\ref{tab:simap_checklist} summarizes how
\textsc{Sigma} instantiates these properties.

\begin{table}[t]
\centering
\small
\setlength{\tabcolsep}{4pt}
\renewcommand{\arraystretch}{1.12}
\caption{
Operational checklist for the \textsc{SI-MAP} protocol. 
Each check corresponds to a required property of executable skill-incidence based multi-agent design, with its concrete realization in \textsc{Sigma} summarized on the right.
}
\begin{tabularx}{\linewidth}{@{}p{0.31\linewidth}Y@{}}
\toprule
\rowcolor{sigmagreen}
\textbf{Property} & \textbf{Instantiation in \textsc{Sigma}} \\
\midrule

\textbf{Compositionality} &
Builds each agent slot from a task-conditioned bundle of reusable skill cards via the incidence matrix $B_q$. \\

\rowcolor{sigmagray}
\textbf{Executability} &
Exposes assigned skills through prompts, tool affordances, retrieval procedures, and skill-specific mailboxes. \\

\textbf{Topology Compatibility} &
Feeds skill-composed agent embeddings $U_q$ into the same graph decoder used for communication topology generation. \\

\rowcolor{sigmagray}
\textbf{Expandability} &
Allows new tool/API skill cards to be appended at test time and encoded by the frozen skill encoder. \\

\bottomrule
\end{tabularx}

\label{tab:simap_checklist}
\end{table}

\subsection{Graph Execution Interface}
\label{app:graph_execution}

The main text keeps only the compact graph definition. For completeness, we state the generic
multi-agent execution interface used by \textsc{Sigma} and the baselines. Given a generated graph
$\mathcal{G}_q$, agents communicate for $T$ rounds. At round $t$, agent $a_k$ receives the task query
and messages produced by its in-neighbors:
\begin{equation}
\begin{aligned}
\mathcal{R}_k^{(t)}
&= a_k\left(\mathcal{P}_{k,\mathrm{sys}}^{(t)},
\mathcal{P}_{k,\mathrm{usr}}^{(t)}\right),\\
\mathcal{P}_{k,\mathrm{usr}}^{(t)}
&= \left\{q, \{\mathcal{R}_j^{(t-1)}:
a_j \in \mathcal{N}_{\mathrm{in}}(a_k)\}\right\},
\end{aligned}
\end{equation}
where $\mathcal{N}_{\mathrm{in}}(a_k)$ is the in-neighborhood of $a_k$ and
$\mathcal{R}_j^{(0)}$ is initialized as an empty message. After the final round, an aggregation
function returns the system answer:
\begin{equation}
\hat{y}_q \leftarrow
\operatorname{Aggregate}\left(\mathcal{R}_1^{(T)}, \ldots, \mathcal{R}_K^{(T)}\right).
\end{equation}
\textsc{Sigma} keeps this interface but replaces atomic role nodes with skill-composed nodes and
routes messages through skill-specific mailboxes.

\subsection{Skill Card Serialization}
\label{app:skill_card_serialization}

Each skill card is serialized into a fixed textual template before being encoded. We use the same
template for training, validation, and test-time skill insertion.
\begin{tcolorbox}[
  enhanced,
  breakable,
  colback=skillbg,
  colframe=skillframe,
  boxrule=0.7pt,
  arc=2mm,
  left=2mm,
  right=2mm,
  top=1.5mm,
  bottom=1.5mm,
  title=\textbf{\textsc{Skill Card Template}} \hfill \skilltag{tag},
  coltitle=black,
  colbacktitle=white,
  fonttitle=\small\bfseries,
  attach boxed title to top left={xshift=2mm,yshift=-1.8mm},
  boxed title style={
    colback=white,
    colframe=skillframe,
    boxrule=0.5pt,
    arc=1mm
  }
]
\small
\begin{tabularx}{\linewidth}{@{}p{0.25\linewidth}Y@{}}
\skillkey{Skill\_id} &
\skillval{"\{unique skill identifier\}"} \\

\skillkey{Name} &
\skillval{"\{skill name\}"} \\

\skillkey{Description} &
\skillval{"\{capability description\}"} \\

\skillkey{Benchmark} &
\skillval{"\{target benchmark or task domain\}"} \\

\skillkey{Inputs} &
\skillval{["\{input field 1\}", "..."]} \\

\skillkey{Outputs} &
\skillval{["\{output field 1\}", "..."]} \\

\skillkey{Tools} &
\skillval{["\{optional tool/API/retrieval resource\}", "..."]} \\

\skillkey{Tags} &
\skillval{["\{retrieval tag 1\}", "..."]} \\
\end{tabularx}
\end{tcolorbox}

\begin{tcolorbox}[
  enhanced,
  breakable,
  colback=skillbg,
  colframe=skillframe,
  boxrule=0.7pt,
  arc=2mm,
  left=2mm,
  right=2mm,
  top=1.5mm,
  bottom=1.5mm,
  title=\textbf{\textsc{Skill Card Example \textsc{I}}} \hfill \skilltag{mmlu},
  coltitle=black,
  colbacktitle=white,
  fonttitle=\small\bfseries,
  attach boxed title to top left={xshift=2mm,yshift=-1.8mm},
  boxed title style={
    colback=white,
    colframe=skillframe,
    boxrule=0.5pt,
    arc=1mm
  }
]
\small
\begin{tabularx}{\linewidth}{@{}p{0.25\linewidth}Y@{}}
\skillkey{Skill\_id} &
\skillval{\texttt{"option\_elimination"}} \\

\skillkey{Name} &
\skillval{\texttt{"Option elimination"}} \\

\skillkey{Description} &
\skillval{Rule out implausible multiple-choice answers before selecting the final answer.} \\

\skillkey{Benchmark} &
\skillval{\texttt{"MMLU"}} \\

\skillkey{Inputs} &
\skillval{\texttt{["query", "messages"]}} \\

\skillkey{Outputs} &
\skillval{\texttt{["analysis", "answer"]}} \\

\skillkey{Tools} &
\skillval{\texttt{[]}} \\

\skillkey{Tags} &
\skillval{\texttt{["mmlu", "multiple-choice", "elimination"]}} \\
\end{tabularx}
\end{tcolorbox}

\begin{tcolorbox}[
  enhanced,
  breakable,
  colback=skillbg,
  colframe=skillframe,
  boxrule=0.7pt,
  arc=2mm,
  left=2mm,
  right=2mm,
  top=1.5mm,
  bottom=1.5mm,
  title=\textbf{\textsc{Skill Card Example \textsc{III}}} \hfill \skilltag{aqua},
  coltitle=black,
  colbacktitle=white,
  fonttitle=\small\bfseries,
  attach boxed title to top left={xshift=2mm,yshift=-1.8mm},
  boxed title style={
    colback=white,
    colframe=skillframe,
    boxrule=0.5pt,
    arc=1mm
  }
]
\small
\begin{tabularx}{\linewidth}{@{}p{0.25\linewidth}Y@{}}
\skillkey{Skill\_id} &
\skillval{\texttt{"synthesize"}} \\

\skillkey{Name} &
\skillval{\texttt{"Synthesize"}} \\

\skillkey{Description} &
\skillval{Produce the final response by consolidating the reasoning trajectory into a concise answer.} \\

\skillkey{Benchmark} &
\skillval{\texttt{"AQuA"}} \\

\skillkey{Inputs} &
\skillval{\texttt{["query", "messages"]}} \\

\skillkey{Outputs} &
\skillval{\texttt{["analysis", "answer"]}} \\

\skillkey{Tools} &
\skillval{\texttt{[]}} \\

\skillkey{Tags} &
\skillval{\texttt{["aqua", "synthesis", "final-answer"]}} \\
\end{tabularx}
\end{tcolorbox}

\begin{tcolorbox}[
  enhanced,
  breakable,
  colback=skillbg,
  colframe=skillframe,
  boxrule=0.7pt,
  arc=2mm,
  left=2mm,
  right=2mm,
  top=1.5mm,
  bottom=1.5mm,
  title=\textbf{\textsc{Skill Card Example \textsc{III}}} \hfill \skilltag{humaneval},
  coltitle=black,
  colbacktitle=white,
  fonttitle=\small\bfseries,
  attach boxed title to top left={xshift=2mm,yshift=-1.8mm},
  boxed title style={
    colback=white,
    colframe=skillframe,
    boxrule=0.5pt,
    arc=1mm
  }
]
\small
\begin{tabularx}{\linewidth}{@{}p{0.25\linewidth}Y@{}}
\skillkey{Skill\_id} &
\skillval{\texttt{"algorithm\_design"}} \\

\skillkey{Name} &
\skillval{\texttt{"Algorithm design"}} \\

\skillkey{Description} &
\skillval{Choose an algorithmic strategy and complexity target before implementing the solution.} \\

\skillkey{Benchmark} &
\skillval{\texttt{"HumanEval"}} \\

\skillkey{Inputs} &
\skillval{\texttt{["query", "messages"]}} \\

\skillkey{Outputs} &
\skillval{\texttt{["analysis", "answer"]}} \\

\skillkey{Tools} &
\skillval{\texttt{[]}} \\

\skillkey{Tags} &
\skillval{\texttt{["humaneval", "algorithm", "complexity"]}} \\
\end{tabularx}
\end{tcolorbox}

For tool or API skills, the executable affordance field contains the callable endpoint, retrieval
operation, or tool-use instruction. For cognitive skills, the affordance is instantiated as a concrete
prompt-level procedure or mailbox-routing target rather than a vague role description. This prevents
purely latent role labels from being treated as reusable executable skills.

\subsection{Hard Incidence Decoding}
\label{app:incidence_decoding}

The incidence generator predicts a soft assignment probability $P_q[m,k]$ for every skill-agent pair.
The executable incidence matrix $B_q$ is obtained by deterministic sparse projection. For each agent
slot $k$, we first keep skills whose probability exceeds $\tau_{\mathrm{skill}}$ and then retain at
most $k_s$ skills. If no skill exceeds the threshold, the highest-scoring skill is used as an argmax
fallback. To avoid repeated agent identities under a fixed crew size, duplicated bundles are repaired
by replacing the lowest-margin duplicated skill with the highest-scoring unused skill.

\begin{algorithm}[h]
\small
\caption{\textsc{Sigma} incidence decoding}
\label{alg:incidence_decoding}
\begin{algorithmic}[1]
\Require Query embedding $z_q$, skill embeddings $E$, slot embeddings $\{p_k\}_{k=1}^{K}$,
incidence generator $F_{\theta}$, threshold $\tau_{\mathrm{skill}}$, bundle size $k_s$
\Ensure Hard incidence matrix $B_q \in \{0,1\}^{M \times K}$
\State Initialize $B_q \leftarrow \mathbf{0}$
\For{$k = 1$ to $K$}
    \For{$m = 1$ to $M$}
        \State $\alpha_{m,k} \leftarrow F_{\theta}([z_q,e_m,p_k,z_q \odot e_m])$
        \State $p_{m,k} \leftarrow \sigma(\alpha_{m,k})$
    \EndFor
    \State $C_{q,k} \leftarrow \{m : p_{m,k} > \tau_{\mathrm{skill}}\}$
    \If{$C_{q,k} \neq \emptyset$}
        \State $I_{q,k} \leftarrow \operatorname{Top}_{k_s}(C_{q,k}; \alpha_{\cdot,k})$
    \Else
        \State $I_{q,k} \leftarrow \{\arg\max_m \alpha_{m,k}\}$
    \EndIf
    \For{$m \in I_{q,k}$}
        \State $B_q[m,k] \leftarrow 1$
    \EndFor
\EndFor
\State Repair duplicated columns of $B_q$ by replacing the lowest-margin duplicated skill with
the best unused skill for the affected slot.
\State \Return $B_q$
\end{algorithmic}
\end{algorithm}

\subsection{Continuous Relaxation and Bundle Attention}
\label{app:continuous_relaxation}

During training, \textsc{Sigma} uses a sparse continuous relaxation aligned with the hard bundle.
For each slot $k$, let
$\mathcal{J}_{q,k}=\operatorname{Top}_{r_c}(\{1,\ldots,M\};\alpha_{\cdot,k})$ be the top-ranked
candidate support with $r_c\geq k_s$. The soft bundle attention is
\begin{equation}
\tilde{\beta}_{m,k}
=
\frac{\mathbf{1}[m\in\mathcal{J}_{q,k}]
p_{m,k}\exp(\alpha_{m,k})}
{\sum_{\ell\in\mathcal{J}_{q,k}}
p_{\ell,k}\exp(\alpha_{\ell,k})},
\end{equation}
where $p_{m,k}=\sigma(\alpha_{m,k})$. We use $\tilde{\beta}_{m,k}$ during differentiable training
and the hard-selected attention $\beta_{m,k}$ during inference. The same sparse support is used when
computing skill-bundle compatibility, so pairwise skill interactions are evaluated over
$\mathcal{J}_{q,i}\times\mathcal{J}_{q,j}$ during training and over hard-selected skill products
during inference.

\subsection{End-to-End Inference Procedure}
\label{app:inference_procedure}

Algorithm~\ref{alg:sigma_inference} gives the complete inference procedure. \textsc{Sigma} first
constructs the task-specific skill-agent incidence matrix, builds incidence-aware agent embeddings,
decodes the communication graph, and then executes the graph with skill-specific mailboxes.

\begin{algorithm}[h]
\small
\caption{\textsc{Sigma} inference}
\label{alg:sigma_inference}
\begin{algorithmic}[1]
\Require Query $q$, skill library $\mathcal{L}$, number of agent slots $K$, encoder $\operatorname{Enc}$,
incidence generator $F_{\theta}$, topology decoder $D_{\phi}$
\Ensure Final answer $\hat{y}_q$
\State $z_q \leftarrow \operatorname{Enc}(q)$
\State $e_m \leftarrow \operatorname{Enc}(s_m)$ for each $s_m \in \mathcal{L}$
\State $B_q \leftarrow$ \textsc{IncidenceDecode}$(z_q,E,F_{\theta})$
\For{$k = 1$ to $K$}
    \State Compute skill attention $\beta_{m,k}$ over selected skills with $B_q[m,k]=1$
    \State $u_{q,k} \leftarrow \operatorname{Norm}(p_k + z_q + \sum_m \beta_{m,k}e_m)$
\EndFor
\State $U_q \leftarrow [u_{q,1},\ldots,u_{q,K}]^\top$
\For{each ordered pair $(i,j)$ with $i \neq j$}
    \State Compute agent affinity $r_{ij}$
    \State Compute skill-bundle compatibility $c_{ij}$
    \State $\ell_{ij} \leftarrow r_{ij} + \lambda_c c_{ij} + \lambda_0 A_0[i,j]$
\EndFor
\State $E_q \leftarrow \{(a_i,a_j): \sigma(\ell_{ij}) > \gamma_e, i \neq j\}$
\State Create one mailbox $M_{q,k,m}$ for every incident pair $(m,k)$ with $B_q[m,k]=1$
\For{$t = 1$ to $T$}
    \For{each agent $a_k$ in the communication schedule}
        \State Route incoming messages to the top-$r_s$ incident skill mailboxes.
        \State Build the system prompt from selected skill cards and state.
        \State Build the user prompt from $q$ and mailbox summaries.
        \State Generate response $R_k^{(t)}$.
    \EndFor
\EndFor
\State $\hat{y}_q \leftarrow \operatorname{Aggregate}(R_1^{(T)},\ldots,R_K^{(T)})$
\State \Return $\hat{y}_q$
\end{algorithmic}
\end{algorithm}

\section{Skill-Mailbox Execution}
\label{app:mailbox_execution}

The mailbox mechanism makes the incidence matrix executable during multi-agent interaction. For
each incident pair $(s_m,a_k)$, \textsc{Sigma} maintains a skill-specific mailbox $M_{q,k,m}$.
When a message arrives at agent $a_k$, the router compares the message embedding with the embeddings
of skills assigned to $a_k$ and routes the message to the top-$r_s$ relevant mailboxes. The agent
prompt then receives compact summaries of these mailboxes instead of a flattened context containing
all messages.

Formally, for every incident pair $(m,k)$ with $B_q[m,k]=1$, the mailbox at round $t$ is
\begin{equation}
\mathcal{M}_{q,k,m}^{(t)}
= \{r: r \rightarrow (a_k,s_m),\; \mathrm{time}(r)<t\},
\end{equation}
where $r \rightarrow (a_k,s_m)$ means that message $r$ is routed to skill $s_m$ inside agent
$a_k$. A frozen router scores each incident skill by
\begin{equation}
\gamma_m(r)=\operatorname{sim}(\operatorname{Enc}(r),\mathbf{e}_m),
\end{equation}
and routes $r$ to the top-$r_s$ skills among those assigned to $a_k$. At round $t$, the prompt is
constructed as
\begin{equation}
\begin{aligned}
\mathbf{s}_{q,k}^{(t)}
&= \operatorname{Summarize}(\mathcal{M}_{q,k}^{(t)}),\\
\mathcal{P}_{k,\mathrm{sys}}^{(t)}
&= \{\mathcal{L}_{q,k}, \texttt{State}_k^{(t)}\},\\
\mathcal{P}_{k,\mathrm{usr}}^{(t)}
&= \left\{q,\mathbf{s}_{q,k}^{(t)}\right\}.
\end{aligned}
\end{equation}
Each mailbox summary is capped at $b$ tokens, and all baselines are evaluated under the same total
context budget.

\subsection{Mailbox Summary Format}
\label{app:mailbox_summary}

We use a fixed summary format for each mailbox:

\begin{tcolorbox}[
  enhanced,
  breakable,
  colback=mailbg,
  colframe=mailframe,
  boxrule=0.7pt,
  arc=2mm,
  left=2mm,
  right=2mm,
  top=1.5mm,
  bottom=1.5mm,
  title=\textbf{\textsc{\textsc{Sigma} Mailbox}} \hfill \mailtag{routed evidence},
  coltitle=black,
  colbacktitle=white,
  fonttitle=\small\bfseries,
  attach boxed title to top left={xshift=2mm,yshift=-1.8mm},
  boxed title style={
    colback=white,
    colframe=mailframe,
    boxrule=0.5pt,
    arc=1mm
  }
]
\small
\begin{tabularx}{\linewidth}{@{}p{0.19\linewidth}Y@{}}
\mailkey{Box} &
\mailval{Mailbox[\{skill\}]} \\

\mailkey{Src} &
\mailval{slot \{id\}, sim=\{score\}} \\

\mailkey{Signal} &
\mailval{\{answer / evidence / tool output\}} \\

\mailkey{Summary} &
\mailval{\{routed message summary\}} \\

\mailkey{Hint} &
\mailval{\{next-step usage\}} \\
\end{tabularx}
\end{tcolorbox}

\begin{tcolorbox}[
  enhanced,
  breakable,
  colback=mailbg,
  colframe=mailframe,
  boxrule=0.7pt,
  arc=2mm,
  left=2mm,
  right=2mm,
  top=1.5mm,
  bottom=1.5mm,
  title=\textbf{\textsc{Sigma Mailbox Example I}} \hfill \mailtag{problem\_decomposition},
  coltitle=black,
  colbacktitle=white,
  fonttitle=\small\bfseries,
  attach boxed title to top left={xshift=2mm,yshift=-1.8mm},
  boxed title style={
    colback=white,
    colframe=mailframe,
    boxrule=0.5pt,
    arc=1mm
  }
]
\small
\begin{tabularx}{\linewidth}{@{}p{0.19\linewidth}Y@{}}

\mailkey{Box} &
\mailval{\texttt{Mailbox[problem\_decomposition]}} \\

\mailkey{Src} &
\mailval{\texttt{slot 0}, \texttt{sim=0.087}} \\

\mailkey{Signal} &
\mailval{\texttt{answer = C}} \\

\mailkey{Summary} &
\mailval{
The predecessor agent identifies option C as the best answer.
Its analysis suggests that the question asks for a skill that is not central to planning.
It contrasts conceptual, analytical, and communication skills with IT and computing skills, arguing that the latter are useful but less fundamental in the planning context.
} \\

\mailkey{Hint} &
\mailval{
Use this routed message to decompose the decision into core planning skills and non-essential supporting skills.
} \\

\end{tabularx}
\end{tcolorbox}

The summary budget is capped by $b$ tokens per mailbox. Baselines are given the same total context
budget, so the mailbox design changes only message organization rather than increasing available
context.

\begin{tcolorbox}[
  enhanced,
  breakable,
  colback=mailbg,
  colframe=mailframe,
  boxrule=0.7pt,
  arc=2mm,
  left=2mm,
  right=2mm,
  top=1.5mm,
  bottom=1.5mm,
  title=\textbf{\textsc{Sigma Mailbox Example II}} \hfill \mailtag{elimination},
  coltitle=black,
  colbacktitle=white,
  fonttitle=\small\bfseries,
  attach boxed title to top left={xshift=2mm,yshift=-1.8mm},
  boxed title style={
    colback=white,
    colframe=mailframe,
    boxrule=0.5pt,
    arc=1mm
  }
]
\small
\begin{tabularx}{\linewidth}{@{}p{0.19\linewidth}Y@{}}

\mailkey{Box} &
\mailval{\texttt{Mailbox[elimination]}} \\

\mailkey{Src} &
\mailval{\texttt{slot 0}, \texttt{sim=0.102}} \\

\mailkey{Signal} &
\mailval{\texttt{answer = C}} \\

\mailkey{Summary} &
\mailval{
The predecessor agent selects option C and supports it by eliminating less plausible alternatives.
The analysis argues that IT and computing skills, although useful, are not the most fundamental management skills for effective planning.
} \\

\mailkey{Hint} &
\mailval{
Use this routed evidence to rule out options describing useful but non-core planning skills, while keeping option C as the strongest candidate.
} \\

\end{tabularx}
\end{tcolorbox}

\subsection{Fallback Routing}
\label{app:fallback_routing}

If all router scores are low or the selected agent has only one assigned skill, the message is routed
to the highest-scoring incident skill. In implementation, we also keep a lightweight general state
for each agent slot, but only skill-specific mailbox summaries are exposed as capability-grounded
execution context. This fallback prevents message loss while preserving the operational meaning of
the incidence matrix.

\section{Pseudo-Label Construction}
\label{app:pseudo_labels}

\subsection{Training Skill Library}
\label{app:training_skill_library}

The training skill library $\mathcal{L}_{\mathrm{train}}$ is constructed only from training-split
sources, including role descriptions, tool/API documents, retrieval procedures, and execution traces.
Held-out queries, held-out answers, held-out execution traces, and held-out tool/API cards are not
used during pseudo-label construction.

\subsection{Deterministic Incidence Pseudo-Labels}
\label{app:teacher_labels}

For each training task, we construct a deterministic pseudo-label incidence matrix
$B_q^\star$ using \texttt{teacher\_light\_label}. These labels are not human annotations and are not
generated by an LLM. The labeler retrieves the top $R=8$ candidate skills from
$\mathcal{L}_{\mathrm{train}}$ and assigns at most $k_s=3$ skills to each of the $K=5$ agent slots.
The selected skills define the columns of $B_q^\star$ and provide supervision for the incidence
predictor.

The label files do not contain ground-truth communication edges. Therefore, when edge supervision is
needed during training, we use a chain prior as structural supervision rather than assuming annotated
edge labels.

\subsection{Pseudo-Label Format}
\label{app:pseudo_label_prompt}

The deterministic pseudo-labeler outputs selected skill names for each slot. The output is stored in
a lightweight JSON-style format:

\begin{tcolorbox}[
  enhanced,
  breakable,
  colback=plbg,
  colframe=plframe,
  boxrule=0.7pt,
  arc=2mm,
  left=2mm,
  right=2mm,
  top=1.5mm,
  bottom=1.5mm,
  title=\textbf{\textsc{Pseudo-label Template}} \hfill \pltag{incidence supervision},
  coltitle=black,
  colbacktitle=white,
  fonttitle=\small\bfseries,
  attach boxed title to top left={xshift=2mm,yshift=-1.8mm},
  boxed title style={
    colback=white,
    colframe=plframe,
    boxrule=0.5pt,
    arc=1mm
  }
]
\small
\begin{tabularx}{\linewidth}{@{}p{0.18\linewidth}Y@{}}
\plkey{Id} &
\plval{\{benchmark split and example id\}} \\

\plkey{Query} &
\plval{\{task input or question text\}} \\

\plkey{S} &
\plval{\{task-relevant skill groups or candidate skill sets\}} \\

\plkey{B} &
\plval{\{binary skill-agent incidence matrix, where $B_{i,k}=1$ assigns skill $i$ to agent slot $k$\}} \\
\end{tabularx}
\end{tcolorbox}

Here is an example: Figure~\ref{fig:pseudo_label_example}

\begin{figure}[t]
\centering
\begin{minipage}{0.96\linewidth}
\begin{tcolorbox}[
  enhanced,
  colback=skillbg,
  colframe=skillframe,
  boxrule=0.7pt,
  arc=2mm,
  left=2mm,
  right=2mm,
  top=1.5mm,
  bottom=1.5mm,
  title=\textbf{\textsc{Pseudo-label Example}},
  coltitle=black,
  colbacktitle=white,
  fonttitle=\small\bfseries,
  attach boxed title to top left={xshift=2mm,yshift=-1.8mm},
  boxed title style={
    colback=white,
    colframe=skillframe,
    boxrule=0.5pt,
    arc=1mm
  }
]
\begin{lstlisting}[style=pseudolabel]
{
  "id": "mmlu-val-abstract_algebra-0",
  "query": "The cyclic subgroup of Z_24 generated by 18 has order
    Option A: 4
    Option B: 8
    Option C: 12
    Option D: ...",
  "S": [
    ["analogy", "counterexample", "domain_translation"],
    ["counterexample", "domain_translation", "problem_decomposition"]
  ],
  "B": [
    [0, 0, 0, 0, 0],
    [0, 0, 0, 0, 1],
    [0, 0, 0, 0, 0],
    [1, 0, 0, 0, 0],
    [0, 0, 0, 0, 0],
    [0, 0, 0, 0, 0],
    [0, 1, 1, 1, 0],
    [0, 0, 1, 1, 1],
    [1, 1, 0, 0, 0],
    [0, 0, 0, 1, 1],
    [0, 0, 0, 0, 0],
    [1, 1, 1, 0, 0]
  ]
}
\end{lstlisting}
\end{tcolorbox}
\end{minipage}
\caption{
Example pseudo-label for skill-agent incidence supervision. 
The field \texttt{S} records task-relevant skill groups, while \texttt{B} is a binary incidence matrix indicating which skills are assigned to each agent slot.
}
\label{fig:pseudo_label_example}
\end{figure}

\subsection{Canonical Slot Ordering}
\label{app:canonical_ordering}

Pseudo-label columns follow the canonical slot order used by the training pipeline. This avoids
introducing an additional bipartite matching objective between predicted slots and target agents.
When multiple assignments are equivalent, ties are resolved deterministically by the labeler.

\section{Experimental Protocol}
\label{app:experimental_protocol}

\subsection{Matched-Crew Evaluation}
\label{app:matched_crew}

All compared methods use the same number of agent slots $K$, the same communication-round budget
$T$, and the same total context budget. This ensures that the comparison isolates how agent nodes are
constructed and how messages are routed, rather than allowing gains from larger crews or longer
contexts.

\subsection{Held-Out Skill-Composition Setting}
\label{app:heldout_composition}

The held-out skill-composition setting evaluates whether a method can solve tasks requiring unseen
combinations of skills. Training tasks and test tasks may share individual skill cards, but the
specific multi-skill bundles required at test time are held out from training. This setting measures
compositional generalization over reusable capabilities. As shown in \ref{app:ood_diagnostic}

\subsection{Test-Time Skill-Insertion Setting}
\label{app:skill_insertion}

The test-time skill-insertion setting evaluates whether a trained designer can use newly appended
tool/API skills without retraining the topology decoder. At test time, new skill cards are serialized
with the same template as training cards and encoded by the frozen encoder. The incidence generator
and skill-aware topology decoder then operate over the expanded library
$\mathcal{L}_{\mathrm{train}} \cup \mathcal{L}_{\mathrm{new}}$.

A method succeeds in this setting only if it can assign the newly inserted skills to appropriate
agent slots and use them during execution. This distinguishes genuine expandability from merely
memorizing training-time role or tool configurations.

\subsection{Baselines and Controlled Variants}
\label{app:baselines}

We compare \textsc{Sigma} with role-based, topology-based, and skill-based controls. The key
controlled variants are summarized in Table~\ref{tab:baseline_controls}.

\begin{table}[t]
\centering
\small
\setlength{\tabcolsep}{4.5pt}
\renewcommand{\arraystretch}{1.15}
\caption{
Baselines and controlled variants used in our experimental protocol. 
}
\begin{tabularx}{\linewidth}{@{}p{0.22\linewidth}p{0.25\linewidth}Y@{}}
\toprule
\rowcolor{sigmagreen}
\textbf{Category} & \textbf{Method} & \textbf{Controlled Design Factor} \\
\midrule

\textbf{Multi-agent baselines} 
& LLM-Debate 
& Uses role-specialized agents for argument exchange, without explicit reusable skill assignment. \\

\rowcolor{sigmagray}
\textbf{Topology baselines} 
& GPTSwarm 
& Optimizes communication structures over predefined agents, but does not predict skill-agent incidence. \\

& $\mathtt{G}$-$\mathtt{Designer}$
& Generates task-adaptive topologies over fixed agent nodes, isolating topology design from capability composition. \\

\rowcolor{sigmagray}
\textbf{Skill assignment variants} 
& Flat-skill designer 
& Retrieves a global skill set for the whole team without assigning skills to individual agent slots. \\

& Embedding-only incidence 
& Assigns skills using frozen embedding similarity, removing the learned incidence generator. \\

\rowcolor{sigmagray}
\textbf{Execution variant} 
& \textsc{Sigma} w/o mailbox 
& Keeps predicted skill assignments but flattens incoming messages into a shared agent context. \\

\textbf{Upper-bound diagnostic} 
& Oracle incidence 
& Uses gold or pseudo-label incidence assignments at test time; included only as a diagnostic upper bound. \\

\bottomrule
\end{tabularx}

\label{tab:baseline_controls}
\end{table}

\subsection{Evaluation Metrics}
\label{app:metrics}

The primary metric is downstream task accuracy under the same evaluator used by each benchmark. We
also report the source-to-unseen performance drop:
\begin{equation}
\Delta = \mathrm{Acc}_{\mathrm{source}} - \mathrm{Acc}_{\mathrm{unseen}}.
\end{equation}
A smaller $\Delta$ indicates stronger robustness when the available skill library changes.

When target incidence labels are available, we report micro-averaged precision, recall, and F1 over
skill-agent assignments. Let $T_{\mathrm{inc}}$, $P_{\mathrm{inc}}$, and $G_{\mathrm{inc}}$ denote the
true-positive, predicted-positive, and gold-positive incidence counts over all $(q,m,k)$, respectively:
\begin{align}
\mathrm{Prec}_{\mathrm{inc}}
&= T_{\mathrm{inc}}/P_{\mathrm{inc}}, \\
\mathrm{Rec}_{\mathrm{inc}}
&= T_{\mathrm{inc}}/G_{\mathrm{inc}}, \\
\mathrm{F1}_{\mathrm{inc}}
&=
2/(\mathrm{Prec}_{\mathrm{inc}}^{-1}+\mathrm{Rec}_{\mathrm{inc}}^{-1}).
\end{align}

When annotated communication graphs are available, we report edge precision, recall, and F1 over directed
edges:
\begin{align}
\mathrm{Prec}_{\mathrm{edge}}
&=
\frac{|E_q \cap E_q^\star|}{|E_q|}, \\
\mathrm{Rec}_{\mathrm{edge}}
&=
\frac{|E_q \cap E_q^\star|}{|E_q^\star|}, \\
\mathrm{F1}_{\mathrm{edge}}
&=
\frac{2\cdot \mathrm{Prec}_{\mathrm{edge}}\cdot \mathrm{Rec}_{\mathrm{edge}}}
{\mathrm{Prec}_{\mathrm{edge}}+\mathrm{Rec}_{\mathrm{edge}}}.
\end{align}

For test-time skill insertion, we additionally report inserted-skill hit rate:
\begin{equation}
\mathrm{Hit}_{\mathrm{new}}
=
\frac{1}{|\mathcal{Q}_{\mathrm{new}}|}
\sum_{q \in \mathcal{Q}_{\mathrm{new}}}
\mathbb{I}\!\left[h_{\mathrm{new}}(q)\right],
\end{equation}
where
\begin{equation}
h_{\mathrm{new}}(q)
=
\exists m \in \mathcal{L}_{\mathrm{new}},\ k
\ \mathrm{s.t.}\ 
B_q[m,k]=1.
\end{equation}

\section{Implementation Details}
\label{app:implementation_details}

\subsection{Hyperparameters}
\label{app:hyperparameters}

Table~\ref{tab:hyperparameters} lists the main hyperparameters used by \textsc{Sigma}. 
Values that depend on a benchmark-specific execution budget are selected on the validation split and then fixed for test evaluation.

\begin{table}[t]
\centering
\footnotesize
\setlength{\tabcolsep}{4pt}
\renewcommand{\arraystretch}{1.18}
\rowcolors{2}{sigmagray}{sigmagray}
\caption{
Main hyperparameters used by \textsc{Sigma}. 
}
\begin{tabularx}{\linewidth}{@{}
>{\raggedright\arraybackslash}p{0.16\linewidth}
>{\raggedright\arraybackslash}X
>{\raggedright\arraybackslash}p{0.31\linewidth}
@{}}
\toprule
\rowcolor{sigmablue}
\textbf{Symbol} & \textbf{Meaning} & \textbf{Setting} \\
\midrule
$K$ & Number of agent slots & 5 \\
$T$ & Communication rounds & Matched to benchmark budget \\
$\tau_{\mathrm{skill}}$ & Skill selection threshold & 0.25 \\
$k_s$ & Maximum skills per agent slot & 3 \\
$R$ & Candidate support size for pseudo-labeling & 8 \\
$r_s$ & Routed skills per incoming message & 2 \\
$\gamma_e$ & Edge prediction threshold & 0.5 \\
$b$ & Mailbox summary budget & Fixed under total context budget \\
$\lambda_e$ & Edge loss weight & 1.0 \\
$\lambda_c$ & Compatibility weight & 1.0 \\
$\lambda_g$ & Graph sparsity weight & $1\times10^{-3}$ \\
\bottomrule
\end{tabularx}
\label{tab:hyperparameters}
\end{table}

\subsection{Training Configuration}
\label{app:training_config}

\begin{figure}[t]
    \centering
    \includegraphics[width=\linewidth]{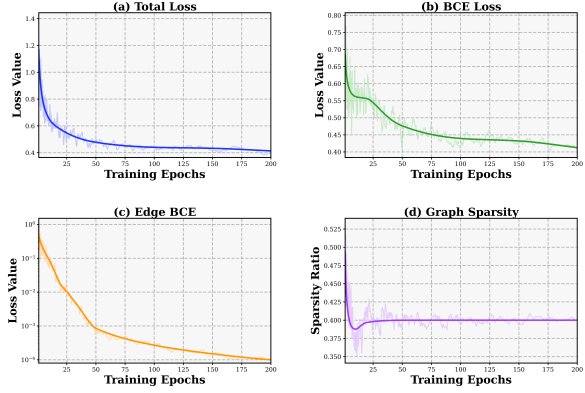}
    \caption{Training dynamic of loss.}
    \label{fig:training_dynamic}
\end{figure}

We summarize the core training configuration of \textsc{Sigma} in
Table~\ref{tab:app_training_config}, as well as the training dynamics of loss-related parameters in Figure~\ref{fig:training_dynamic}. The incidence predictor and topology decoder are trained
jointly with full-batch optimization. Unless otherwise specified, all source-library and
unseen-library experiments use the same architecture, objective, and decoding thresholds.

\begin{table}[t]
\centering
\footnotesize
\setlength{\tabcolsep}{4pt}
\renewcommand{\arraystretch}{1.14}
\rowcolors{2}{sigmagray}{sigmagray}
\caption{
Core training configuration of \textsc{Sigma}.
}
\begin{tabularx}{\linewidth}{@{}
>{\raggedright\arraybackslash}p{0.42\linewidth}
>{\raggedright\arraybackslash}X
@{}}
\toprule
\rowcolor{sigmablue}
\textbf{Item} & \textbf{Setting} \\
\midrule
Training target & Joint training of $F_{\theta}$ and $D_{\phi,\psi}$ \\
Optimizer & AdamW \\
Epochs & 50 for MMLU experiments \\
Learning rate & $1\times 10^{-3}$ \\
Text encoder & \texttt{HashingTextEncoder} \\
Embedding dimension & 384 \\
Hidden dimension of $F_{\theta}$ & 256 \\
Hidden dimension of $D_{\phi,\psi}$ & 128 \\
Number of agent slots $K$ & 5 \\
Top-$r$ candidate skills & 8 \\
Candidate support size & 8 \\
Maximum skills per slot & 3 \\
Edge label mode & \texttt{chain} \\
Anchor mode & \texttt{chain} \\
Acyclic graph constraint & \texttt{true} \\
Edge loss weight & 1.0 \\
Graph sparsity weight & $1\times 10^{-3}$ \\
Compatibility weight & 1.0 \\
Anchor loss weight & 1.0 \\
Random seed & 0 \\
Incidence threshold & 0.25 \\
Edge prediction threshold & 0.5 \\
Temperature & 0.0 \\
\bottomrule
\end{tabularx}
\label{tab:app_training_config}
\end{table}

\subsection{Training Objective}
\label{app:training_objective}

The training objective combines incidence supervision, optional edge reconstruction, and sparsity
regularization. The incidence generator is trained with binary cross-entropy over skill-agent
assignments:
\begin{equation}
\mathcal{L}_{\mathrm{inc}}
=
\sum_q \sum_{m=1}^{M}\sum_{k=1}^{K}
\operatorname{BCE}\left(\mathbf{B}_q^\star[m,k], p_{m,k}\right),
\end{equation}
where $p_{m,k}=\sigma(\alpha_{m,k})$. When annotated communication graphs are available, the topology
decoder is trained with
\begin{equation}
\mathcal{L}_{\mathrm{edge}}
=
\sum_q\sum_{i\neq j}
\operatorname{BCE}(\mathbf{A}_q^\star[i,j],\sigma(\ell_{ij})).
\end{equation}
If annotated target graphs are unavailable, this term is computed against the shared structural prior
described in Appendix~\ref{app:teacher_labels} or omitted in the matched-decoder control. We further
use a soft bundle-size cap and a graph sparsity penalty:
\begin{equation}
\begin{aligned}
\mathcal{L}_{\mathrm{sparse}}
=& \sum_q \sum_{k=1}^{K}
\max\left(0,\sum_{m=1}^{M} p_{m,k} - k_s\right)\\
&+\lambda_g\sum_q\sum_{i\neq j}\sigma(\ell_{ij}).
\end{aligned}
\end{equation}
The final objective is
\begin{equation}
\mathcal{L}
=
\mathcal{L}_{\mathrm{inc}}
+
\lambda_e \mathcal{L}_{\mathrm{edge}}
+
\lambda_s \mathcal{L}_{\mathrm{sparse}}.
\end{equation}
The incidence term trains the skill-agent assignment predictor. The edge term is used only when
annotated communication graphs or a shared structural prior are available. The sparsity term prevents degenerate solutions in which
every slot receives too many skills or the decoded communication graph becomes overly dense.

\subsection{Complexity}
\label{app:complexity}

Let $M$ be the number of skills and $K$ the number of agent slots. Incidence scoring requires
$O(MK)$ skill-agent pair evaluations. The skill-bundle compatibility term is not computed over all
$M^2$ skill pairs. Instead, it is evaluated over sparse active supports. During inference, each agent
has at most $k_s$ selected skills, so pairwise skill compatibility across agent pairs costs
$O(K^2 k_s^2)$. Therefore, the active compatibility computation is controlled by the bundle-size
budget rather than the full skill-library size.

\section{Additional Results}
\label{app:additional_results}

\subsection{Robustness to Unfamiliar Skill Libraries}
\label{app:unseen_skill_library}

Table~\ref{tab:unseen_skill_library} reports the full source-to-unseen library results across
six benchmarks. \textsc{Sigma} achieves the best unseen-library accuracy on every benchmark and shows
the smallest performance drop in all cases. On average, \textsc{Sigma} drops by only 0.95 points,
compared with 2.99 for $\mathtt{G}$-$\mathtt{Designer}$, 4.06 for GPTSwarm, and 4.02 for LLM-Debate. This suggests
that incidence-based skill assignment is less dependent on a fixed skill inventory and can more
reliably compose newly introduced skill cards into effective agent slots.

\subsection{Detailed Token and Runtime Analysis}
\label{app:detailed_token_runtime}

To further quantify the computational cost of \textsc{Sigma}, we report detailed token consumption and wall-clock runtime across six benchmarks and three base LLMs. 
As shown in Table~\ref{tab:token_runtime_by_dataset}, the overall cost varies across benchmarks, mainly depending on the number of evaluated questions and the difficulty of the task. 
GSM8K contributes the largest token usage for all three backbones due to its larger test set, while HumanEval and MMLU require substantially fewer total tokens. 
Despite these differences, \textsc{Sigma} remains practical across model scales: all benchmark evaluations finish within a few hours, and most individual benchmark runs complete within roughly one hour.

\begin{table}[H]
\centering
\footnotesize
\setlength{\tabcolsep}{4.0pt}
\renewcommand{\arraystretch}{1.14}
\caption{
Skill sets used by each dataset.
}
\label{tab:dataset_skill_list}
\begin{tabularx}{\linewidth}{@{}l>{\raggedright\arraybackslash}X@{}}
\toprule
\rowcolor{sigmablue}
\textbf{Dataset} & \textbf{Skills} \\
\midrule

MMLU 
& \texttt{final\_synthesis}, \texttt{problem\_decomposition}, \texttt{evidence\_check}, \texttt{analogy}, \texttt{counterexample}, \texttt{elimination}, \texttt{concept\_recall}, \texttt{constraint\_tracking}, \texttt{causal\_reasoning}, \texttt{uncertainty\_calibration}, \texttt{calculation}, \texttt{domain\_translation} \\

\midrule

GSM8K 
& \texttt{synthesize}, \texttt{verify}, \texttt{retrieve}, \texttt{reason} \\

\midrule

MultiArith 
& \texttt{synthesize}, \texttt{reason}, \texttt{retrieve}, \texttt{verify} \\

\midrule

SVAMP 
& \texttt{synthesize}, \texttt{retrieve}, \texttt{reason}, \texttt{verify} \\

\midrule

AQuA 
& \texttt{synthesize}, \texttt{retrieve}, \texttt{reason}, \texttt{verify} \\

\midrule

HumanEval 
& \texttt{edge\_cases}, \texttt{implementation}, \texttt{final\_code\_review}, \texttt{spec\_parse}, \texttt{complexity\_check} \\

\bottomrule
\end{tabularx}
\end{table}

\section{Prompt Templates}
\label{app:prompts}

\subsection{Agent Execution Prompt}
\label{app:execution_prompt}

\begin{figure}[H]
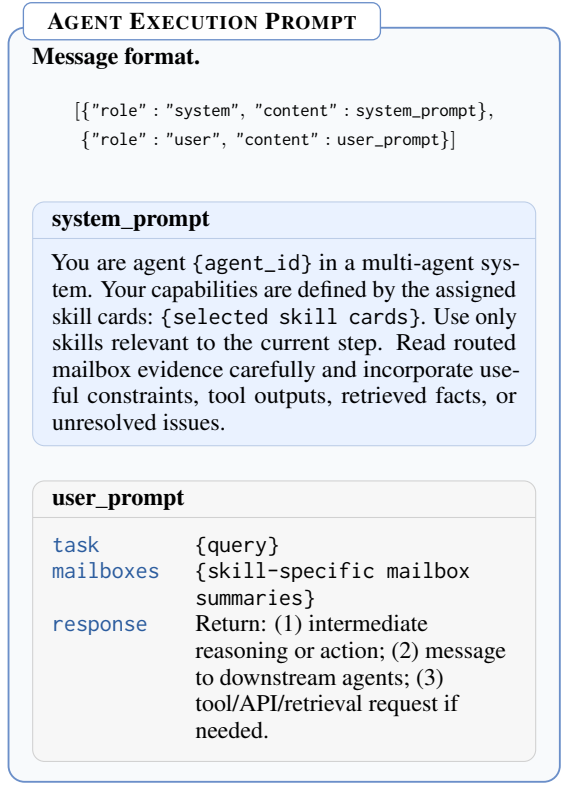

\centering
\begin{minipage}{0.95\linewidth}
\begin{tcolorbox}[
  enhanced,
  colback=promptbg,
  colframe=promptframe,
  boxrule=0.7pt,
  arc=2mm,
  left=2mm,
  right=2mm,
  top=1.5mm,
  bottom=1.5mm,
  title=\textbf{\textsc{Agent Execution Prompt}},
  coltitle=black,
  colbacktitle=white,
  fonttitle=\small\bfseries,
  attach boxed title to top left={xshift=2mm,yshift=-1.8mm},
  boxed title style={
    colback=white,
    colframe=promptframe,
    boxrule=0.5pt,
    arc=1mm
  }
]

\small
\textbf{Message format.}
{\scriptsize
\[
\begin{aligned}
\relax[
&\{\texttt{"role"}:\texttt{"system"},\ \texttt{"content"}:\texttt{system\_prompt}\},\\
&\{\texttt{"role"}:\texttt{"user"},\ \texttt{"content"}:\texttt{user\_prompt}\}
\relax]
\end{aligned}
\]
}

\vspace{1mm}

\begin{tcolorbox}[
  colback=systembg,
  colframe=promptframe!45,
  boxrule=0.4pt,
  arc=1.5mm,
  left=1.5mm,
  right=1.5mm,
  top=1mm,
  bottom=1mm,
  title=\textbf{system\_prompt},
  fonttitle=\footnotesize\bfseries,
  coltitle=black,
  colbacktitle=systembg
]
\footnotesize
You are agent \promptplaceholder{agent\_id} in a multi-agent system. 
Your capabilities are defined by the assigned skill cards:
\promptplaceholder{selected skill cards}. 
Use only skills relevant to the current step. 
Read routed mailbox evidence carefully and incorporate useful constraints, tool outputs, retrieved facts, or unresolved issues.
\end{tcolorbox}

\vspace{1mm}

\begin{tcolorbox}[
  colback=userbg,
  colframe=black!18,
  boxrule=0.4pt,
  arc=1.5mm,
  left=1.5mm,
  right=1.5mm,
  top=1mm,
  bottom=1mm,
  title=\textbf{user\_prompt},
  fonttitle=\footnotesize\bfseries,
  coltitle=black,
  colbacktitle=userbg
]
\footnotesize
\begin{tabularx}{\linewidth}{@{}p{0.24\linewidth}Y@{}}
\promptkey{task} &
\promptplaceholder{query} \\

\promptkey{mailboxes} &
\promptplaceholder{skill-specific mailbox summaries} \\

\promptkey{response} &
Return: 
(1) intermediate reasoning or action; 
(2) message to downstream agents; 
(3) tool/API/retrieval request if needed. \\
\end{tabularx}
\end{tcolorbox}

\end{tcolorbox}
\end{minipage}
\caption{
Prompt format used for agent execution. 
}
\label{fig:agent_execution_prompt}
\end{figure}

During execution, each agent receives a two-part prompt consisting of a system message and a user message. 
As shown in Figure~\ref{fig:agent_execution_prompt}, the system message specifies the agent identity, its assigned skill cards, and the rule that only task-relevant skills should be used. 
The user message provides the current task, routed mailbox summaries from predecessor agents, and the required response format. 
This prompt design ensures that skill assignments are not only used for node construction, but are also exposed to the LLM as executable instructions during multi-agent interaction.
Here is an example: Figure~\ref{fig:agent_execution_prompt_example}

\subsection{Baseline Prompt Templates}
\label{app:baseline_prompts}

For reproducibility, we report the concrete \textbf{Vanilla} and \textbf{CoT} prompt templates used for
single-agent baselines. Each template uses the same two-message format as the agent execution prompt
above. For a given dataset, \texttt{system\_prompt} is constructed by concatenating the dataset-specific
\textit{System base} with either the \textit{Vanilla suffix} or the \textit{CoT suffix}; the
\texttt{user\_prompt} contains the benchmark instance. All methods use the same benchmark split,
deterministic decoding with \texttt{temperature}=0.0, and the same dataset-specific answer parser.

\begin{tcolorbox}[
  enhanced,
  breakable,
  colback=promptbg,
  colframe=promptframe,
  boxrule=0.7pt,
  arc=2mm,
  left=2mm,
  right=2mm,
  top=1.5mm,
  bottom=1.5mm,
  title=\textbf{\textsc{Baseline Prompt Template: MMLU}},
  coltitle=black,
  colbacktitle=white,
  fonttitle=\small\bfseries,
  boxed title style={colback=white,colframe=promptframe,boxrule=0.5pt,arc=1mm}
]
\footnotesize
\textbf{Message format.}
{\scriptsize
\[
\begin{aligned}
\relax[
&\{\texttt{"role"}:\texttt{"system"},\ \texttt{"content"}:\texttt{system\_prompt}\},\\
&\{\texttt{"role"}:\texttt{"user"},\ \texttt{"content"}:\texttt{user\_prompt}\}
\relax]
\end{aligned}
\]
}

\begin{tcolorbox}[
  colback=systembg,
  colframe=promptframe!45,
  boxrule=0.4pt,
  arc=1.5mm,
  left=1.5mm,
  right=1.5mm,
  top=1mm,
  bottom=1mm,
  title=\textbf{system\_prompt},
  fonttitle=\footnotesize\bfseries,
  coltitle=black,
  colbacktitle=systembg
]
\begin{tabularx}{\linewidth}{@{}p{0.23\linewidth}Y@{}}
\pkey{System base} &
I will ask you a multiple-choice question.
There are 4 answer options enumerated as A, B, C, and D.
Only one option is correct.
Always select the best available option from A, B, C, and D, even if none seems perfect.
Do not answer that none of the options is correct. \\

\pkey{Vanilla suffix} &
Reply with only one letter: A, B, C, or D.
Do not include any analysis. \\

\pkey{CoT suffix} &
Reason step by step before choosing the answer.
Use at most 5 short sentences.
Do not write tables, exhaustive cases, or long derivations.
Put your final answer on the last line exactly in this format:
\texttt{Final answer: X}
where X is one of A, B, C, or D. \\
\end{tabularx}
\end{tcolorbox}

\begin{tcolorbox}[
  colback=userbg,
  colframe=black!18,
  boxrule=0.4pt,
  arc=1.5mm,
  left=1.5mm,
  right=1.5mm,
  top=1mm,
  bottom=1mm,
  title=\textbf{user\_prompt},
  fonttitle=\footnotesize\bfseries,
  coltitle=black,
  colbacktitle=userbg
]
\begin{tabularx}{\linewidth}{@{}p{0.23\linewidth}Y@{}}
\pkey{task} &
\texttt{The task is:\textbackslash n\textbackslash n} \\
& \promptplaceholder{question-with-options} \\
\end{tabularx}
\end{tcolorbox}
\end{tcolorbox}

\begin{tcolorbox}[
  enhanced,
  breakable,
  colback=promptbg,
  colframe=promptframe,
  boxrule=0.7pt,
  arc=2mm,
  left=2mm,
  right=2mm,
  top=1.5mm,
  bottom=1.5mm,
  title=\textbf{\textsc{Baseline Prompt Template: AQuA}},
  coltitle=black,
  colbacktitle=white,
  fonttitle=\small\bfseries,
  boxed title style={colback=white,colframe=promptframe,boxrule=0.5pt,arc=1mm}
]
\footnotesize
\textbf{Message format.}
{\scriptsize
\[
\begin{aligned}
\relax[
&\{\texttt{"role"}:\texttt{"system"},\ \texttt{"content"}:\texttt{system\_prompt}\},\\
&\{\texttt{"role"}:\texttt{"user"},\ \texttt{"content"}:\texttt{user\_prompt}\}
\relax]
\end{aligned}
\]
}

\begin{tcolorbox}[
  colback=systembg,
  colframe=promptframe!45,
  boxrule=0.4pt,
  arc=1.5mm,
  left=1.5mm,
  right=1.5mm,
  top=1mm,
  bottom=1mm,
  title=\textbf{system\_prompt},
  fonttitle=\footnotesize\bfseries,
  coltitle=black,
  colbacktitle=systembg
]
\begin{tabularx}{\linewidth}{@{}p{0.23\linewidth}Y@{}}
\pkey{System base} &
I will ask you a multiple-choice math question.
There are 5 answer options enumerated as A, B, C, D, and E.
Only one option is correct.
Always select the best available option from A, B, C, D, and E. \\

\pkey{Vanilla suffix} &
Reply with only one letter: A, B, C, D, or E.
Do not include any analysis. \\

\pkey{CoT suffix} &
Reason step by step before choosing the answer.
Use at most 5 short sentences.
Do not write tables, exhaustive cases, or long derivations.
Put your final answer on the last line exactly in this format:
\texttt{Final answer: X}
where X is one of A, B, C, D, or E. \\
\end{tabularx}
\end{tcolorbox}

\begin{tcolorbox}[
  colback=userbg,
  colframe=black!18,
  boxrule=0.4pt,
  arc=1.5mm,
  left=1.5mm,
  right=1.5mm,
  top=1mm,
  bottom=1mm,
  title=\textbf{user\_prompt},
  fonttitle=\footnotesize\bfseries,
  coltitle=black,
  colbacktitle=userbg
]
\begin{tabularx}{\linewidth}{@{}p{0.23\linewidth}Y@{}}
\pkey{task} &
\texttt{The task is:\textbackslash n\textbackslash n} \\
& \promptplaceholder{question-with-A-E-options} \\
\end{tabularx}
\end{tcolorbox}
\end{tcolorbox}

\begin{tcolorbox}[
  enhanced,
  breakable,
  colback=promptbg,
  colframe=promptframe,
  boxrule=0.7pt,
  arc=2mm,
  left=2mm,
  right=2mm,
  top=1.5mm,
  bottom=1.5mm,
  title=\textbf{\textsc{Baseline Prompt Template: GSM8K, MultiArith, and SVAMP}},
  coltitle=black,
  colbacktitle=white,
  fonttitle=\small\bfseries,
  boxed title style={colback=white,colframe=promptframe,boxrule=0.5pt,arc=1mm}
]
\footnotesize
\textbf{Message format.}
{\scriptsize
\[
\begin{aligned}
\relax[
&\{\texttt{"role"}:\texttt{"system"},\ \texttt{"content"}:\texttt{system\_prompt}\},\\
&\{\texttt{"role"}:\texttt{"user"},\ \texttt{"content"}:\texttt{user\_prompt}\}
\relax]
\end{aligned}
\]
}

\begin{tcolorbox}[
  colback=systembg,
  colframe=promptframe!45,
  boxrule=0.4pt,
  arc=1.5mm,
  left=1.5mm,
  right=1.5mm,
  top=1mm,
  bottom=1mm,
  title=\textbf{system\_prompt},
  fonttitle=\footnotesize\bfseries,
  coltitle=black,
  colbacktitle=systembg
]
\begin{tabularx}{\linewidth}{@{}p{0.23\linewidth}Y@{}}
\pkey{GSM8K base} &
I will ask you a grade-school math word problem.
Solve the problem and provide the numeric answer without units.
Always put your final answer on the last line exactly in this format:
\texttt{The answer is X}
where X is the numeric answer. \\

\pkey{M-Arith base} &
I will ask you a multi-step arithmetic word problem.
Solve the problem and provide the numeric answer without units.
Always put your final answer on the last line exactly in this format:
\texttt{The answer is X}
where X is the numeric answer. \\

\pkey{SVAMP base} &
I will ask you an arithmetic word problem.
Solve the problem and provide the numeric answer without units.
Always put your final answer on the last line exactly in this format:
\texttt{The answer is X}
where X is the numeric answer. \\

\pkey{Vanilla suffix} &
Reply with only the final answer line.
Do not include analysis. \\

\pkey{CoT suffix} &
Reason step by step before giving the final answer.
Keep the reasoning concise. \\
\end{tabularx}
\end{tcolorbox}

\begin{tcolorbox}[
  colback=userbg,
  colframe=black!18,
  boxrule=0.4pt,
  arc=1.5mm,
  left=1.5mm,
  right=1.5mm,
  top=1mm,
  bottom=1mm,
  title=\textbf{user\_prompt},
  fonttitle=\footnotesize\bfseries,
  coltitle=black,
  colbacktitle=userbg
]
\begin{tabularx}{\linewidth}{@{}p{0.23\linewidth}Y@{}}
\pkey{task} &
\texttt{The task is:\textbackslash n\textbackslash n} \\
& \promptplaceholder{math-word-problem} \\
\end{tabularx}
\end{tcolorbox}
\end{tcolorbox}

\begin{tcolorbox}[
  enhanced,
  breakable,
  colback=promptbg,
  colframe=promptframe,
  boxrule=0.7pt,
  arc=2mm,
  left=2mm,
  right=2mm,
  top=1.5mm,
  bottom=1.5mm,
  title=\textbf{\textsc{Baseline Prompt Template: HumanEval}},
  coltitle=black,
  colbacktitle=white,
  fonttitle=\small\bfseries,
  boxed title style={colback=white,colframe=promptframe,boxrule=0.5pt,arc=1mm}
]
\footnotesize
\textbf{Message format.}
{\scriptsize
\[
\begin{aligned}
\relax[
&\{\texttt{"role"}:\texttt{"system"},\ \texttt{"content"}:\texttt{system\_prompt}\},\\
&\{\texttt{"role"}:\texttt{"user"},\ \texttt{"content"}:\texttt{user\_prompt}\}
\relax]
\end{aligned}
\]
}

\begin{tcolorbox}[
  colback=systembg,
  colframe=promptframe!45,
  boxrule=0.4pt,
  arc=1.5mm,
  left=1.5mm,
  right=1.5mm,
  top=1mm,
  bottom=1mm,
  title=\textbf{system\_prompt},
  fonttitle=\footnotesize\bfseries,
  coltitle=black,
  colbacktitle=systembg
]
\begin{tabularx}{\linewidth}{@{}p{0.23\linewidth}Y@{}}
\pkey{System base} &
You will be given a Python function signature and docstring.
Write a correct, concise Python implementation.
Do not change the function name, arguments, or expected return type.
The final answer must include a complete Python code block with the full function definition. \\

\pkey{Vanilla suffix} &
Reply with only one Python code block.
Do not include analysis outside the code block. \\

\pkey{CoT suffix} &
Reason briefly before writing the implementation.
Keep the reasoning concise.
Put the final implementation in the last Python code block.
Do not write anything after the final code block. \\
\end{tabularx}
\end{tcolorbox}

\begin{tcolorbox}[
  colback=userbg,
  colframe=black!18,
  boxrule=0.4pt,
  arc=1.5mm,
  left=1.5mm,
  right=1.5mm,
  top=1mm,
  bottom=1mm,
  title=\textbf{user\_prompt},
  fonttitle=\footnotesize\bfseries,
  coltitle=black,
  colbacktitle=userbg
]
\begin{tabularx}{\linewidth}{@{}p{0.23\linewidth}Y@{}}
\pkey{task} &
\texttt{The task is:\textbackslash n\textbackslash n} \\
& \promptplaceholder{function spec} \\
\end{tabularx}
\end{tcolorbox}
\end{tcolorbox}

\begin{tcolorbox}[
  enhanced,
  breakable,
  colback=promptbg,
  colframe=promptframe,
  boxrule=0.7pt,
  arc=2mm,
  left=2mm,
  right=2mm,
  top=1.5mm,
  bottom=1.5mm,
  title=\textbf{\textsc{Answer Extraction Used with the Prompt Templates}},
  coltitle=black,
  colbacktitle=white,
  fonttitle=\small\bfseries,
  boxed title style={colback=white,colframe=promptframe,boxrule=0.5pt,arc=1mm}
]
\footnotesize
\begin{tabularx}{\linewidth}{@{}p{0.23\linewidth}Y@{}}
\pkey{MMLU/AQuA} &
The parser first ignores any hidden reasoning before a closing \texttt{</think>} tag, then reads a leading standalone option letter, explicit final-answer markers such as \texttt{Final answer: X}, \texttt{correct answer is X}, and \texttt{answer: X}, or option-selection phrases such as \texttt{choose option X}. AQuA additionally supports matching emitted option text back to A--E. \\

\pkey{Numeric math} &
The parser matches explicit answer markers including \texttt{final answer}, \texttt{the answer is}, \texttt{answer is}, \texttt{answer}, and \texttt{\#\#\#\#}. It also supports boxed answers. If no marker is found, it uses the last numeric expression and normalizes commas, currency symbols, percent signs, braces, fractions, decimals, and surrounding punctuation. \\

\pkey{HumanEval} &
We report pass@1. The evaluator extracts the last valid Python code block containing a function or class definition and scores it once with the benchmark unit tests, with no retries, no execution-feedback loop, and no access to held-out tests during generation. \\
\end{tabularx}
\end{tcolorbox}

\section{Additional Details}
\label{app:additional_detail}

\subsection{Dataset description}
\label{app:dataset_description}

We evaluate \textsc{Sigma} on six benchmarks covering general reasoning, mathematical reasoning, and code generation. 
MMLU is used to test broad multiple-choice reasoning ability, while GSM8K, MultiArith, SVAMP, and AQuA evaluate mathematical reasoning under both numerical-answer and multiple-choice formats. 
HumanEval is used to assess code generation ability with pass@1 as the evaluation metric.
Table~\ref{tab:dataset_statistics} summarizes the answer type, metric, test-set size, and license information for each benchmark.
For MMLU, we follow the cost-controlled protocol of prior graph-based MAS designers, including $\mathtt{G}$-$\mathtt{Designer}$ and ARG-Designer, and use the same fixed 153-question shuffled evaluation subset for all methods and base LLMs.
This subset is fixed before method-specific tuning; optimization and pseudo-label construction use separate training/dev examples and do not access held-out answers.

We use all existing benchmark artifacts only for their intended research evaluation purposes: MMLU for multiple-choice reasoning, GSM8K, MultiArith, SVAMP, and AQuA for mathematical reasoning, and HumanEval for code-generation evaluation. 
We do not redistribute modified benchmark data or use these artifacts outside research evaluation settings.

\begin{table}[t]
\centering
\scriptsize
\setlength{\tabcolsep}{2pt}
\renewcommand{\arraystretch}{1.10}
\caption{
Dataset descriptions and statistics. 
}
\begin{tabularx}{\linewidth}{@{}
p{0.20\linewidth}
X
p{0.16\linewidth}
p{0.13\linewidth}
r
p{0.16\linewidth}
@{}}
\toprule
\rowcolor{sigmablue}
\textbf{Category} & \textbf{Dataset} & \textbf{Type} & \textbf{Metric} & \textbf{\#Test} & \textbf{Lic.} \\
\midrule

General
& MMLU & Multi-choice & Acc. & 153 & MIT \\

\midrule
\multirow{4}{*}{Math}
& GSM8K      & Num.       & Acc. & 1,319 & MIT \\
& M-Arith & Num.       & Acc. & 600   & -- \\
& SVAMP      & Num.       & Acc. & 1,000 & MIT \\
& AQuA       & Multi-choice & Acc. & 254   & Apache-2.0 \\

\midrule
Code
& H-Eval & Code & Pass@1 & 164 & MIT \\

\bottomrule
\end{tabularx}

\label{tab:dataset_statistics}
\end{table}

\subsection{Skill distribution}
\label{app:skill_distribution}

\begin{figure}[H]
    \centering
    \includegraphics[width=\linewidth]{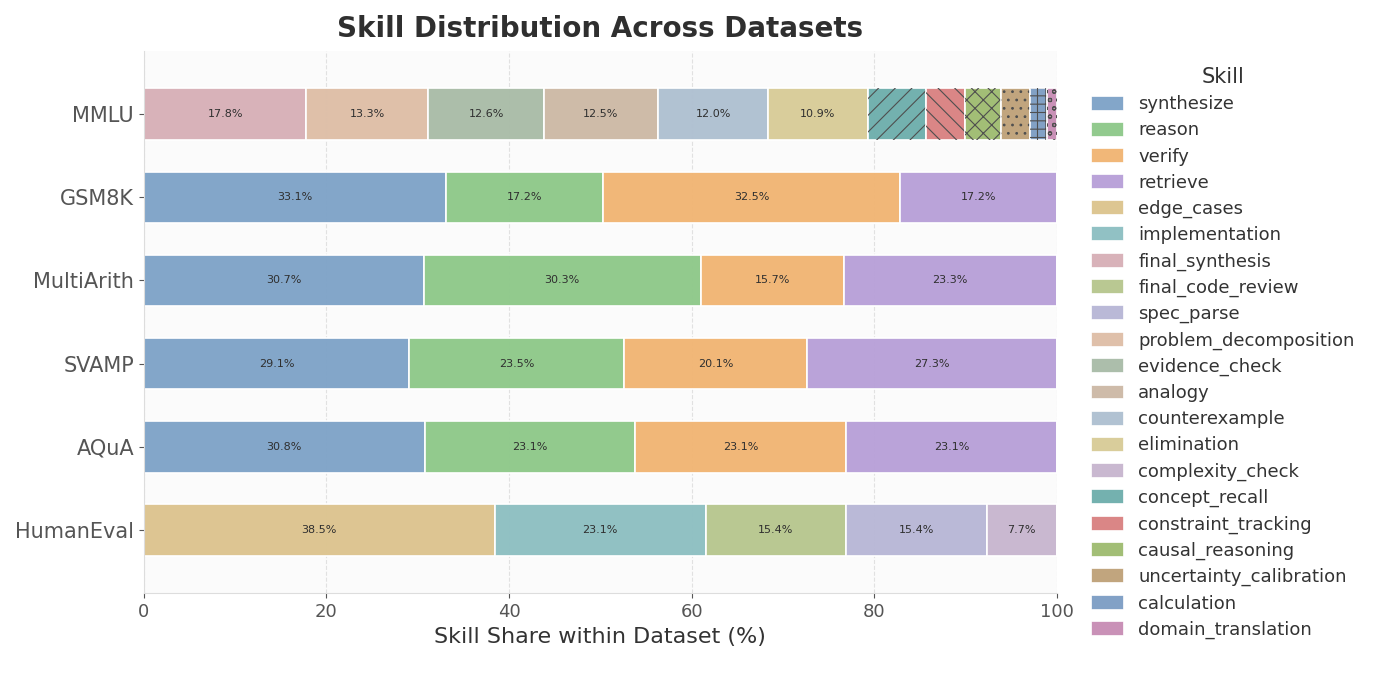}
    \caption{
    Skill distribution across datasets.
    }
    \label{fig:skill_distribution}
\end{figure}

Figure~\ref{fig:skill_distribution} and Table~\ref{tab:dataset_skill_list} show the normalized skill distribution across different datasets.
Overall, the skill distributions reveal clear task-dependent specialization.
To complement this distributional view, Table~\ref{tab:skill_library_statistics} reports the corresponding skill-library statistics, including the number of source and unseen skill cards, the cognitive/tool-API split, the average number of skills assigned to each agent where available, and the source--unseen library overlap.
Together, these results show not only which skills are frequently selected for each dataset, but also how the underlying skill libraries are constructed and controlled across source and unseen settings.

\begin{table*}[t]
\centering
\footnotesize
\setlength{\tabcolsep}{3.8pt}
\renewcommand{\arraystretch}{1.12}
\caption{
Controlled comparison between \textsc{Single-Agent+Skills} and \textsc{Sigma} using \textit{GPT-OSS-120B}.
\textsc{Single-Agent+Skills} receives the same union of selected skill cards as \textsc{Sigma}, but removes multi-agent execution, topology decoding, and mailbox routing.
}
\label{tab:single_agent_skills_control}
\resizebox{\textwidth}{!}{%
\begin{tabular}{lccccccccccc}
\toprule
\rowcolor{sigmablue}
\textbf{Method}
& \textbf{Mul.}
& \textbf{Topo.}
& \textbf{Comp.}
& \textbf{Mailbox}
& \textbf{MMLU}
& \textbf{GSM8K}
& \textbf{MultiArith}
& \textbf{SVAMP}
& \textbf{AQuA}
& \textbf{HumanEval}
& \textbf{Avg.} \\
\midrule
CoT
& \xmark
& \xmark
& \xmark
& \xmark
& 83.39
& 84.81
& 95.63
& 95.19
& 76.02
& 90.87
& 87.65 \\
\midrule

\textsc{Single-Agent+Skills}
& \xmark
& \xmark
& \cmark
& \xmark
& 85.41
& 87.65
& 95.93
& 96.73
& 88.37
& 93.24
& 91.22 \\

\rowcolor{sigmablue!8}
\textsc{Sigma}
& \cmark
& \cmark
& \cmark
& \cmark
& 87.28
& 96.25
& 98.34
& 95.94
& 90.16
& 95.83
& 93.97 \\

\bottomrule
\end{tabular}
}
\end{table*}

\begin{table*}[t]
\centering
\footnotesize
\setlength{\tabcolsep}{4pt}
\renewcommand{\arraystretch}{1.12}
\caption{
Skill-library statistics for the source and unseen skill libraries.
$M_{\mathrm{src}}$ and $M_{\mathrm{unseen}}$ denote the number of skill cards in the source and unseen libraries.
Cog. and Tool/API report the number of cognitive and tool/API skills in the source/unseen libraries.
}
\resizebox{\textwidth}{!}{%
\begin{tabular}{lcccccc>{\raggedright\arraybackslash}p{0.39\textwidth}}
\toprule
\rowcolor{sigmablue}
\textbf{Dataset}
& $\boldsymbol{M_{\mathrm{src}}}$
& $\boldsymbol{M_{\mathrm{unseen}}}$
& \textbf{Cog. S/U}
& \textbf{Tool/API S/U}
& \textbf{Avg. skills/agent S/U}
& \textbf{Overlap}
& \textbf{Examples: source $\rightarrow$ unseen} \\
\midrule

MMLU, half split
& 6
& 6
& 6 / 6
& 0 / 0
& 2.80 / 2.88
& 0
& \makecell[l]{
\texttt{concept\_recall} $\rightarrow$ \texttt{analogy};\\
\texttt{calculation} $\rightarrow$ \texttt{uncertainty\_calibration}
} \\

\midrule

GSM8K
& 12
& 12
& 12 / 12
& 0 / 0
& -- / --
& 0
& \texttt{source\_extract\_numbers} $\rightarrow$ \texttt{target\_quantity\_schema} \\

\midrule

MultiArith
& 12
& 12
& 12 / 12
& 0 / 0
& -- / --
& 0
& \texttt{source\_choose\_operation} $\rightarrow$ \texttt{target\_operation\_router} \\

\midrule

SVAMP
& 12
& 12
& 12 / 12
& 0 / 0
& -- / --
& 0
& \texttt{source\_unit\_tracking} $\rightarrow$ \texttt{target\_answer\_unit\_check} \\

\midrule

AQuA
& 12
& 12
& 12 / 12
& 0 / 0
& -- / --
& 0
& \texttt{source\_consistency\_check} $\rightarrow$ \texttt{target\_estimate\_validate} \\

\midrule

HumanEval
& 12
& 12
& 12 / 10
& 0 / 2
& -- / 2.42
& 0
& \makecell[l]{
\texttt{source\_spec\_parse} $\rightarrow$ \texttt{target\_contract\_reader}\\
\texttt{source\_test\_design} $\rightarrow$ \texttt{target\_local\_test}\\
\phantom{\texttt{source\_test\_design} $\rightarrow$ }\texttt{\_designer}
} \\

\bottomrule
\end{tabular}
}
\label{tab:skill_library_statistics}
\end{table*}

\subsection{Model Size and Computation Budget}
\label{app:model_size_budget}

Table~\ref{tab:model_size} reports the size of the base executors and the trainable components introduced by \textsc{Sigma}. 
The base LLMs are used only as frozen executors during multi-agent inference, while \textsc{Sigma} adds a lightweight controller consisting of the incidence predictor $F_{\theta}$ and the topology decoder $D_{\phi,\psi}$.
The controller has fewer than one million trainable parameters in total, making the additional training cost small compared with the base LLM executors.

For execution, we use three representative backbones with different serving configurations.
\textit{Qwen3-8B} and \textit{GPT-OSS-120B} are served locally with vLLM, where \textit{Qwen3-8B} disables the thinking template and \textit{GPT-OSS-120B} uses reasoning effort.
\textit{GPT-4o-mini} is accessed through an OpenAI-compatible backend.
All \textsc{Sigma} controller training runs are conducted on $4$ $\times$ NVIDIA RTX A6000 GPUs 48GB.

\begin{table}[t]
\centering
\footnotesize
\setlength{\tabcolsep}{3.2pt}
\renewcommand{\arraystretch}{1.15}
\caption{
Model and artifact sizes used by \textsc{Sigma}. 
}
\begin{tabularx}{\linewidth}{@{}
>{\raggedright\arraybackslash}p{0.35\linewidth}
>{\raggedright\arraybackslash}p{0.20\linewidth}
>{\raggedright\arraybackslash}X
@{}}
\toprule
\rowcolor{sigmablue}
\textbf{Component} & \textbf{Size} & \textbf{Role} \\
\midrule

\rowcolor{sigmagray}
\multicolumn{3}{@{}l}{\textbf{Base LLM executors}} \\
Qwen3-8B 
& 8B class 
& Local vLLM executor for the fixed MMLU-153 evaluation protocol. \\

GPT-OSS-120B 
& 120B class 
& Executor for MMLU, GSM8K, MultiArith, SVAMP, AQuA, HumanEval, and ablations. \\

GPT-4o-mini 
& Undisclosed 
& Auxiliary executor for additional benchmark artifacts. \\

\midrule
\rowcolor{sigmagray}
\multicolumn{3}{@{}l}{\textbf{\textsc{Sigma} trainable controller}} \\
$F_{\theta}$ incidence predictor 
& 461,953 params 
& Predicts the skill-agent incidence matrix $B$. \\

$D_{\phi,\psi}$ topology decoder 
& 427,266 params 
& Predicts the learned communication topology. \\

Full \textsc{Sigma} checkpoint 
& $\approx$3.57 MB 
& Contains $F_{\theta}+D_{\phi,\psi}$, with 889,219 trainable parameters in total. \\

Incidence-only checkpoint 
& 164,997 params / 664 KB 
& Early variant using only $F_{\theta}$ without learned topology decoding. \\

\bottomrule
\end{tabularx}

\label{tab:model_size}
\end{table}

\subsection{OOD Diagnostic Analysis}
\label{app:ood_diagnostic}

In the MMLU Split-A setting, ID examples contain observed skill-bundle patterns, whereas OOD examples contain the held-out bundle 
\{\texttt{elimination}, \texttt{evidence\_check}, \texttt{final\_synthesis}\}, 
testing whether \textsc{Sigma} generalizes to unseen skill compositions within the same benchmark.
In the skill-library transfer setting, ID uses the source skill library, while OOD replaces it with semantically different unseen skill cards at test time, testing whether the controller can transfer across changed capability resources.

Table~\ref{tab:ood_diagnostic} reports ID/OOD accuracy, OOD performance change, and OOD retention.
Across both diagnostics, \textsc{Sigma} maintains strong OOD accuracy and non-negative OOD change.
These results suggest that its gains are not merely due to memorizing in-distribution skill assignments, but also to recomposing useful capabilities under shifted skill-composition and skill-library conditions.

\begin{table}[t]
\centering
\small
\setlength{\tabcolsep}{4.0pt}
\renewcommand{\arraystretch}{1.12}
\caption{
OOD diagnostic results.
ID denotes the in-distribution or source-library setting, while OOD denotes the held-out or unseen-library setting.
$\Delta_{\text{OOD}}$ is computed as OOD accuracy minus ID accuracy, where higher values indicate better OOD retention.
}
\label{tab:ood_diagnostic}
\begin{tabularx}{\linewidth}{@{}l>{\raggedright\arraybackslash}Xcccc@{}}
\toprule
\rowcolor{sigmablue}
\textbf{Diag.}
& \textbf{Method}
& \textbf{ID}
& \textbf{OOD}
& $\boldsymbol{\Delta_{\text{OOD}}}$ $\uparrow$
& \textbf{Ret.} $\uparrow$ \\
\midrule

\multirow{2}{*}{\makecell{MMLU\\Split-A}}
& Fixed-chain
& 75.00
& 75.00
& $+0.00$
& 100.00 \\

& \cellcolor{sigmagray}\textbf{\textsc{Sigma}}
& \cellcolor{sigmagray}\textbf{82.50}
& \cellcolor{sigmagray}\textbf{85.00}
& \cellcolor{sigmagray}\textbf{$+2.50$}
& \cellcolor{sigmagray}\textbf{103.03} \\

\midrule

\multirow{2}{*}{\makecell{Skill-\\Lib.}}
& $\mathtt{G}$-$\mathtt{Designer}$
& 77.12
& 77.78
& $+0.65$
& 100.85 \\

& \cellcolor{sigmagray}\textbf{\textsc{Sigma}}
& \cellcolor{sigmagray}\textbf{77.12}
& \cellcolor{sigmagray}\textbf{79.08}
& \cellcolor{sigmagray}\textbf{$+1.96$}
& \cellcolor{sigmagray}\textbf{102.54} \\

\bottomrule
\end{tabularx}

\vspace{0.6mm}
\begin{minipage}{0.96\linewidth}
\footnotesize
\textit{Split details.}
MMLU Split-A uses 
\{\texttt{elimination}, \texttt{evidence\_check}, \texttt{final\_synthesis}\}
as the held-out OOD bundle.
Skill-Lib. evaluates source-to-unseen skill-card transfer at test time.
\end{minipage}
\end{table}

\begin{figure}[t]
\centering
\begin{minipage}{0.96\linewidth}
\begin{tcolorbox}[
  enhanced,
  colback=promptbg,
  colframe=promptframe,
  boxrule=0.7pt,
  arc=2mm,
  left=2mm,
  right=2mm,
  top=1.5mm,
  bottom=1.5mm,
  title=\textbf{\textsc{Agent Execution Prompt Example}},
  coltitle=black,
  colbacktitle=white,
  fonttitle=\small\bfseries,
  attach boxed title to top left={xshift=2mm,yshift=-1.8mm},
  boxed title style={
    colback=white,
    colframe=promptframe,
    boxrule=0.5pt,
    arc=1mm
  }
]

\footnotesize
\textbf{Message format.}
{\scriptsize
\[
\begin{aligned}
\relax[
&\{\texttt{"role"}:\texttt{"system"},\ \texttt{"content"}:\texttt{system\_prompt}\},\\
&\{\texttt{"role"}:\texttt{"user"},\ \texttt{"content"}:\texttt{user\_prompt}\}
\relax]
\end{aligned}
\]
}

\vspace{1mm}

\begin{tcolorbox}[
  colback=systembg,
  colframe=promptframe!45,
  boxrule=0.4pt,
  arc=1.5mm,
  left=1.5mm,
  right=1.5mm,
  top=1mm,
  bottom=1mm,
  title=\textbf{system\_prompt},
  fonttitle=\footnotesize\bfseries,
  coltitle=black,
  colbacktitle=systembg
]
\begin{tabularx}{\linewidth}{@{}p{0.27\linewidth}Y@{}}
\pkey{skill bundle} &
\texttt{Evidence checking}: verify that the answer follows from stated evidence.
Inputs: \texttt{query, messages}; outputs: \texttt{analysis, answer}; tools: \texttt{none}; tags: \texttt{mmlu}. \\

\pkey{profile state} &
No accumulated interaction history before execution. \\

\pkey{format constraint} &
The agent receives one question with four options \texttt{A--D}; exactly one option is correct.
It must select the best available option, use other agents' reasoning only as critical advice, reply in fewer than 100 words, and put only one letter on the first line. \\
\end{tabularx}
\end{tcolorbox}

\vspace{1mm}

\begin{tcolorbox}[
  colback=userbg,
  colframe=black!18,
  boxrule=0.4pt,
  arc=1.5mm,
  left=1.5mm,
  right=1.5mm,
  top=1mm,
  bottom=1mm,
  title=\textbf{user\_prompt},
  fonttitle=\footnotesize\bfseries,
  coltitle=black,
  colbacktitle=userbg
]
\begin{tabularx}{\linewidth}{@{}p{0.27\linewidth}Y@{}}
\pkey{task} &
What one of the following is \emph{not} a key management skill in planning?\\
& \texttt{Option A: Conceptual skills}\\
& \texttt{Option B: Analytical skills}\\
& \texttt{Option C: IT and computing skills}\\
& \texttt{Option D: Communication skills} \\

\pkey{mailbox context} &
\texttt{Mailbox[evidence\_check]} receives routed evidence from \textsc{Sigma} slot 0 with similarity score \texttt{0.032}. 
The predecessor selects \texttt{C} and argues that conceptual, analytical, and communication skills are core managerial competencies in planning, while technical/IT skills are less central. \\

\pkey{expected output} &
Return the answer letter and a concise evidence-based explanation. \\
\end{tabularx}
\end{tcolorbox}

\end{tcolorbox}
\end{minipage}
\caption{
Example runtime prompt used by \textsc{Sigma}. 
The system message injects the assigned skill card, profile state, and answer-format constraints, while the user message provides the concrete question and routed mailbox evidence.
}
\label{fig:agent_execution_prompt_example}
\end{figure}

\section{Failure Analysis}
\label{app:failure_analysis}

We further inspect representative failure cases of \textsc{Sigma} to understand when skill-incidence based multi-agent design is less effective. 
Overall, the failures mainly come from five sources: overly sparse topology decoding, insufficient skill diversity, biased final aggregation, benchmark-specific answer priors, and imperfect mailbox routing.

\paragraph{Overly sparse communication topology.}
A notable failure mode is that the decoded communication graph can become empty when the edge threshold is too strict or when the learned topology scores are overly sparse. 
In an over-sparse MMLU-153 diagnostic variant, all 153 evaluated examples were decoded as zero-edge graphs, which effectively prevents agents from exchanging information.
In this case, \textsc{Sigma} degenerates into isolated skill-composed agents, losing the benefit of multi-agent collaboration. 
This suggests that topology sparsity should be regularized but not over-constrained. 
A practical fix is to calibrate the edge threshold on the validation split, enforce a minimum number of edges, or add a topology loss that discourages fully disconnected graphs.

\paragraph{Correct minority lost during aggregation.}
In 9 out of 27 MMLU wrong cases, at least one agent produces the correct answer, but the final decision still follows the wrong majority. 
This suggests that simple majority-style aggregation can suppress useful minority evidence. 
The issue is not necessarily that \textsc{Sigma} fails to generate the right reasoning, but that the final decision mechanism does not sufficiently distinguish high-quality evidence from repeated but weak agreement. 
A natural improvement is evidence-weighted aggregation, where final answers are weighted by skill relevance, confidence, mailbox evidence quality, or verifier judgments rather than raw vote count.

\paragraph{Answer-option bias.}
We also observe that wrong predictions sometimes overuse particular answer options, especially A or D. 
This points to a possible option-prior or prompt-order bias in multiple-choice tasks. 
Such bias may come from the base LLM, the answer-format constraint, or the distribution of predecessor messages in the mailbox. 
To diagnose this issue, we recommend option-order robustness tests, where the same question is evaluated under different permutations of answer choices. 
If predictions change substantially under permutation, the system should use option-normalized prompting or answer re-mapping during evaluation.


\section{Impact Statement}
\label{app:impact}
\paragraph{Positive impact.}
This work studies skill-incidence graphs for compositional multi-agent system design.
By constructing agents from task-conditioned skill bundles rather than fixed role descriptions, \textsc{Sigma} may improve the modularity, reusability, and adaptability of multi-agent systems.

\paragraph{Risks and mitigation.}
Flexible skill composition may introduce risks when unsafe, irrelevant, or conflicting skills are assigned to agents, especially in tool-augmented, code-execution, or real-world decision-making settings.
Incorrect skill routing can propagate errors across agents and make system behavior harder to audit.
To mitigate these risks, \textsc{Sigma} should be used with bounded skill libraries, explicit execution constraints, assignment logging, and human oversight in high-stakes applications.
Our experiments are limited to controlled benchmarks, and we do not recommend direct deployment in safety-critical domains without further validation and robustness evaluation.

\clearpage
\begin{table}[t]
\centering
\footnotesize
\setlength{\tabcolsep}{3pt}
\renewcommand{\arraystretch}{1.08}
\caption{
Token comparison across representative benchmarks and multi-agent methods.
}
\resizebox{\linewidth}{!}{%
\begin{tabular}{llcr}
\toprule
\rowcolor{sigmablue}
\textbf{Dataset} & \textbf{Method} & \textbf{Perf.} & \textbf{Tokens / Runtime} \\
\midrule

\multirow{6}{*}{MMLU}
& \textsc{Sigma}      & \textbf{87.28} & $1.18{\times}10^{5}$ / $18.0min$ \\
& \textsc{CARD}       & 85.78          & $7.23{\times}10^{5}$ / $20.5min$ \\
& ARG-Designer        & 85.63          & $5.57{\times}10^{5}$ / $41.2min$ \\
& $\mathtt{G}$-$\mathtt{Designer}$ & 85.06 & $5.25{\times}10^{5}$ / $67.9min$ \\
& GPTSwarm            & 83.58          & $2.61{\times}10^{6}$ / $55.1min$ \\
& LLM-Debate          & 82.92          & $1.54{\times}10^{6}$ / $75.4min$ \\

\midrule

\multirow{6}{*}{HumanEval}
& \textsc{Sigma}      & \textbf{95.83} & $4.24{\times}10^{5}$ / $30.9min$ \\
& \textsc{CARD}       & 93.98          & $7.28{\times}10^{5}$ / $70.2min$ \\
& ARG-Designer        & 93.47          & $8.71{\times}10^{5}$ / $51.6min$ \\
& $\mathtt{G}$-$\mathtt{Designer}$ & 92.88 & $7.05{\times}10^{5}$ / $50.8min$ \\
& GPTSwarm            & 92.30          & $4.57{\times}10^{6}$ / $56.9min$ \\
& LLM-Debate          & 91.43          & $6.90{\times}10^{6}$ / $68.6min$ \\

\midrule

\multirow{6}{*}{GSM8K}
& \textsc{Sigma}      & \textbf{96.25} & $5.48{\times}10^{6}$ / $84.1min$ \\
& \textsc{CARD}       & 93.44          & $9.91{\times}10^{6}$ / $154.1min$  \\
& ARG-Designer        & 91.08          & $1.06{\times}10^{7}$ / $149.4min$ \\
& $\mathtt{G}$-$\mathtt{Designer}$ & 88.62 & $8.09{\times}10^{6}$ / $201.3min$ \\
& GPTSwarm            & 86.88          & $1.44{\times}10^{7}$ / $241.2min$ \\
& LLM-Debate          & 85.53          & $2.63{\times}10^{7}$ / $267.7min$ \\

\midrule

\multirow{6}{*}{AQuA}
& \textsc{Sigma}      & \textbf{90.16} & $8.22{\times}10^{5}$ / $62.9min$ \\
& \textsc{CARD}       & 82.76          & $1.76{\times}10^{6}$ / $72.9min$ \\
& ARG-Designer        & 78.62          & $1.09{\times}10^{6}$ / $67.7min$ \\
& $\mathtt{G}$-$\mathtt{Designer}$ & 78.13 & $9.01{\times}10^{5}$ / $92.1min$ \\
& GPTSwarm            & 77.41          & $2.07{\times}10^{6}$ / $109.2min$ \\
& LLM-Debate          & 75.38          & $3.12{\times}10^{6}$ / $144.8min$ \\

\midrule

\multirow{6}{*}{SVAMP}
& \textsc{Sigma}      & \textbf{95.94} & $2.14{\times}10^{6}$ / $85.5min$ \\
& \textsc{CARD}       & 95.48          & $8.01{\times}10^{6}$ / $117.5min$ \\
& ARG-Designer        & 94.81          & $7.17{\times}10^{6}$ / $104.8min$ \\
& $\mathtt{G}$-$\mathtt{Designer}$ & 95.21 & $3.07{\times}10^{6}$ / $101.2min$ \\
& GPTSwarm            & 94.77          & $2.97{\times}10^{7}$ / $101.3min$ \\
& LLM-Debate          & 94.89          & $3.55{\times}10^{7}$ / $165.9min$ \\

\midrule

\multirow{6}{*}{MultiArith}
& \textsc{Sigma}      & 98.34          & $8.62{\times}10^{5}$ / $48.1min$ \\
& \textsc{CARD}       & 98.21          & $2.38{\times}10^{6}$ / $64.2min$ \\
& ARG-Designer        & \textbf{98.67} & $1.14{\times}10^{6}$ / $78.7min$ \\
& $\mathtt{G}$-$\mathtt{Designer}$ & 96.86 & $1.09{\times}10^{6}$ / $95.4min$ \\
& GPTSwarm            & 96.50          & $2.00{\times}10^{6}$ / $125.5min$ \\
& LLM-Debate          & 95.72          & $3.10{\times}10^{6}$ / $173.4min$ \\

\bottomrule
\end{tabular}
}

\label{tab:token_runtime_by_dataset}
\end{table}

\begin{table}[t]
\centering
\small
\setlength{\tabcolsep}{4.0pt}
\renewcommand{\arraystretch}{1.12}
\caption{
Robustness to unfamiliar skill libraries across benchmarks using GPT-4o-mini as the base model. 
$\Delta$ denotes the performance drop from the source skill library to unseen skill cards, 
with lower values indicating stronger generalization under library changes.
}
\label{tab:unseen_skill_library}
\begin{tabularx}{\linewidth}{@{}l>{\raggedright\arraybackslash}Xccc@{}}
\toprule
\rowcolor{sigmablue}
\textbf{Bench.}
& \textbf{Method}
& \textbf{Source Lib.}
& \textbf{Unseen Lib.}
& $\boldsymbol{\Delta}$ $\downarrow$ \\
\midrule

\multirow{4}{*}{MMLU}
& LLM-Debate
& 68.96 & 63.73 & 5.23 \\
& GPTSwarm
& 71.37 & 64.84 & 6.53 \\
& $\mathtt{G}$-$\mathtt{Designer}$
& 73.32 & 70.05 & 3.27 \\
& \textsc{CARD}
& 75.11 & 73.68 & 1.43 \\
& \cellcolor{sigmagray}\textbf{\textsc{Sigma}}
& \cellcolor{sigmagray}\textbf{75.68}
& \cellcolor{sigmagray}\textbf{75.03}
& \cellcolor{sigmagray}\textbf{0.65} \\

\midrule

\multirow{4}{*}{GSM8K}
& LLM-Debate
& 87.32 & 85.29 & 2.03 \\
& GPTSwarm
& 87.83 & 84.47 & 3.36 \\
& $\mathtt{G}$-$\mathtt{Designer}$
& 89.26 & 87.28 & 1.98 \\
& \textsc{CARD}
& 93.21 & 91.44 & 1.77 \\
& \cellcolor{sigmagray}\textbf{\textsc{Sigma}}
& \cellcolor{sigmagray}\textbf{92.96}
& \cellcolor{sigmagray}\textbf{91.86}
& \cellcolor{sigmagray}\textbf{1.10} \\

\midrule

\multirow{4}{*}{AQuA}
& LLM-Debate
& 72.86 & 66.96 & 5.90 \\
& GPTSwarm
& 74.65 & 69.14 & 5.51 \\
& $\mathtt{G}$-$\mathtt{Designer}$
& 76.60 & 70.3 & 6.30 \\
& \textsc{CARD}
& 80.41 & 76.85 & 3.56 \\
& \cellcolor{sigmagray}\textbf{\textsc{Sigma}}
& \cellcolor{sigmagray}\textbf{83.67}
& \cellcolor{sigmagray}\textbf{82.50}
& \cellcolor{sigmagray}\textbf{1.17} \\

\midrule

\multirow{4}{*}{Human.}
& LLM-Debate
& 79.4 & 72.58 & 6.82 \\
& GPTSwarm
& 81.26 & 76.41 & 4.85 \\
& $\mathtt{G}$-$\mathtt{Designer}$
& 82.48 & 78.45 & 4.03 \\
& \textsc{CARD}
& 85.07 & 81.48 & 3.59 \\
& \cellcolor{sigmagray}\textbf{\textsc{Sigma}}
& \cellcolor{sigmagray}\textbf{88.71}
& \cellcolor{sigmagray}\textbf{86.29}
& \cellcolor{sigmagray}\textbf{2.42} \\

\midrule

\multirow{4}{*}{SVAMP}
& LLM-Debate
& 87.43 & 85.46 & 1.97 \\
& GPTSwarm
& 90.12 & 88.14 & 1.98 \\
& $\mathtt{G}$-$\mathtt{Designer}$
& 90.32 & 89.17 & 1.15 \\
& \textsc{CARD} 
& 91.65 & 90.79 & 0.86 \\
& \cellcolor{sigmagray}\textbf{\textsc{Sigma}}
& \cellcolor{sigmagray}\textbf{93.64}
& \cellcolor{sigmagray}\textbf{93.43}
& \cellcolor{sigmagray}\textbf{0.21} \\

\midrule

\multirow{4}{*}{Multi.}
& LLM-Debate
& 94.11 & 91.96 & 2.15 \\
& GPTSwarm
& 96.07 & 93.93 & 2.14 \\
& $\mathtt{G}$-$\mathtt{Designer}$
& 96.75 & 95.50 & 1.25 \\
& \textsc{CARD}
& 96.94 & 95.95 & 0.99 \\
& \cellcolor{sigmagray}\textbf{\textsc{Sigma}}
& \cellcolor{sigmagray}\textbf{99.64}
& \cellcolor{sigmagray}\textbf{99.46}
& \cellcolor{sigmagray}\textbf{0.18} \\

\bottomrule
\end{tabularx}
\end{table}

\clearpage
\onecolumn
\raggedbottom
\section{Case Study}
\label{app:case_study}
\subsection{HumanEval: Communication-Linked Code Generation }

\begin{normalcasebox}{Qualitative Study Setting}
\caseblock{Evaluation Setting}
\begin{tabularx}{\linewidth}{@{}L R@{}}
Method
& \texttt{SIGMA-humaneval}. \\[0.25em]

Dataset / split
& \texttt{HumanEval} test. \\[0.25em]

Base LLM
& \texttt{gpt-4o-mini}. \\[0.25em]

Decision method
& \texttt{FinalWriteCode}. \\[0.25em]

Run summary
& $124$ executed problems, $110$ solved problems, $88.71\%$ accuracy. \\[0.25em]

Interpretation focus
& This case illustrates how communication exposes edge cases and turns them into an executable final implementation. \\
\end{tabularx}

\vspace{0.8em}
\hrule
\vspace{0.55em}

\caseblock{Learned Topology}

\begin{tabularx}{\linewidth}{@{}L R@{}}
Node ids
& \texttt{5euR} (slot 0),
  \texttt{5Rg5} (slot 1),
  \texttt{7b2z} (slot 2),
  \texttt{7Tja} (slot 3),
  \texttt{4mKN} (slot 4). \\[0.25em]

Spatial chain
& \texttt{5euR -> 5Rg5 -> 7b2z -> 7Tja -> 4mKN}. \\[0.25em]

Temporal edges
& None. \\[0.25em]

Edge probabilities
& $0{\to}1=0.9987$,
  $1{\to}2=0.9981$,
  $2{\to}3=0.9982$,
  $3{\to}4=0.9986$. \\[0.25em]

Graph constraint
& \texttt{anchor=chain};
  \texttt{acyclic=true}. \\
\end{tabularx}

\vspace{0.8em}
\hrule
\vspace{0.55em}

\caseblock{Reading Guide}

\begin{tabularx}{\linewidth}{@{}L R@{}}
\ding{182}Query
& Function signature, docstring, examples, ground-truth implementation, and final prediction. \\[0.25em]

\ding{183}Trace
& Selected skill bundles, chain topology, execution order, and round-level message flow. \\[0.25em]

\ding{184}Agent cards
& Each card summarizes one skill-composed node and its main contribution. \\[0.25em]

\ding{185}Final
& The \texttt{FinalWriteCode} decision node that writes the final Python solution. \\
\end{tabularx}
\end{normalcasebox}

\begin{normalcasebox}{Case: Edge-case repair for \texttt{iscube}}

\caseentryword{Query}{querybg}{queryframe}{Function specification and expected behavior}
{
\textbf{Record id:} \texttt{HumanEval/77}. \quad
\textbf{Solved:} yes.\\
\textbf{Task:} implement \texttt{iscube(a)}, which returns \texttt{True} if integer \texttt{a} is a cube of some integer.\\
\textbf{Examples:}
\texttt{iscube(1) -> True},
\texttt{iscube(2) -> False},
\texttt{iscube(-1) -> True},
\texttt{iscube(64) -> True},
\texttt{iscube(0) -> True},
\texttt{iscube(180) -> False}.\\
\textbf{Ground-truth pattern:} use the absolute value or signed cube-root handling, then check whether the rounded cube root cubed equals the original magnitude.\\
\textbf{Final prediction:} a negative-aware cube-root check with \texttt{root ** 3 == a}.
}

\caseentryword{Trace}{querybg}{queryframe}{Skill-incidence and message flow}
{
\textbf{Execution order:}
\texttt{5euR -> 5Rg5 -> 7b2z -> 7Tja -> 4mKN}.\\
\textbf{Slot 0 skills:}
\{\texttt{implementation}, \texttt{edge\_cases}, \texttt{final\_code\_review}\}.\\
\textbf{Slot 1 skills:}
\{\texttt{edge\_cases}, \texttt{final\_code\_review}\}.\\
\textbf{Slot 2 skills:}
\{\texttt{spec\_parse}, \texttt{implementation}, \texttt{edge\_cases}\}.\\
\textbf{Slot 3 skills:}
\{\texttt{implementation}, \texttt{edge\_cases}\}.\\
\textbf{Slot 4 skills:}
\{\texttt{spec\_parse}, \texttt{edge\_cases}, \texttt{complexity\_check}\}.\\
\textbf{Interpretation:}
the learned chain routes the task from planning and algorithm description to code generation, testing feedback, and final bug fixing.
}
\casecardsplit

\caseentry{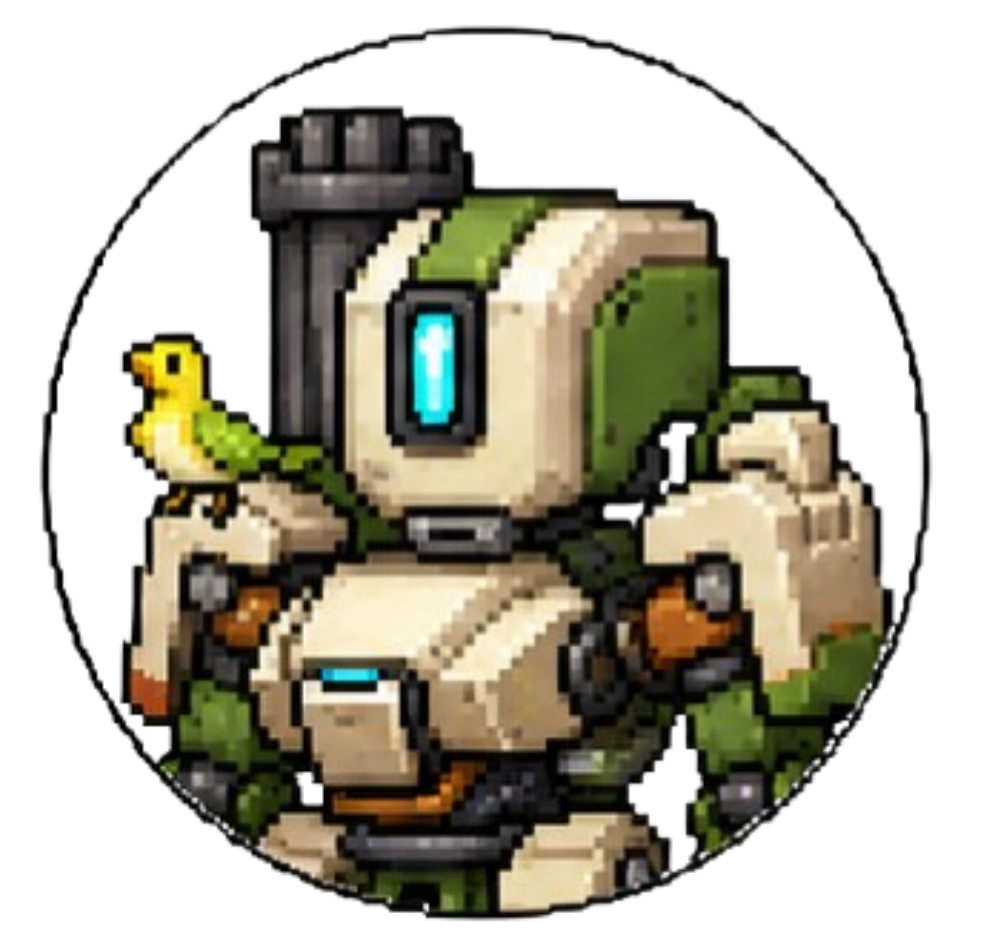}{agent0bg}{agent0frame}{\texttt{5euR}: implementation + edge-case analysis + final review}
{
\textbf{Role/profile state:} \texttt{SIGMA slot 0}; no accumulated history before execution.\\
\textbf{Mailbox input:} empty, because this is the first node in the chain.\\
\textbf{Main response:} proposes a single-function design and suggests checking whether the rounded cube root cubed equals the input.\\
\textbf{Representative code pattern:} \texttt{return round(a ** (1/3)) ** 3 == a}.\\
\textbf{Diagnosis:} useful high-level implementation plan, but the initial pattern is incomplete for negative inputs and may be numerically fragile.
}

\caseentry{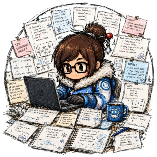}{agent1bg}{agent1frame}{\texttt{5Rg5}: edge-case analysis + final review}
{
\textbf{Mailbox input:} receives the slot-0 design through the chain edge \texttt{5euR -> 5Rg5}.\\
\textbf{Main response:} restates the cube-root algorithm, documents expected usage, and emphasizes the equality check after rounding.\\
\textbf{Evidence contribution:} preserves the concise algorithmic structure and keeps the implementation simple.\\
\textbf{Diagnosis:} useful algorithmic clarification, but it still inherits the same cube-root risk from the initial design.
}

\caseentry{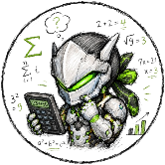}{agent2bg}{agent2frame}{\texttt{7b2z}: specification parsing + implementation + edge-case analysis}
{
\textbf{Mailbox input:} receives the slot-1 algorithm description through \texttt{5Rg5 -> 7b2z}.\\
\textbf{Main response:} writes an executable implementation using \texttt{round(a ** (1/3))}.\\
\textbf{Internal-test signal:} the generated implementation passes internal testing in the trace.\\
\textbf{Diagnosis:} partially useful because it produces runnable code, but risky because the round-then-cube method can mishandle negative inputs or large integers.
}

\caseentry{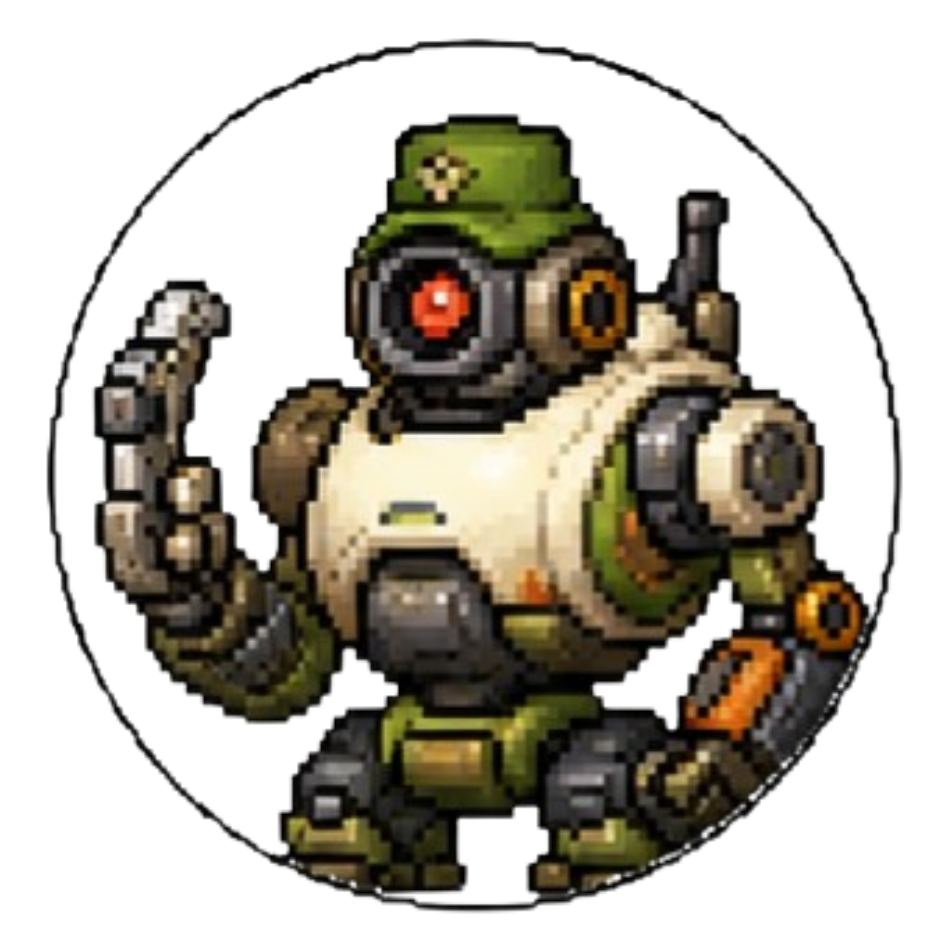}{agent3bg}{agent3frame}{\texttt{7Tja}: implementation + edge-case analysis}
{
\textbf{Mailbox input:} receives the slot-2 code through \texttt{7b2z -> 7Tja}.\\
\textbf{Main response:} warns that the implementation may fail due to floating-point rounding and highlights negative numbers and large integers as important edge cases.\\
\textbf{Concrete checks suggested:} \texttt{iscube(-8)}, \texttt{iscube(-1)}, and large cube values such as \texttt{729} or \texttt{1000000000}.\\
\textbf{Diagnosis:} this is the key corrective signal: the edge-case skill identifies the hidden weakness of the earlier implementation.
}

\caseentry{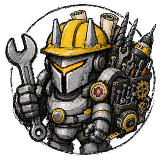}{agent4bg}{agent4frame}{\texttt{4mKN}: specification parsing + edge-case analysis + complexity check}
{
\textbf{Mailbox input:} receives the tester feedback through \texttt{7Tja -> 4mKN}.\\
\textbf{Main response:} revises the implementation to handle negative inputs by computing a signed cube-root candidate before checking \texttt{root ** 3 == a}.\\
\textbf{Complexity:} constant-time arithmetic with no auxiliary data structures.\\
\textbf{Diagnosis:} integrates the upstream warning into a bug-fixed implementation, showing how the final slot acts as a repair node rather than another independent generator.
}
\casecardsplit

\caseentry{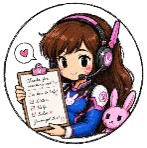}{finalbg}{finalframe}{\texttt{4sMD}: \texttt{FinalWriteCode} decision node}
{
\textbf{System role:} final code writer; must output the completed Python function.\\
\textbf{Final input pattern:} task specification plus the chain-produced implementation and edge-case feedback.\\
\textbf{Final behavior:} outputs a negative-aware implementation: if \texttt{a < 0}, compute a signed cube-root candidate using \texttt{-(-a) ** (1/3)}; otherwise use \texttt{a ** (1/3)}; finally check \texttt{root ** 3 == a}.\\
\textbf{Final result:} solved.\\
\textbf{Interpretation:} the case shows a successful repair path. The early implementation nodes provide a simple candidate solution, the edge-case node exposes the negative-input risk, and the final writer incorporates that correction into the submitted code.
}
\end{normalcasebox}
\casecaption{
A HumanEval case where \textsc{Sigma}'s chain topology turns a simple but risky cube-root implementation into a negative-aware final solution.
}{fig:case_humaneval_iscube}

\subsection{MMLU: Skill-Conditioned Evidence Aggregation}

\begin{normalcasebox}{Qualitative Study Setting}
\caseblock{Evaluation Setting}
\begin{tabularx}{\linewidth}{@{}L R@{}}
Base LLM and split
& \texttt{Qwen3-8B} on the fixed \texttt{MMLU-153} evaluation subset. \\[0.25em]

Overall run
& $153$ executed questions, $94$ solved questions, $61.44\%$ accuracy. \\[0.25em]

Execution regime
& Five \texttt{AnalyzeAgent} nodes produce skill-conditioned answers, followed by one \texttt{FinalRefer} decision node. \\[0.25em]

Interpretation focus
& These cases examine whether skill-composed agents provide complementary evidence and whether the final decision can repair or amplify noisy intermediate answers. \\
\end{tabularx}

\vspace{0.8em}
\hrule
\vspace{0.55em}

\caseblock{Skill-conditioned Aggregation Regime}

\begin{tabularx}{\linewidth}{@{}L R@{}}
Node representation
& Each case constructs $U_q \in \mathbb{R}^{5 \times 384}$ from the selected skill bundles before graph decoding. \\[0.25em]

Observed graph state
& The MMLU examples activate no pairwise spatial edge after thresholding, so the final decision node aggregates independently generated skill-conditioned evidence. \\[0.25em]

What this shows
& The analysis isolates the benefit and limitation of skill composition and evidence aggregation, rather than attributing the MMLU examples to agent-to-agent message passing. \\
\end{tabularx}

\vspace{0.8em}
\hrule
\vspace{0.55em}

\caseblock{Reading Guide}

\begin{tabularx}{\linewidth}{@{}L R@{}}
\ding{182}Query
& Question, options, record id, ground-truth answer, and final prediction. \\[0.25em]

\ding{183}Trace
& Concrete node ids, execution order, skill-incidence bundles, activated edges, and maximum decoded edge probability. \\[0.25em]

\ding{184}Agent cards
& The five executable columns selected by the skill-incidence matrix $B_q$. Each card lists the node id, role, injected skill profile, mailbox state, and agent output. \\[0.25em]

\ding{185}Final
& The \texttt{FinalRefer} node. Its user prompt contains the task plus the five node outputs shown immediately above the final card; the final response is restricted to one letter. \\
\end{tabularx}

\end{normalcasebox}

\begin{normalcasebox}{Case: Answer-label repair on MMLU}

\caseentryword{Query}{querybg}{queryframe}{Question and expected answer}
{
\textbf{Record id:} \texttt{mmlu-test-miscellaneous-39}. \quad
\textbf{Solved:} yes.\\
\textbf{Question:} In the film \textit{The Talented Mr. Ripley}, who plays Mr. Ripley?\\
\textbf{Options:}
A. Jude Law \quad
B. Matt Damon \quad
C. Dustin Hoffman \quad
D. Ben Affleck.\\
\textbf{Ground truth:} B.\\
\textbf{Final prediction:} B.\\
\textbf{Case focus:} the compact answer prefixes are noisy, but the rationales mostly identify Matt Damon; the final decision node repairs the answer-label mismatch.
}

\caseentryword{Trace}{querybg}{queryframe}{Topology and skill-incidence trace}
{
\textbf{Nodes:}
\texttt{mLV2} (slot 0),
\texttt{hj9X} (slot 1),
\texttt{wxdu} (slot 2),
\texttt{3M9o} (slot 3),
\texttt{d7XF} (slot 4).\\
\textbf{Execution order:}
\texttt{mLV2 -> hj9X -> wxdu -> 3M9o -> d7XF}.\\
\textbf{Slot 0 skills:}
\{\texttt{elimination}, \texttt{analogy}, \texttt{evidence\_check}\}.\\
\textbf{Slot 1 skills:}
\{\texttt{evidence\_check}\}.\\
\textbf{Slot 2 skills:}
\{\texttt{final\_synthesis}\}.\\
\textbf{Slot 3 skills:}
\{\texttt{counterexample}, \texttt{final\_synthesis}\}.\\
\textbf{Slot 4 skills:}
\{\texttt{counterexample}\}.\\
\textbf{Activated edges:}
\texttt{spatial\_edges=[]}, \texttt{temporal\_edges=[]};
maximum decoded edge probability $=3.16{\times}10^{-3}<0.5$.\\
\textbf{Interpretation:}
the case isolates skill-conditioned evidence aggregation rather than agent-to-agent message passing.
}
\casecardsplit

\caseentry{picture/case/A0.png}{agent0bg}{agent0frame}{\texttt{mLV2}: elimination + analogy + evidence check}
{
\textbf{Role/profile state:} \texttt{SIGMA slot 0}; no accumulated interaction history before execution.\\
\textbf{Injected skill profile:} option elimination rules out implausible choices; analogy mapping connects the query to a known pattern; evidence checking verifies that the answer follows from stated evidence.\\
\textbf{Mailbox input:} empty because no predecessor output arrived.\\
\textbf{Agent output:} A. The rationale recalls that \textit{The Talented Mr. Ripley} stars Matt Damon as the titular character, eliminates other options by filmography, and confirms Matt Damon as the correct answer.\\
\textbf{Diagnosis:} useful evidence, but the leading answer prefix conflicts with the rationale and the option mapping.
}

\caseentry{picture/case/A1.png}{agent1bg}{agent1frame}{\texttt{hj9X}: evidence check}
{
\textbf{Role/profile state:} \texttt{SIGMA slot 1}; no accumulated interaction history before execution.\\
\textbf{Injected skill profile:} evidence checking verifies that the answer follows from stated evidence.\\
\textbf{Mailbox input:} empty because no predecessor output arrived.\\
\textbf{Agent output:} A. The rationale identifies the target role, recalls that Matt Damon portrayed Mr. Ripley in the 1999 film, and selects the answer based on this information.\\
\textbf{Diagnosis:} useful evidence, again with a wrong leading letter.
}

\caseentry{picture/case/A2.png}{agent2bg}{agent2frame}{\texttt{wxdu}: final synthesis}
{
\textbf{Role/profile state:} \texttt{SIGMA slot 2}; no accumulated interaction history before execution.\\
\textbf{Injected skill profile:} final synthesis combines partial arguments into a concise final answer.\\
\textbf{Mailbox input:} empty because no predecessor output arrived.\\
\textbf{Agent output:} A. The response identifies the film, recalls the actor who portrayed Mr. Ripley, but incorrectly states that Jude Law played the role.\\
\textbf{Diagnosis:} wrong factual hypothesis; this is the main misleading card in the aggregation set.
}

\caseentry{picture/case/A3.png}{agent3bg}{agent3frame}{\texttt{3M9o}: counterexample + final synthesis}
{
\textbf{Role/profile state:} \texttt{SIGMA slot 3}; no accumulated interaction history before execution.\\
\textbf{Injected skill profile:} counterexample search looks for edge cases that falsify an answer; final synthesis combines partial arguments.\\
\textbf{Mailbox input:} empty because no predecessor output arrived.\\
\textbf{Agent output:} A. The rationale identifies Matt Damon as the actor who portrayed Mr. Ripley in \textit{The Talented Mr. Ripley}.\\
\textbf{Diagnosis:} useful corrective evidence despite the noisy answer prefix.
}

\caseentry{picture/case/A4.png}{agent4bg}{agent4frame}{\texttt{d7XF}: counterexample}
{
\textbf{Role/profile state:} \texttt{SIGMA slot 4}; no accumulated interaction history before execution.\\
\textbf{Injected skill profile:} counterexample search looks for edge cases that falsify an answer.\\
\textbf{Mailbox input:} empty because no predecessor output arrived.\\
\textbf{Agent output:} A. The rationale states that Matt Damon portrayed the character in the 1999 film and explicitly ends with \texttt{Answer: B}.\\
\textbf{Diagnosis:} explicitly exposes the mismatch between the noisy prefix A and the correct textual answer B.
}
\casecardsplit

\caseentry{picture/case/A5.png}{finalbg}{finalframe}{\texttt{UmWw}: \texttt{FinalRefer} decision node}
{
\textbf{System role:} top decision-maker; must output exactly one of A, B, C, or D.\\
\textbf{FinalRefer input:} the prompt contains the task/options plus all five node outputs above: \texttt{mLV2: A}, \texttt{hj9X: A}, \texttt{wxdu: A}, \texttt{3M9o: A}, and \texttt{d7XF: A ... Answer: B}.\\
\textbf{Final response:} B.\\
\textbf{Final result:} solved.\\
\textbf{Interpretation:} the final node is not a raw majority vote over answer prefixes. It reads the rationales and repairs the answer-label noise because most cards point to Matt Damon and one card explicitly states \texttt{Answer: B}.
}

\end{normalcasebox}
\casecaption{
A complete case where the ffnal decision node repairs answer-label noise by reading the agent rationales.

}{fig:case_mmlu_answer_label_repair}

\begin{normalcasebox}{Case B: Counterexample agents correct a noisy majority}

\caseentryword{Query}{querybg}{queryframe}{Question and expected answer}
{
\textbf{Record id:} \texttt{mmlu-test-abstract\_algebra-167}. \quad
\textbf{Solved:} yes.\\
\textbf{Question:} Statement 1: If a finite group has order $n$ then the group contains a subgroup of order $d$ for every positive divisor $d$ of $n$. Statement 2: If $a$ belongs to a finite group then $|a|$ divides $|G|$.\\
\textbf{Options:}
A. True, True \quad
B. False, False \quad
C. True, False \quad
D. False, True.\\
\textbf{Ground truth:} D.\\
\textbf{Final prediction:} D.\\
\textbf{Case focus:} three agents produce noisy compact prefixes A, but most rationales identify the correct false/true mapping; counterexample agents provide the decisive correction.
}

\caseentryword{Trace}{querybg}{queryframe}{Topology and skill-incidence trace}
{
\textbf{Nodes/execution order:}
\texttt{mLV2 -> hj9X -> wxdu -> 3M9o -> d7XF}.\\
\textbf{Slot 0 skills:}
\{\texttt{elimination}, \texttt{analogy}, \texttt{evidence\_check}\}.\\
\textbf{Slot 1 skills:}
\{\texttt{evidence\_check}\}.\\
\textbf{Slot 2 skills:}
\{\texttt{final\_synthesis}\}.\\
\textbf{Slot 3 skills:}
\{\texttt{counterexample}, \texttt{final\_synthesis}\}.\\
\textbf{Slot 4 skills:}
\{\texttt{counterexample}\}.\\
\textbf{Compact answers:} A, A, A, D, D.\\
\textbf{Activated edges:}
\texttt{spatial\_edges=[]}, \texttt{temporal\_edges=[]};
maximum decoded edge probability $=1.71{\times}10^{-3}<0.5$.\\
\textbf{Interpretation:}
the final decision follows the mathematical evidence rather than the noisy compact-answer majority.
}
\casecardsplit

\caseentry{picture/case/A0.png}{agent0bg}{agent0frame}{\texttt{mLV2}: elimination + analogy + evidence check}
{
\textbf{Mailbox input:} empty.\\
\textbf{Agent output:} A. The rationale says Statement 1 is false because Lagrange's theorem only states that subgroup orders divide the group order, not that every divisor appears as a subgroup order. It also says Statement 2 is true because the order of an element divides the group order.\\
\textbf{Diagnosis:} the reasoning supports false/true, but maps this evidence to option A instead of D.
}

\caseentry{picture/case/A1.png}{agent1bg}{agent1frame}{\texttt{hj9X}: evidence check}
{
\textbf{Mailbox input:} empty.\\
\textbf{Agent output:} A. The rationale identifies Statement 1 as false and Statement 2 as true, and explicitly notes that A is incorrect while D is the correct false/true option.\\
\textbf{Diagnosis:} noisy prefix, but the rationale explicitly identifies D.
}

\caseentry{picture/case/A2.png}{agent2bg}{agent2frame}{\texttt{wxdu}: final synthesis}
{
\textbf{Mailbox input:} empty.\\
\textbf{Agent output:} A. The response again states that Statement 1 is false and Statement 2 is true, then explains that the correct choice should be D.\\
\textbf{Diagnosis:} another prefix/rationale mismatch with correct mathematical evidence.
}

\caseentry{picture/case/A3.png}{agent3bg}{agent3frame}{\texttt{3M9o}: counterexample + final synthesis}
{
\textbf{Mailbox input:} empty.\\
\textbf{Agent output:} D. Statement 1 is false because not all finite groups have subgroups for every divisor of their order; Statement 2 is true by Lagrange's theorem.\\
\textbf{Diagnosis:} clean counterexample-style correction that directly supports the ground-truth label.
}

\caseentry{picture/case/A4.png}{agent4bg}{agent4frame}{\texttt{d7XF}: counterexample}
{
\textbf{Mailbox input:} empty.\\
\textbf{Agent output:} D. The response independently confirms the same false/true mapping: subgroup orders must divide the group order, but not every divisor must occur, while element order divides group order.\\
\textbf{Diagnosis:} independently confirms the correction and reinforces the minority compact answer.
}
\casecardsplit

\caseentry{picture/case/A5.png}{finalbg}{finalframe}{\texttt{UmWw}: \texttt{FinalRefer} decision node}
{
\textbf{System role:} top decision-maker; must output exactly one of A, B, C, or D.\\
\textbf{FinalRefer input:} task/options plus five node outputs: three compact prefixes A and two compact prefixes D, with most rationales supporting false/true.\\
\textbf{Final response:} D.\\
\textbf{Final result:} solved.\\
\textbf{Interpretation:} the final node follows the consistent mathematical evidence rather than the noisy compact-answer majority, showing that skill cards are treated as evidence sources rather than plain role labels.
}

\end{normalcasebox}
\casecaption{
A MMLU case where counterexample skills provide the decisive corrective signal and allow the final decision node to override a noisy compact-answer majority.
}{fig:case_mmlu_counterexample_corrects_majority}

\begin{normalcasebox}{Case C: Recoverable disagreement among skill-composed agents}

\caseentryword{Query}{querybg}{queryframe}{Question and expected answer}
{
\textbf{Record id:} \texttt{mmlu-test-public\_relations-3}. \quad
\textbf{Solved:} yes.\\
\textbf{Question:} Which of these organizations is most effective in engaging with customers online?\\
\textbf{Options:}
A. Starbucks \quad
B. Coca-Cola \quad
C. Wholefoods \quad
D. Redbull.\\
\textbf{Ground truth:} A.\\
\textbf{Final prediction:} A.\\
\textbf{Case focus:} heterogeneous skill-composed agents surface both Starbucks-supporting evidence and a plausible Red Bull distractor, but the final decision selects the better-supported answer.
}

\caseentryword{Trace}{querybg}{queryframe}{Topology and skill-incidence trace}
{
\textbf{Nodes/execution order:}
\texttt{mLV2 -> hj9X -> wxdu -> 3M9o -> d7XF}.\\
\textbf{Slot 0 skills:}
\{\texttt{elimination}, \texttt{analogy}, \texttt{evidence\_check}\}.\\
\textbf{Slot 1 skills:}
\{\texttt{evidence\_check}\}.\\
\textbf{Slot 2 skills:}
\{\texttt{final\_synthesis}\}.\\
\textbf{Slot 3 skills:}
\{\texttt{counterexample}, \texttt{final\_synthesis}\}.\\
\textbf{Slot 4 skills:}
\{\texttt{counterexample}\}.\\
\textbf{Compact answers:} A, A, D, A, D.\\
\textbf{Activated edges:}
\texttt{spatial\_edges=[]}, \texttt{temporal\_edges=[]};
maximum decoded edge probability $=1.96{\times}10^{-3}<0.5$.\\
\textbf{Interpretation:}
the aggregation remains stable despite disagreement because the final input contains enough grounded evidence for Starbucks.
}
\casecardsplit

\caseentry{picture/case/A0.png}{agent0bg}{agent0frame}{\texttt{mLV2}: elimination + analogy + evidence check}
{
\textbf{Mailbox input:} empty.\\
\textbf{Agent output:} A. The response argues that Starbucks has a strong online presence, including a user-friendly website, mobile app, and social media engagement.\\
\textbf{Evidence contribution:} supports Starbucks through integrated digital strategy and customer engagement.\\
\textbf{Diagnosis:} useful Starbucks-supporting evidence.
}

\caseentry{picture/case/A1.png}{agent1bg}{agent1frame}{\texttt{hj9X}: evidence check}
{
\textbf{Mailbox input:} empty.\\
\textbf{Agent output:} A. The rationale emphasizes Starbucks' mobile app with rewards, active social media engagement, and integrated online customer experience.\\
\textbf{Evidence contribution:} independently reinforces the same answer with additional customer-engagement evidence.\\
\textbf{Diagnosis:} useful aligned evidence.
}

\caseentry{picture/case/A2.png}{agent2bg}{agent2frame}{\texttt{wxdu}: final synthesis}
{
\textbf{Mailbox input:} empty.\\
\textbf{Agent output:} D. The response argues that Red Bull has a strong online presence, interactive content, and community engagement through platforms such as YouTube and social media.\\
\textbf{Evidence contribution:} surfaces a plausible distractor.\\
\textbf{Diagnosis:} useful for stress-testing the answer, but ultimately less supported than Starbucks.
}

\caseentry{picture/case/A3.png}{agent3bg}{agent3frame}{\texttt{3M9o}: counterexample + final synthesis}
{
\textbf{Mailbox input:} empty.\\
\textbf{Agent output:} A. The rationale argues that Starbucks excels through its mobile app, personalized offers, and social media interaction, while Coca-Cola and Red Bull also have strong digital presences.\\
\textbf{Evidence contribution:} compares the distractors against Starbucks and supports the correct option.\\
\textbf{Diagnosis:} useful synthesis that acknowledges alternatives without switching away from A.
}

\caseentry{picture/case/A4.png}{agent4bg}{agent4frame}{\texttt{d7XF}: counterexample}
{
\textbf{Mailbox input:} empty.\\
\textbf{Agent output:} D. The response favors Red Bull because of interactive content, community engagement, and digital innovation.\\
\textbf{Evidence contribution:} makes the plausible Red Bull distractor visible.\\
\textbf{Diagnosis:} noisy but informative disagreement.
}
\casecardsplit

\caseentry{picture/case/A5.png}{finalbg}{finalframe}{\texttt{UmWw}: \texttt{FinalRefer} decision node}
{
\textbf{System role:} top decision-maker; must output exactly one of A, B, C, or D.\\
\textbf{FinalRefer input:} task/options plus five node outputs containing three Starbucks-supporting cards and two Red Bull distractor cards.\\
\textbf{Final response:} A.\\
\textbf{Final result:} solved.\\
\textbf{Interpretation:} disagreement is recoverable because the final node receives enough grounded evidence for Starbucks while still seeing a plausible distractor. This is the desired behavior of skill diversity: disagreement is visible but not automatically decisive.
}

\end{normalcasebox}
\casecaption{
A MMLU case where disagreement among skill-composed agents is recoverable because the final mailbox contains enough grounded evidence for the correct answer.
}{fig:case_mmlu_recoverable_disagreement}

\begin{normalcasebox}{Case D: Shared misconception in a domain-specific legal question}

\caseentryword{Query}{querybg}{queryframe}{Question and expected answer}
{
\textbf{Record id:} \texttt{mmlu-test-international\_law-1}. \quad
\textbf{Solved:} no.\\
\textbf{Question:} Who is an ``injured State'' in the law of international responsibility?\\
\textbf{Options:}
A. A State is ``injured'' in case that it has suffered a damage from the internationally wrongful conduct.
B. A State is ``injured'' in cases that there has been a violation of a peremptory norm of international law.
C. A State is ``injured'' should it acknowledge the existence of the internationally wrongful conduct.
D. A State is ``injured'' if the obligation breached was owed to it individually or if it was owed to a group of States, including that State, and it was specially affected.\\
\textbf{Ground truth:} D.\\
\textbf{Final prediction:} A.\\
\textbf{Case focus:} all agents share the same false legal premise, so aggregation has no corrective evidence to follow.
}

\caseentryword{Trace}{querybg}{queryframe}{Topology and skill-incidence trace}
{
\textbf{Nodes/execution order:}
\texttt{mLV2 -> hj9X -> wxdu -> 3M9o -> d7XF}.\\
\textbf{Slot 0 skills:}
\{\texttt{elimination}, \texttt{analogy}, \texttt{evidence\_check}\}.\\
\textbf{Slot 1 skills:}
\{\texttt{evidence\_check}\}.\\
\textbf{Slot 2 skills:}
\{\texttt{final\_synthesis}\}.\\
\textbf{Slot 3 skills:}
\{\texttt{counterexample}, \texttt{final\_synthesis}\}.\\
\textbf{Slot 4 skills:}
\{\texttt{problem\_decomposition}, \texttt{final\_synthesis}\}.\\
\textbf{Compact answers:} A, A, A, A, A.\\
\textbf{Activated edges:}
\texttt{spatial\_edges=[]}, \texttt{temporal\_edges=[]};
maximum decoded edge probability $=1.70{\times}10^{-3}<0.5$.\\
\textbf{Interpretation:}
this failure reflects insufficient domain-specific evidence coverage rather than an answer-format issue.
}
\casecardsplit

\caseentry{picture/case/A0.png}{agent0bg}{agent0frame}{\texttt{mLV2}: elimination + analogy + evidence check}
{
\textbf{Mailbox input:} empty.\\
\textbf{Agent output:} A. The rationale defines an injured State as one that has suffered damage from an internationally wrongful act and treats option A as directly correct.\\
\textbf{Diagnosis:} shares the false premise that actual damage is the defining legal standard.
}

\caseentry{picture/case/A1.png}{agent1bg}{agent1frame}{\texttt{hj9X}: evidence check}
{
\textbf{Mailbox input:} empty.\\
\textbf{Agent output:} A. The response repeats that an injured State is a State that suffered damage from wrongful conduct, rejecting the other options as related but incorrect.\\
\textbf{Diagnosis:} aligned with the wrong premise; no corrective legal standard is introduced.
}

\caseentry{picture/case/A2.png}{agent2bg}{agent2frame}{\texttt{wxdu}: final synthesis}
{
\textbf{Mailbox input:} empty.\\
\textbf{Agent output:} A. The response synthesizes the same definition and chooses A as the most direct description.\\
\textbf{Diagnosis:} fluent synthesis amplifies the shared misconception.
}

\caseentry{picture/case/A3.png}{agent3bg}{agent3frame}{\texttt{3M9o}: counterexample + final synthesis}
{
\textbf{Mailbox input:} empty.\\
\textbf{Agent output:} A. The response argues that option A correctly defines the concept, while the other options misstate criteria or add unnecessary conditions.\\
\textbf{Diagnosis:} the counterexample skill does not recover the missing Article-style legal condition.
}

\caseentry{picture/case/A4.png}{agent4bg}{agent4frame}{\texttt{d7XF}: problem decomposition + final synthesis}
{
\textbf{Mailbox input:} empty.\\
\textbf{Agent output:} A. The response decomposes the definition, says option A captures damage from wrongful conduct, and rejects option D as overly complex.\\
\textbf{Diagnosis:} additional decomposition changes the prompt shape but still fails to retrieve the correct legal criterion.
}
\casecardsplit

\caseentry{picture/case/A5.png}{finalbg}{finalframe}{\texttt{UmWw}: \texttt{FinalRefer} decision node}
{
\textbf{System role:} top decision-maker; must output exactly one of A, B, C, or D.\\
\textbf{FinalRefer input:} task/options plus five aligned but wrong A-supporting node outputs.\\
\textbf{Final response:} A.\\
\textbf{Final result:} not solved.\\
\textbf{Interpretation:} this is a failure of evidence coverage. No card introduces the missing legal standard in option D, so the final node has no corrective signal to follow.
}

\end{normalcasebox}
\casecaption{
A MMLU failure case where skill diversity does not help because every card shares the same false legal premise.
}{fig:case_mmlu_shared_legal_misconception}

\begin{normalcasebox}{Case E: Correct majority lost during final aggregation}

\caseentryword{Query}{querybg}{queryframe}{Question and expected answer}
{
\textbf{Record id:} \texttt{mmlu-test-jurisprudence-37}. \quad
\textbf{Solved:} no.\\
\textbf{Question:} Which of the following criticisms of Llewellyn's distinction between the grand and formal styles of legal reasoning is the most compelling?\\
\textbf{Options:}
A. There is no distinction between the two forms of legal reasoning.
B. Judges are appointed to interpret the law, not to make it.
C. It is misleading to pigeon-hole judges in this way.
D. Judicial reasoning is always formal.\\
\textbf{Ground truth:} C.\\
\textbf{Final prediction:} A.\\
\textbf{Case focus:} three agents correctly support C, but the final node overweights a salient minority counterexample and outputs A.
}

\caseentryword{Trace}{querybg}{queryframe}{Topology and skill-incidence trace}
{
\textbf{Nodes/execution order:}
\texttt{mLV2 -> hj9X -> wxdu -> 3M9o -> d7XF}.\\
\textbf{Slot 0 skills:}
\{\texttt{elimination}, \texttt{analogy}, \texttt{evidence\_check}\}.\\
\textbf{Slot 1 skills:}
\{\texttt{evidence\_check}\}.\\
\textbf{Slot 2 skills:}
\{\texttt{final\_synthesis}\}.\\
\textbf{Slot 3 skills:}
\{\texttt{counterexample}, \texttt{final\_synthesis}\}.\\
\textbf{Slot 4 skills:}
\{\texttt{counterexample}\}.\\
\textbf{Compact answers:} C, C, C, A, D.\\
\textbf{Activated edges:}
\texttt{spatial\_edges=[]}, \texttt{temporal\_edges=[]};
maximum decoded edge probability $=2.06{\times}10^{-3}<0.5$.\\
\textbf{Interpretation:}
this case exposes a final-aggregation limitation: correct majority evidence can be lost when a wrong minority rationale appears forceful.
}
\casecardsplit

\caseentry{picture/case/A0.png}{agent0bg}{agent0frame}{\texttt{mLV2}: elimination + analogy + evidence check}
{
\textbf{Mailbox input:} empty.\\
\textbf{Agent output:} C. The rationale says option C criticizes the categorization of judicial reasoning as oversimplified, while A and D contradict each other and B is less relevant.\\
\textbf{Diagnosis:} useful correct evidence.
}

\caseentry{picture/case/A1.png}{agent1bg}{agent1frame}{\texttt{hj9X}: evidence check}
{
\textbf{Mailbox input:} empty.\\
\textbf{Agent output:} C. The response argues that option C challenges the validity of pigeon-holing judges into reasoning styles, while B concerns judicial role rather than the distinction itself.\\
\textbf{Diagnosis:} independently supports the correct option with aligned rationale.
}

\caseentry{picture/case/A2.png}{agent2bg}{agent2frame}{\texttt{wxdu}: final synthesis}
{
\textbf{Mailbox input:} empty.\\
\textbf{Agent output:} C. The response selects C as the most nuanced and relevant criticism because it argues the categorization oversimplifies judges' approaches.\\
\textbf{Diagnosis:} third aligned correct card.
}

\caseentry{picture/case/A3.png}{agent3bg}{agent3frame}{\texttt{3M9o}: counterexample + final synthesis}
{
\textbf{Mailbox input:} empty.\\
\textbf{Agent output:} A. The response claims option A directly challenges the core of Llewellyn's distinction by denying its existence and treats this as the strongest counterexample.\\
\textbf{Diagnosis:} salient but wrong minority rationale.
}

\caseentry{picture/case/A4.png}{agent4bg}{agent4frame}{\texttt{d7XF}: counterexample}
{
\textbf{Mailbox input:} empty.\\
\textbf{Agent output:} D. The response argues that option D challenges the distinction by asserting that judicial reasoning is always formal.\\
\textbf{Diagnosis:} another distractor card; it exposes an alternative challenge but does not match the best criticism.
}
\casecardsplit

\caseentry{picture/case/A5.png}{finalbg}{finalframe}{\texttt{UmWw}: \texttt{FinalRefer} decision node}
{
\textbf{System role:} top decision-maker; must output exactly one of A, B, C, or D.\\
\textbf{FinalRefer input:} task/options plus five node outputs: three C-supporting cards, one A-supporting counterexample card, and one D-supporting counterexample card.\\
\textbf{Final response:} A.\\
\textbf{Final result:} not solved.\\
\textbf{Interpretation:} this is an aggregation failure. The final node overweights a forceful but wrong minority counterexample and ignores that Agents 0--2 independently agree on C with aligned rationales.
}

\end{normalcasebox}
\casecaption{
A MMLU failure case where three agents identify the correct answer, but the final decision node follows a minority distractor.
}{fig:case_mmlu_correct_majority_lost}

\end{document}